\documentclass[reprint,amsmath,amssymb,aip,jcp]{revtex4-1}

\pdfoutput=1

\usepackage{hyperref,url}
\usepackage{amsmath}
\usepackage{amsfonts}
\usepackage{mathtools}
\usepackage[capitalise]{cleveref}
\usepackage{placeins}
\usepackage{mhchem}
\hypersetup{breaklinks,colorlinks=true,linkcolor=blue,citecolor=blue,urlcolor=blue,filecolor=blue}
\usepackage{graphicx}
\usepackage{dcolumn}
\usepackage{rotating}
\usepackage{etoolbox}

\newcommand\mat\mathbf

\newcommand{\rev}[1]{{\textcolor{black}{#1}}}

\newcommand{\Columbia}{\affiliation{Department of Chemistry, Columbia University, New York, NY, USA}}
\newcommand{\Cal}{\affiliation{Department of Chemistry, University of California, Berkeley, CA, USA}}
\newcommand{\QChem}{\affiliation{Q-Chem Inc., Pleasanton, CA, USA}}

\begin{document}

\author{Joonho Lee}
\email{jl5653@columbia.edu}
\Columbia
\author{Xintian Feng}
\QChem
\author{Leonardo A. Cunha}
\Cal
\author{J{\'e}r{\^o}me F. Gonthier}
\Cal
\author{Evgeny Epifanovsky}
\QChem
\author{Martin Head-Gordon}
\Cal
\title{Approaching the Basis Set Limit in Gaussian-Orbital-Based Periodic Calculations with Transferability: 
Performance of Pure Density Functionals for Simple Semiconductors}

\begin{abstract}
Simulating solids with quantum chemistry methods and Gaussian-type orbitals (GTOs) has been gaining popularity. Nonetheless, there are few systematic studies that assess the basis set incompleteness error (BSIE) in these GTO-based simulations over a variety of solids. In this work, we report a GTO-based implementation for solids, and apply it to address the basis set convergence issue. We employ a simple strategy to generate large uncontracted (unc) GTO basis sets, that we call the unc-def2-GTH sets. These basis sets exhibit systematic improvement towards the basis set limit as well as good transferability based on application to a total of 43 simple semiconductors. Most notably, we found the BSIE of unc-def2-QZVP-GTH to be smaller than 0.7 m$E_h$ per atom in total energies and 20 meV in band gaps for all systems considered here. Using  unc-def2-QZVP-GTH, we report band gap benchmarks of a combinatorially designed meta generalized gradient functional (mGGA), B97M-rV, and show that B97M-rV performs similarly (a root-mean-square-deviation (RMSD) of 1.18 eV) to other modern mGGA functionals, M06-L (1.26 eV), MN15-L (1.29 eV), and SCAN (1.20 eV). This represents a clear improvement over older pure functionals such as LDA (1.71 eV) and PBE (1.49 eV) though all these mGGAs are still far from being quantitatively accurate. We also provide several cautionary notes on the use of our uncontracted bases and on future research on GTO basis set development for solids. 
\end{abstract}
\maketitle
\newpage
\section{Introduction}
Condensed phase simulations 
using quantum chemistry tools originally developed for molecules
have gained popularity over many years,\cite{Ayala2001Dec,Katagiri2005Jun,Izmaylov2008, Pisani2008Oct, Pisani2012, DelBen2012Nov,Usvyat2015Sep,McClain2017Mar,Shang2020Nov}
with the hope of enabling development of
new systematically improvable
tools that can go beyond
standard density functional approaches,\cite{Maurer2019Jul} as well as existing Green's function methods\cite{Hybertsen1986Oct,Aryasetiawan1998Mar} in the field.
These simulations can
be broadly categorized into two classes:
(1) large $\Gamma$-point calculations to describe spatial inhomegeneity as found in gas, liquid, and surface simulations
and
(2) calculations with a relatively small unit cell and a large number of $\mathbf k$-points as relevant for simulations of solids.
The former category resembles
large cluster calculations that are routinely performed in the molecular community
and the use of Gaussian-type orbitals (GTOs) as a computational basis is not uncommon \rev{and numerically well-behaved}.
The use of GTOs to reach the thermodynamic limit (TDL) of \rev{(dense)} solids 
often faces numerical difficulties associated with overcompleteness of GTOs that leads to a severe linear dependency among basis functions towards the TDL.\cite{Lowdin1970Jan,Klahn1977Jun,VandeVondele2007Sep,Peintinger2013Mar}
Nonetheless, 
many studies have employed
Gaussian basis sets
either using those developed for molecular calculations,
those developed for periodic mean-field calculations,\cite{VandeVondele2007Sep, Peintinger2013Mar, Laun2018Jul, Oliveira2019Oct}
or
those optimized system-specifically without much in the way of transferability guarantees.\cite{Morales2020Nov,Daga2020Apr,Zhou2021May}
\rev{The use of GTOs for solid-state calculations has been growing as well exemplified by 
many existing GTO-based quantum chemistry programs with the periodic boundary condition capability.\cite{Li1995Feb,Soler2002Mar,Artacho2008Jan,Verzijl2012Nov,Kudin2000Jun,Balasubramani2020May,Dovesi2020May,Kuhne2020May,Prentice2020May} }

The development of compact GTO basis sets\cite{jensen2013atomic,nagy2017basis} has a long history in molecular quantum chemistry.\cite{Shavitt1993Jan,Pritchard2019Nov}
Since McWeeny's first proposal\cite{Mcweeny1950Jul} and Boys' early attempt\cite{Boys1950Feb} to use GTOs for molecular systems,
many developments on contracted Gaussian basis sets such as atomic natural orbital\cite{Almlof1987Apr}, correlation-consistent\cite{Dunning1989Jan} and polarization-consistent\cite{jensen2001polarization,jensen2002polarization} basis sets have
made high-accuracy quantum chemistry calculations practical.
However, these highly optimized contracted basis sets are usually not considered applicable to solids due to emerging linear dependencies.\cite{Peintinger2013Mar}
In the early days of basis set development, 
even-tempered\cite{Bardo1973Dec,Feller1979Sep} and well-tempered\cite{Huzinaga2011Feb} bases 
were explored
as a means to obtain high-quality results using only primitive GTOs reducing the complications in sophisticated optimization procedures for exponents and contraction coefficients.
In the even-tempered bases, one employs three parameters for each angular momentum shell $l$ to define a set of ``even-tempered'' primitive GTOs by
\begin{equation}
\phi_{lmk} (\mathbf r ) \propto \exp(-\zeta_{lk}r^2) r^l S_{lm}(\Omega)
\end{equation}
where $\phi_{lmk}$ is an atomic orbital, $l$ and $m$ are angular momentum quantum number, $S_{lm}(\Omega)$ are the real spherical harmonics at a solid angle $\Omega$,
$k$ sets the total number of primitive GTOs for $l,m$,
and $\zeta_{lk}$ is parameterized by a geometric series,
\begin{equation}
\zeta_{lk} = \alpha_l \beta_l^{k-1},\:\:\alpha_l,\beta_l >0,\:\:\beta_l\ne1
\label{eq:et}
\end{equation}
In the well-tempered variants, a more sophisticated form is used for $\zeta_{lk}$. 
In the even-tempered basis, one needs to pick a total of three parameters $k$, $\alpha_{lk}$, and $\beta_{lk}$.
The appropriate values may be obtained by looking at atoms and small molecules though finding these values can generally be tedious.\cite{Feller1979Sep}
Even-tempered basis sets are generally much larger than contracted GTOs and thus they are rarely used in modern quantum chemistry calculations.
Nonetheless, these bases have not yet been explored in the context of solid-state applications.

In this work, we propose an even simpler basis set generation protocol than that of even-tempered bases which does not involve any optimizations. 
Our procedure is to generate large uncontracted GTO bases that yield density functional theory (DFT) total energies per cell within 0.7 m$E_h$ per atom in the unit cell from the complete basis set limit obtained by planewave (PW) basis.
The idea is to take two existing GTO bases (one from the def2-series\cite{Weigend2005Aug} and SZV-MOLOPT-SR-GTH\cite{VandeVondele2007Sep}), uncontract these bases, and take the union of the resulting primitive GTOs while removing core orbitals that are treated by the underlying GTH pseudopotential.
Like the even-tempered bases, our sets are much larger than typical contracted GTOs available in the literature, but
they are not optimized for specific systems and/or mean-field methods so they should naturally bear transferability.

As an application of these bases, we focus on the basic goal of quantifying the basis set error of Gaussian-based DFT calculations.
This goal is even more important to reach when considering correlated wavefunction calculations. However, the basis set incompleteness error (BSIE) in correlation energies can be quantified and characterized only after the underlying mean-field energy is converged to the basis set limit.
The BSIE was directly quantified by employing the same pseudopotential proposed by Hutter and co-workers (called the GTH pseudopotential)\cite{Goedecker1996Jul,Hartwigsen1998Aug} in
both the new Gaussian-based program developed in this work 
and a PW-based code, Quantum Espresso (QE).\cite{giannozzi2020quantum}

Furthermore, we also apply our basis set to validating the performance of ten selected pure exchange-correlation (XC) functionals.
These ten XC functionals consist of 
one local density approximation (LDA) functional,\cite{slater1951simplification,perdew1981self} 
five generalized gradient approximation (GGA) functionals (PBE,\cite{perdew1996generalized} PBEsol,\cite{Perdew2008Apr} revPBE,\cite{Zhang1998Jan} BLYP,\cite{becke1988density,lee1988development} B97-D\cite{Grimme2006Nov}), and
four meta GGA (mGGA) functionals (SCAN,\cite{sun2015strongly} M06-L,\cite{Zhao2006Nov} MN15-L,\cite{Yu2016Mar} B97M-rV\cite{mardirossian2015mapping,mardirossian2017use}).
Our benchmark set has a total of 43 semiconductors where 40 of them were taken from the SC40 set\cite{heyd2005energy} and the remaining 3 (LiH,\cite{vidal1986accurate,nolan2009calculation,hoang2013lih} LiF,\cite{matsushita2011comparative,garza2016predicting} and LiCl\cite{matsushita2011comparative,garza2016predicting}) were taken from other places.
The performance of LDA and PBE on the majority of these systems using GTOs was already documented in ref. \citenum{heyd2005energy} though the underlying BSIE of the associated GTO basis sets is unclear.
Many PW-based codes including QE have LDA, GGA, and SCAN functionals available so it is not very difficult to assess their performance using PW-based codes.\cite{giannozzi2020quantum} In fact, the performance of LDA and GGA functionals, as well as the SCAN mGGA, is relatively well understood for band gap problems.\cite{perdew2017understanding,yang2016more}
However, the recently developed functionals that were combinatorially optimized for main group molecular chemistry, $\omega$B97X-V,\cite{mardirossian2014omegab97x} $\omega$B97M-V,\cite{mardirossian2016omega} and B97M-V,\cite{mardirossian2015mapping,mardirossian2017use} have rarely appeared in condensed phase studies\cite{pestana2017ab,ruiz2018quest,pestana2019diels,lininger2021challenges,li2021critical} and are relatively less common and less used in PW-based codes. The same is true for the Minnesota functionals (M06-L and MN15-L).
Replacing the -V tail with the -rV tail (the rVV10 van der Waals  (vdW) correction\cite{sabatini2013nonlocal} instead of the VV10 vdW correction\cite{vydrov2010nonlocal}), an efficient implementation of the -rV tail is now available in some planewave-based codes.\cite{giannozzi2020quantum}
Aside from the computational cost associated with the long-range exact exchange, an efficient implementation of these functionals should be readily possible.
These combinatorially optimized functionals were found to be statistically the best XC functionals at each rung of Jacob's ladder for main group chemistry problems,\cite{mardirossian2017thirty}, and they have performed very well in other molecular benchmarks also.\cite{Goerigk:2017,Najibi:2018b} In the condensed phase, the mGGA, B97M-rV appears to describe properties of liquid water as accurately as far more computationally demanding hybrid functions.\cite{ruiz2018quest} However, the performance of B97M-rV for band gap problems is largely unknown at present.
Motivated by this, we report the performance of B97M-rV for band gaps here.

This paper is organized as follows:
(1) we first review basic formalisms of periodic mean-field calculations, the gaussian planewave (GPW) density fitting scheme, and an efficient implementation of rVV10,
(2) we then describe our strategies for generating transferable GTO bases for simulating solids towards the TDL,
(3) we discuss computational details,
(4) we present results for basis set convergence of DFT total energies and band gaps using the proposed bases,
(5) we assess the performance of pure XC functionals comparing against experimental band gaps,
(6) we deliver cautionary notes on using our bases and on the future basis set development for solids featuring striking failures of existing GTH bases,
and
(7) we then conclude.

\section{Theory}
Periodic mean-field calculations using a linear combination of atomic orbitals have been well-documented in many places.\cite{Evarestov2007,Evarestov2012}
Nonetheless, we
aim to give a pedagogical review of the relevant theories on periodic DFT calculations within the GPW implementation and the implementation of rVV10 since these are the key compute kernels in our new implementation.
Experienced readers may skip some of the subsequent sections and start from \cref{sec:gto}.

\subsection{Periodic Mean-Field Calculations}
As a consequence of real-space translational symmetry, crystal momentum ($\mathbf{k}$) is a good quantum number. Periodic mean-field (PMF) calculations with GTOs are hence done with
crystalline molecular orbitals (CMOs), $\{\psi_i^{\mathbf k}\}$,\cite{kudin1998fast}
\begin{equation}
\psi_i^{\mathbf k}(\mathbf r) = 
\sum_\mu C_{\mu i}^{\mathbf k} \phi_\mu^{\mathbf k} (\mathbf r)
\end{equation}
where crystalline atomic orbitals (CAOs) are defined with a lattice summation,
\begin{equation}
\phi_\mu^{\mathbf k} (\mathbf r)
=
\sum_{\mathbf R}
\phi_\mu^{\mathbf R} (\mathbf r) e^{i\mathbf k \cdot \mathbf R}
\end{equation}
\rev{where $\mathbf R$ denotes a lattice vector represented by a sum of integer multiples of primitive vectors of the direct lattice.}
In PMF calculations, analogously to their molecular counterparts, the PMF energy is minimized when the CMO coefficient matrix obeys a self-consistent Roothaan-Hall equation,
\begin{equation}
\mathbf F^{\mathbf k}
\mathbf C^{\mathbf k}
=
\mathbf S^{\mathbf k}
\mathbf C^{\mathbf k}
\mathbf \epsilon^{\mathbf k}
\end{equation}
where $\mathbf F^{\mathbf k}$ is the Fock matrix at $\mathbf k$,
$\mathbf S^{\mathbf k}$ is the overlap matrix of CAOs at $\mathbf k$ defined as
\begin{equation}
S_{\mu\nu}^{\mathbf k}
=
\sum_{\mathbf R}\langle{\phi}_{\mu}^\mathbf{0}|{\phi}_{\nu}^\mathbf{R}\rangle e^{i\mathbf k \cdot \mathbf R}
=
\sum_{\mathbf R}S_{\mu\nu}^{\mathbf 0\mathbf R}e^{i\mathbf k \cdot \mathbf R},
\end{equation}
and $\mathbf \epsilon^{\mathbf k}$ is the band energy at $\mathbf k$.

In periodic calculations with GTOs, it is very common to 
observe linear dependencies of the CAOs which makes the metric (overlap) matrix $\mathbf S^{\mathbf k}$ poorly conditioned.
Within finite precision computer arithmetic, the resulting truncation error in the inverse metric can lead to numerical instability, convergence issues, and non-trivial errors in the PMF energies.
Therefore, handling exact and near linear dependencies is crucial in GTO-based periodic calculations
especially when one attempts to get to the basis set limit where linear dependencies become progressively severe.
In this work, we adopted the canonical orthogonalization procedure.\cite{Szabo1996Jul}
The canonical orthogonalization procedure is defined as follows:
\begin{enumerate}
\item The diagonalization of $\mathbf S^{\mathbf{k}}$ is performed for each $\mathbf k$:
\begin{equation}
\mathbf S^{\mathbf{k}}
=
\mathbf U^{\mathbf{k}} \mathbf s^\mathbf{k} (\mathbf U^{\mathbf{k}})^\dagger
\end{equation}
\item For a given threshold $\epsilon_\text{lindep}$, one retains the $N_\text{CMO}^{\mathbf k}$ 
eigenvalues in $\mathbf s^\mathbf{k}$ above $\epsilon_\text{lindep}$ along with their corresponding eigenvectors.
We refer these subsets of eigenvalues and eigenvectors to as $\tilde{\mathbf s}^\mathbf{k}$ and $\tilde{\mathbf U}^\mathbf{k}$, respectively.
\item We then define the orthogonalization matrix $\mathbf X^{\mathbf k}$,
\begin{equation}
\mathbf X^{\mathbf k} = 
\tilde{\mathbf U}^{\mathbf{k}} (\tilde{\mathbf s}^\mathbf{k})^{-1/2}
\end{equation}
The dimension of $\mathbf X^{\mathbf k}$ is $N_\text{CAO}$-by-$N_\text{CMO}^{\mathbf k}$
and $N_\text{CMO}^{\mathbf k}$ is the dimension of the effective variational space after removing numerical linear dependencies.
We note that we then have
\begin{equation}
(\mathbf X^{\mathbf k})^\dagger
\mathbf S^{\mathbf k}
\mathbf X^{\mathbf k}
=
\mathbf I_{N_\text{CMO}^{\mathbf k}}
\end{equation}
\end{enumerate}
The choice of $\epsilon_\text{lindep}$ should be made so as to balance between
numerical stability (i.e., removing enough basis functions to avoid excessive roundoff error and precision loss) and quality of the resulting basis set (i.e., keeping as many basis functions as possible).
We picked $\epsilon_\text{lindep}$ to be $10^{-6}$ which is the default value of our molecular computations in Q-Chem.\cite{Shao2015Jan,Epifanovsky2021Aug}
We note that this linear dependency threshold is chosen to be reasonable for double precision, and could be tightened up if one could afford quadruple or higher precision arithmetic.

\subsection{Review of the GPW algorithm}
The GPW density fitting algorithm was first proposed by Hutter and co-workers\cite{lippert1997hybrid} and has been popularized via 
the implementation in CP2K.\cite{vandevondele2005quickstep,Kuhne2020May}
The central idea of the algorithm is that
one employs
planewaves as the auxiliary basis set
for density-fitting while using GTOs as the primary computational basis set.
This strategy is particularly well-suited for solid-state calculations
since periodic boundary conditions are naturally imposed and planewave density fitting can be done efficiently.
\rev{While applying GPW to three-dimensional (3D) systems is the most straightforward, lower-dimensional systems 
(0D, 1D, and 2D) need special attention to remove spurious image-image interactions.
The application of GPW was successfully carried out by F{\"u}sti-Moln{\'ar} and Pulay\cite{Fusti-Molnar2002May,Fusti-Molnar2002Nov} for molecules (i.e., 0D)
where spurious image-image interactions were removed exactly by using a truncated Coulomb potential. A similar idea can be generalized to 1D and 2D.\cite{Rozzi2006May}}

Among various terms in $\mathbf F^{\mathbf k}$, in this work, we focus on the Coulomb matrix, $\mathbf J^{\mathbf k}$,
because this contribution is typically the computational bottleneck in pure DFT calculations.
We want to compute
the Coulomb matrix element between
a basis function $\phi_\mu$ located in a unit cell $\mathbf R = \mathbf{0}$ (denoted as $\phi_\mu^{\mathbf 0}$) and 
a basis function $\phi_\nu$ located in a unit cell $\mathbf R$ (denoted as $\phi_\nu^{\mathbf R}$),
\begin{align}\nonumber
J_{\mu\nu}^{\mathbf 0\mathbf R}
&\equiv
\int_{\mathbf{r}}
\phi_\mu^\mathbf{0}(\mathbf{r})
V_J(\mathbf{r})
\phi_\nu^\mathbf{R}(\mathbf{r})\\
&=
\sum_{\mathbf{R'}}
\int_{\mathbf{r}\in\mathbf{R'}}
\phi_\mu^\mathbf{0}(\mathbf{r})
V_J(\mathbf{r}-\mathbf{R'})
\phi_\nu^\mathbf{R}(\mathbf{r})
\end{align}
where $V_J(\mathbf{r})$ is the Coulomb potential defined as
\begin{equation}
V_{J}(\mathbf{r}) = 
\int_{\mathbf{r'}}
\frac{\rho(\mathbf{r'})}
{\mathbf{r}-\mathbf{r'}}
\end{equation}
and we used the fact that $V_J(\mathbf{r})$ is periodic in the unit cell displacements.
We note that $\mathbf r\in \mathbf R'$ implies that the domain of the integration is restricted to the unit cell centered at $\mathbf R'$.

The evaluation of $V_{J}(\mathbf{r})$ can be done with $\mathcal O(N_g \log N_g)$ complexity via the fast Fourier transform (FFT) algorithm for discrete Fourier transform where $N_g$ is the number of grid points within the simulation cell.
In reciprocal space,
\begin{equation}
V_{J}(\mathbf{G})
=
\frac{4\pi}{|\mathbf G|^2}
\tilde{\rho}(\mathbf G)
\label{eq:kernel}
\end{equation}
where
\begin{equation}
\tilde{\rho}(\mathbf G) = \frac1\Omega\int_{\mathbf r} \rho(\mathbf r) e^{i \mathbf G\cdot \mathbf r}
\end{equation}
with $\Omega$ being the volume of the computational unit cell.
Using these, the GPW algorithm computes $\mathbf J^{\mathbf k}$ as follows:
\begin{enumerate}
\item
Compute $\rho(\mathbf r)$ within a unit cell via
\begin{equation}
\rho(\mathbf r)
=
\frac{1}{N_k}\sum_{\mathbf k}\sum_i \sum_{\mu\nu}
C_{\mu i}^{\mathbf k}
(C_{\nu i}^{\mathbf k})^*
\phi_\mu^{\mathbf k}(\mathbf r)
(\phi_\nu^{\mathbf k}(\mathbf r))^*
\end{equation}
where $N_k$ is the number of k-points.
\item Fourier transform $\rho(\mathbf r)$ to obtain $\tilde{\rho}(\mathbf G)$.
This is the ``density-fitting'' step using a planewave auxiliary basis set.
\item Compute the Coulomb potential in reciprocal space via \cref{eq:kernel} and inverse Fourier transform to obtain $V_{J}(\mathbf r)$.
Note that we ignore the $|\mathbf G| = 0$ component.
\item Compute $\mathbf J^{\mathbf k}$ via
\begin{equation}
J_{\mu\nu}^{\mathbf k}
=
\int_{\mathbf r \in \text{U.C.}}
(\phi_{\mu}^{\mathbf k}(\mathbf r))^*
V_J(\mathbf r)
\phi_{\nu}^{\mathbf k}(\mathbf r)
\end{equation}
where the quadrature is performed only within the unit cell (U.C.).
\end{enumerate}
Our implementation computes $\phi_\mu^{\mathbf k}(\mathbf r)$ once in the beginning and stores these in memory.
Therefore, our GPW implementation for the J-build has
$\mathcal O(N_k N_g)$ storage cost (due to storing $\phi_{\mu}^{\mathbf k}(\mathbf r)$)
and $\mathcal O(N_k N_g + N_g \log N_g)$ compute cost assuming sparsity of CAOs. Since $N_g$ scales with the unit cell volume while $N_k$ does not, this algorithm approaches $\mathcal O(N)$ scaling. Diagonalization is performed by dense linear algebra with $\mathcal O(N^3)$ scaling. 

\subsection{Summary of implementation of rVV10}
Some of the more modern density functionals use the VV10 vdW correction, but the cost of evaluating VV10 scales quadratically with system size.
Using ideas from the work of Rom{\'a}n-P{\'e}rez and Soler,\cite{roman2009efficient} Sabatini and others proposed an alternative functional form called rVV10\cite{sabatini2013nonlocal} which can be implemented efficiently with linear complexity
for planewave codes while retaining similar accuracy as VV10.
Subsequently, the use of rVV10 was verified for combinatorially optimized density functionals (B97M-V, $\omega$B97X-V, and $\omega$B97M-V)
leading to B97M-rV, $\omega$B97X-rV, and $\omega$B97M-rV.\cite{mardirossian2017use}
We are interested in investigating the performance of these combinatorially optimized functionals for band gaps so an efficient implementation of
rVV10 is highly desirable.

The rVV10 energy functional reads\cite{vydrov2010nonlocal,sabatini2013nonlocal}
\begin{equation}
    E_\text{rVV10} = E^\text{local}_\text{rVV10} + E^\text{non-local}_\text{rVV10}
\end{equation}
where the local part can be absorbed into the local density approximation terms and the non-local part 
poses implementational challenges with a na{\'i}ve quadratic scaling cost.
The non-local contribution
is defined as
\begin{equation}
    E^\text{non-local}_\text{rVV10} = \frac{1}{2} \int_{\mathbf{r}}\int_{\mathbf{r'}} 
    \rho(\mathbf{r})\kappa(\mathbf{r})^{-3/2} 
    \rho(\mathbf{r'})\kappa(\mathbf{r'})^{-3/2}
    \Phi(\mathbf{r, r'})
\label{eq:rvvnlc}
\end{equation}
where $\rho(\mathbf r)$ is the electron density, and 
the kernel $\Phi(\mathbf r, \mathbf r')$ is
\begin{equation}
    \Phi(\mathbf{r, r'}) = \frac{-1.5}{(q(\mathbf{r})R^2+1)(q(\mathbf{r'})R^2+1)(q(\mathbf{r})R^2+q(\mathbf{r'})R^2+2)}
\end{equation}
with $R = |\mathbf r - \mathbf r'|$. The remaining terms are 
\begin{equation}
    q(\mathbf{r}) = \kappa(\mathbf{r})^{-1}\sqrt{C|\frac{\nabla\rho(\mathbf{r})}{\rho(\mathbf{r})}|^4+\frac{4}{3}\pi\rho(\mathbf{r})},
\end{equation}
and
\begin{equation}
    \kappa(\mathbf{r}) = 1.5b\pi(\frac{\rho(\mathbf{r})}{9\pi})^{1/6}.
\end{equation}
The fixed parameters $b$ and $C$ are a part of the definition of
each XC functional that includes the rVV10 contribution.
The evaluation of this 
leads to an overall quadratic scaling in $N_g$ due to its six-dimensional double integral in \cref{eq:rvvnlc}.

As discussed in ref. \citenum{roman2009efficient}, we first use cubic splines to interpolate $\Phi$ such that
\begin{equation}
    \Phi(\mathbf{r, r'}) \approx \sum_{\alpha,\beta} \Phi(q_\alpha,q_\beta,R)
    p_\alpha(q(\mathbf{r})) p_\beta(q(\mathbf{r'}))
\end{equation}
where $q_\alpha$ and $q_\beta$ are interpolation points and $p_\alpha$ and $p_\beta$ are interpolating polynomials.
This makes the evaluation of $\Phi$ computationally convenient because $\Phi$ becomes a function of only $|\mathbf{r-r'}|$.
Its dependence on $q_\alpha$ and $q_\beta$ is easy to handle as $q_\alpha$ and $q_\beta$ are fixed interpolation points.
The number of the interpolation points is also very manageable as it is typically set to 20.\cite{sabatini2013nonlocal}
We now define an intermediate,
\begin{equation}
    \theta_\alpha(\mathbf{r}) =  \rho(\mathbf{r})\kappa(\mathbf{r})^{-3/2}p_\alpha(q(\mathbf{r}))
\end{equation}
and use it to recast the non-local energy contribution into a convolution form:
\begin{align}\nonumber
    E^\text{non-local}_\text{rVV10} 
    &= 
    \frac{1}{2} \sum_{\alpha,\beta}  \int_{\mathbf{r}}\int_{\mathbf{r'}} 
    \theta_\alpha(\mathbf{r}) \theta_\beta(\mathbf{r'})
    \Phi_{\alpha\beta}(|\mathbf{r-r'}|)\\
&    = 
    \frac{1}{2} \sum_{\alpha,\beta} \int_{\mathbf{G}} \tilde{\theta}_\alpha(\mathbf{G})
    \tilde{\theta}_\beta(\mathbf{G}) \tilde{\Phi}_{\alpha\beta}(|\mathbf{G}|)
 \label{eq:nonlocal}
\end{align}
Since $\Phi_{\alpha\beta}(|\mathbf{r-r'}|) = \Phi_{\alpha\beta}(R)$ is spherically symmetric, its Fourier transform can be computed by one-dimensional Fourier-sine transformation. 
The values of $\tilde{\Phi}_{\alpha\beta}(|\mathbf G|)$ on a pre-defined set of $|\mathbf{G}|$ points can be precomputed and tabulated.
These tabulated values are then used for interpolation to perform the convolution in \cref{eq:nonlocal} for a specific set of $|\mathbf G|$.
We note that the convolution in \cref{eq:nonlocal} can be performed in $\mathcal O(N_g\log N_g)$ time as opposed to the quadratic-scaling runtime of the na{\'i}ve algorithm.
A similar approach can be used to compute the Fock matrix contribution associated with rVV10.

\subsection{Strategies for assessing the basis set error for simple solids and generating transferable GTOs} \label{sec:gto}
Our goal in this work is to
access the near basis set limit of pure density functionals for solids using GTOs. 
For this purpose, it is critical to have 
well-defined basis set limit reference values.
A popular planewave code, Quantum Espresso (QE), also implements the GTH pseudopotential\cite{Goedecker1996Jul} developed by Hutter and co-workers, which was originally developed for use in the CP2K program.\cite{Kuhne2020May} We have adopted the same GTH pseudopotential for use in our code as well.
This allows for a direct comparison between QE and our code, which is particularly useful because QE can converge the total energy to the basis set limit almostly completely by increasing the planewave cutoff.

We considered the 40 semiconductors benchmark set (SC40) first proposed by Scuseria and co-workers\cite{heyd2005energy} along with three rocksalt solids
(LiH,\cite{vidal1986accurate,nolan2009calculation,hoang2013lih} LiF,\cite{matsushita2011comparative,garza2016predicting} LiCl\cite{matsushita2011comparative,garza2016predicting}).
For these compounds, 
all-electron GTO basis sets have been proposed
but
their accuracy remains largely unknown.\cite{heyd2005energy}
Moreover, to be used with the GTH pseudopotential, we need a basis set without core electrons.
Unfortunately,
the standard GTH basis set series does not have a broad coverage of the periodic table beyond its minimal basis set (SZV-GTH).\cite{VandeVondele2007Sep}
To access the basis set limit for a variety of solids considered in this work, we propose a simple way to 
generate a large basis set which yields the total energy per cell close to the basis set limit (errors smaller than 0.7 $mE_h$ per atom for DFT calculations performed here, as will be shown later).
We also note that this same strategy of uncontracting existing GTO bases can be applied to the generation of all-electron bases as well.

To generate the basis set, we follow a straightforward procedure:
\begin{enumerate}
\item
We take the existing def2-bases and uncontract the contracted GTOs therein.
We then remove GTOs with an exponent greater than 20 since they correspond to core electrons that are 
already covered by the GTH pseudopotential.
This cutoff of 20 was empirically determined and we expect that the results are not sensitive to the precise value of the cutoff given the large size of our final basis set (see below for more discussion).
\item 
We take the union of these uncontracted def2 bases and the uncontracted SZV-MOLOPT-SR-GTH basis set
to enhance the resolution within the minimal basis set space defined by the GTH pseudopotential.
This final basis set will be referred to as
unc-def2-X-GTH where X can be SVP, TZVP, QZVP, etc.
\end{enumerate}
One may think that having a fixed cutoff of 20 for all elements could be unphysical because increasing the atomic number tends to increase all of GTO exponents. In our case, however, the GTOs from def2 bases with an exponent larger than the largest exponent in SZV-MOLOPT-SR-GTH belong to the core region that is already treated by the GTH pseudopotential. 
Inspecting the range of exponents in SZV-MOLOPT-SR-GTH basis, one finds that the largest ones are smaller than 20 with the exception of Na (23.5) and Mg (30.7) up to atomic number 86. Based on our results on solids involving Mg, the cutoff of 20 works well for this element as well.
Overall, the contraction of electron density due to the increase in the nuclear charge is reflected appropriately and there is no sensitivity stemming from this cutoff. 
We also note that
when taking the union of two bases some of the
exponents can be very close in value, but 
for simplicity
we do not remove those obvious near-linear-dependencies. Instead we let the canonical orthogonalization procedure take care of them.
We report these unc-def-GTH bases (unc-def2-SVP-GTH,
unc-def2-SVPD-GTH,
unc-def2-TZVP-GTH,
unc-def2-TZVPD-GTH,
unc-def2-TZVPP-GTH,
unc-def2-TZVPPD-GTH,
unc-def2-QZVP-GTH,
unc-def2-QZVPD-GTH,
unc-def2-QZVPP-GTH,
unc-def2-QZVPPD-GTH) through the Zenodo repository,\cite{zenodo} as well as in the text files included in the final publication.

With regard to the existing GTH-based contracted GTO basis sets, at present neither the
range of Gaussian exponents
nor the 
contraction coefficients
have been specifically optimized to approach the basis set limit: rather they have been designed to offer a good trade-off between compute cost and accuracy for solid-state applications.
The use of uncontracted basis functions in this work
is an attempt to probe
the suitability of using a broad range of Gaussian exponents and angular momenta
while obtaining the contraction coefficients
via
variational energy minimization (i.e., the MO coefficients are the contraction coefficients in our case).
As a consequence of decontraction, our proposed basis sets range from quite large to very large and are heavily linearly dependent.
Nonetheless, this brute force approach will permit us to
assess systematic convergence of our total energies towards the basis set limit energies obtained through QE.
We emphasize that potential numerical instability issues are quite well handled by the simple canonical orthogonalization procedure.

Last but not least, we note that our Gaussian basis set generation procedure does not utilize any system-dependent parameters or optimization protocols.
As evidenced by even-tempered bases,\cite{Bardo1973Dec,Feller1979Sep} this is particularly important for ensuring transferability.
Our exponents retain both tight exponents that are effective for condensed phase and
relatively diffuse exponents that are effective for atomic (or molecular) limits. 
Therefore, we expect that the BSIE is relatively insensitive to the underlying geometry.
Nonetheless, when atoms come too close together, GTOs are expected to perform more poorly due to the higher degree of linear dependence as will be shown later in \cref{sec:cold}.

\section{Computational details}\label{sec:comp}
We implemented a GPW-based periodic DFT code in a development version of Q-Chem.\cite{Shao2015Jan,Epifanovsky2021Aug}
Our implementation assumes overall $\mathcal O(N^2)$ memory storage that amounts to storing Fock, density, and CMO coefficient matrices. For the systems examined in this work, our memory strorage is dominated by keeping the
GTO basis function values evaluated on the FFT grid despite its formal linear scaling based on sparsity.
In the future, this practical memory bottleneck can be removed by computing these on the fly.
We also note that our GPW algorithms scale linearly with the system size to produce the Fock matrix and our SCF program scales cubically with system size due to linear algebra functions such as matrix diagonalization.
We control the resolution of the PW density fitting basis with a single parameter: the kinetic energy cutoff ($E_\text{cut}$).
Each auxiliary basis PW can be indexed by 3 integers, $(n_1,n_2,n_3)$ which reside on a $(2n_1^\text{max}-1)\times (2n_2^\text{max}-1) \times (2n_3^\text{max}-1)$ grid where each integer $n_i\in \{-n_i^\text{max}, \cdots, n_i^\text{max}\}$ with
\begin{equation}
n_i^\text{max} = \frac{\sqrt{8 E_\text{cut}}} { ||\mathbf{b}_i||}
\end{equation}
where $\mathbf{b}_i$ denotes one of the reciprocal vectors.
For our GPW calculations, we used $E_\text{cut}$ of 1500 eV for everything except
those that contain Ba (2000 eV) and Mg (4500 eV). The resulting density fitting error was found to be smaller than 100 $\mu E_h$ per atom in the unit cell, which is negligible for the purpose of this paper.

The reference planewave basis calculations were all performed with QE where
we used $E_\text{cut}$ (for the wavefunction itself) of 1200 Ry for total energy calculations.
For the band structure calculations, 
we used $E_\text{cut}$ of 1200 Ry for systems containing Mg
and
$E_\text{cut}$ of 750 Ry for everything else.

The lattice constants were fixed at experimental values \cite{heyd2005energy} and
experimental band gaps for the SC40 set were taken from \citenum{heyd2005energy}.
The experimental band gaps and lattice constants of LiH, LiF, and LiCl were taken from refs. \citenum{vidal1986accurate,nolan2009calculation,hoang2013lih,matsushita2011comparative,garza2016predicting}.
We used the GTH-LDA pseudopotential in all calculations for both GPW and PW (through QE) calculations to enable direct comparison of total energies.
We used the Monkhorst-Pack\cite{monkhorst1976special} $\mathbf k$-mesh to sample the first Brillouin zone
and ensured the convergence of the total energy per cell to the TDL for all solids examined here.
We found that a $6\times6\times6$ $\mathbf k$-mesh is enough to converge the total energy per cell to an error of smaller than 0.1 mH for all solids considered. 
Therefore, for band structure calculations and cold curve calculations, we used a $6\times6\times6$ Monkhorst-Pack $\mathbf k$-mesh.
Since the GPW implementation is also available in other open-source packages such as CP2K\cite{Kuhne2020May} and PySCF,\cite{sun2020recent} 
we also used these two packages to validate our implementation in the initial stage of this work.

We examined a total of ten XC functionals, 
LDA (Slater exchange\cite{slater1951simplification} and PZ81 correlation\cite{perdew1981self}), PBE,\cite{perdew1996generalized} PBEsol,\cite{Perdew2008Apr} revPBE,\cite{Zhang1998Jan} BLYP,\cite{becke1988density,lee1988development} B97-D,\cite{Grimme2006Nov} SCAN,\cite{sun2015strongly} M06-L,\cite{Zhao2006Nov} MN15-L,\cite{Yu2016Mar}  and B97M-rV.\cite{mardirossian2015mapping,mardirossian2017use}
For 11 solids in our benchmark set (C, Si, SiC, BN, BP, AlN, MgO, MgS, LiH, LiF, LiCl), 
widely used GTH basis sets\cite{VandeVondele2007Sep} are available:
DZVP-GTH, TZVP-GTH, TZV2P-GTH, QZV2P-GTH, and QZV3P-GTH.
We therefore assessed the accuracy of those existing bases only over
a smaller subset of our benchmark set, but our proposed basis sets were examined for all 43 solids considered in this work.
The basis set convergence study against PW was carried out only for LDA and PBE while the overall band gap accuracy was examined for all ten functionals.

\section{Results and Discussion}
\subsection{Basis set convergence of total DFT energies}\label{sec:total}
\begin{figure}[!ht]
    \centering
\includegraphics[scale=0.43]{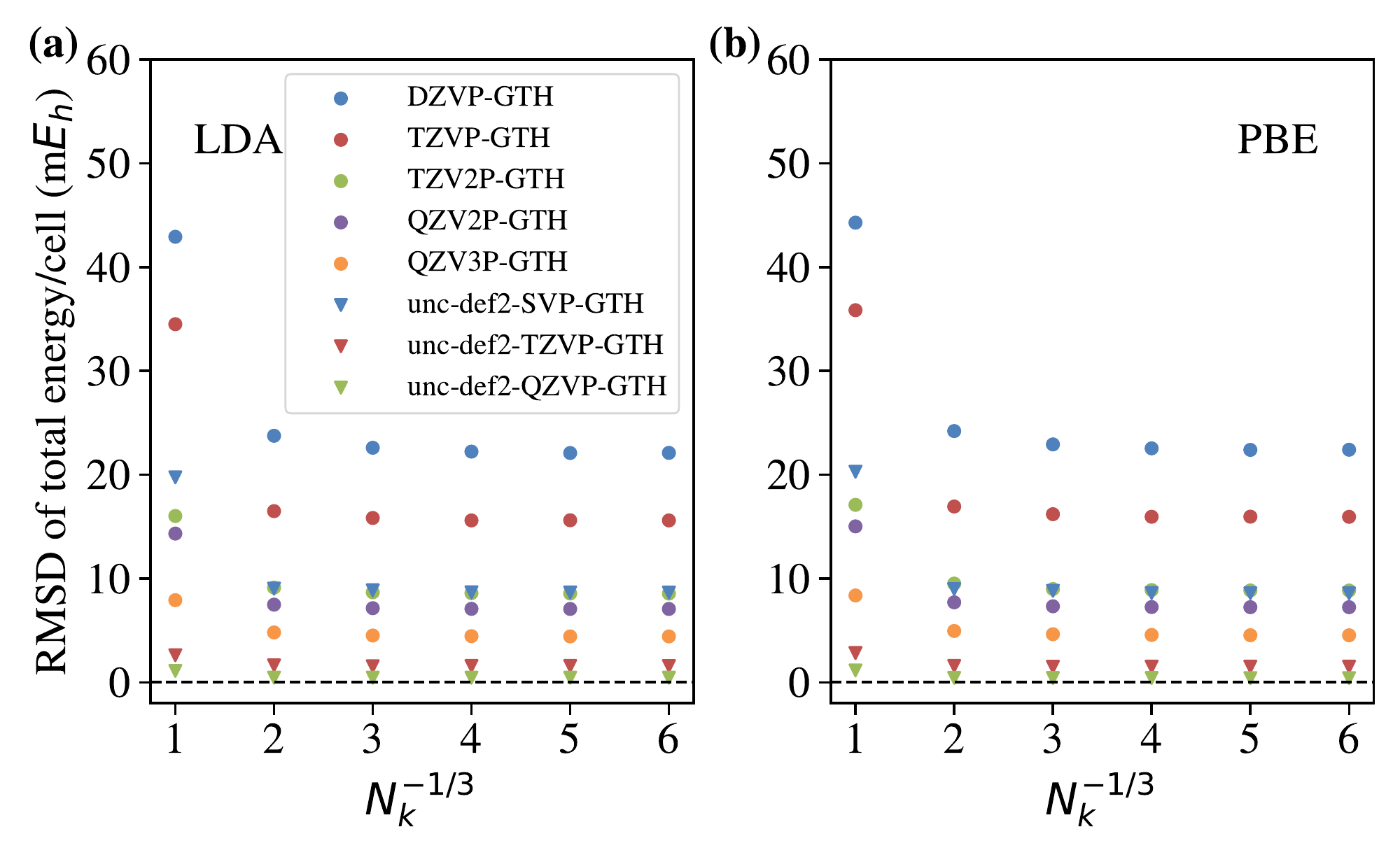}
    \caption{Root mean square deviation (RMSD) of DFT total energies (m$E_h$) per cell with respect to that of QE over 11 solids as a function of the number of $\mathbf k$-points for (a) LDA and (b) PBE functionals using GTH and unc-def2-GTH bases. }
    \label{fig:fig1}
\end{figure}
We first examine 
the subset of 11 solids for LDA and PBE functionals as presented in \cref{fig:fig1}.
In particular, \cref{fig:fig1} shows  
the root-mean-square-deviation (RMSD) of total energies compared to QE total energies (namely total energies in the basis set limit) as function of the size of the $\mathbf k$-mesh. 
$N_k = 216$ ($6\times6\times6$) is enough to reach the TDL.
\rev{
While smaller $\mathbf k$-mesh calculations such as the $\Gamma$-point only calculations are unphysical,
we are interested in how the basis set error changes as a function of the $\mathbf k$-mesh size.
Since both PW and our GTO results share the same Hamiltonian for each size of $\mathbf k$-mesh,
we can assess the basis set error of GTO calculations as long as the reference PW energies are fully converged to the basis set limit.}
For all $\mathbf k$-mesh sizes, the GTH basis set series shows systematically more accurate results relative to the basis set limit as cardinality (and the size of the basis set) increases.
We note that $N_k = 1$ (just including the $\Gamma$-point) shows the largest basis set error in all examples considered here. 
This is because the local expressive power of GTOs also increases
as one increases the size of $\mathbf k$-mesh due to the non-orthogonality of GTOs.
While the systematic improvement of GTH bases is very appealing, we note that the residual basis set error with QZV3P-GTH is still about 5 m$E_h$ which is quite large considering how small the simulation cells are (2 or 4 atoms total).

The unc-def2-GTH basis series also shows a systematic improvement with cardinal number, with much smaller errors than the corresponding contracted GTH basis results. 
As an example, the performance of unc-def2-SVP-GTH is nearly on par with TZV2P-GTH except at the $\Gamma$-point.
The larger bases, unc-def2-TZVP-GTH and unc-def2-QZVP-GTH, both perform excellently on this set, including the $\Gamma$-point result.
In particular, unc-def2-QZVP-GTH is able to deliver total energies in the TDL that are all within 1 m$E_h$ of the basis set limit.
This shows the completeness of our proposed bases though of course these are much bigger in size than standard GTH bases, and therefore far more computationally demanding. We provide more detailed information on selected elements in \cref{subsec:size}.
Finally, we note that \cref{fig:fig1} (a) for LDA and (b) for PBE show virtually no difference, which suggests that our conclusions do not depend on functional (of course functionals that depend particularly strongly on fine details of the density may be far harder to converge to the basis set limit using GTOs\cite{mardirossian2013characterizing}).

\begin{figure}[!ht]
    \centering
\includegraphics[scale=0.43]{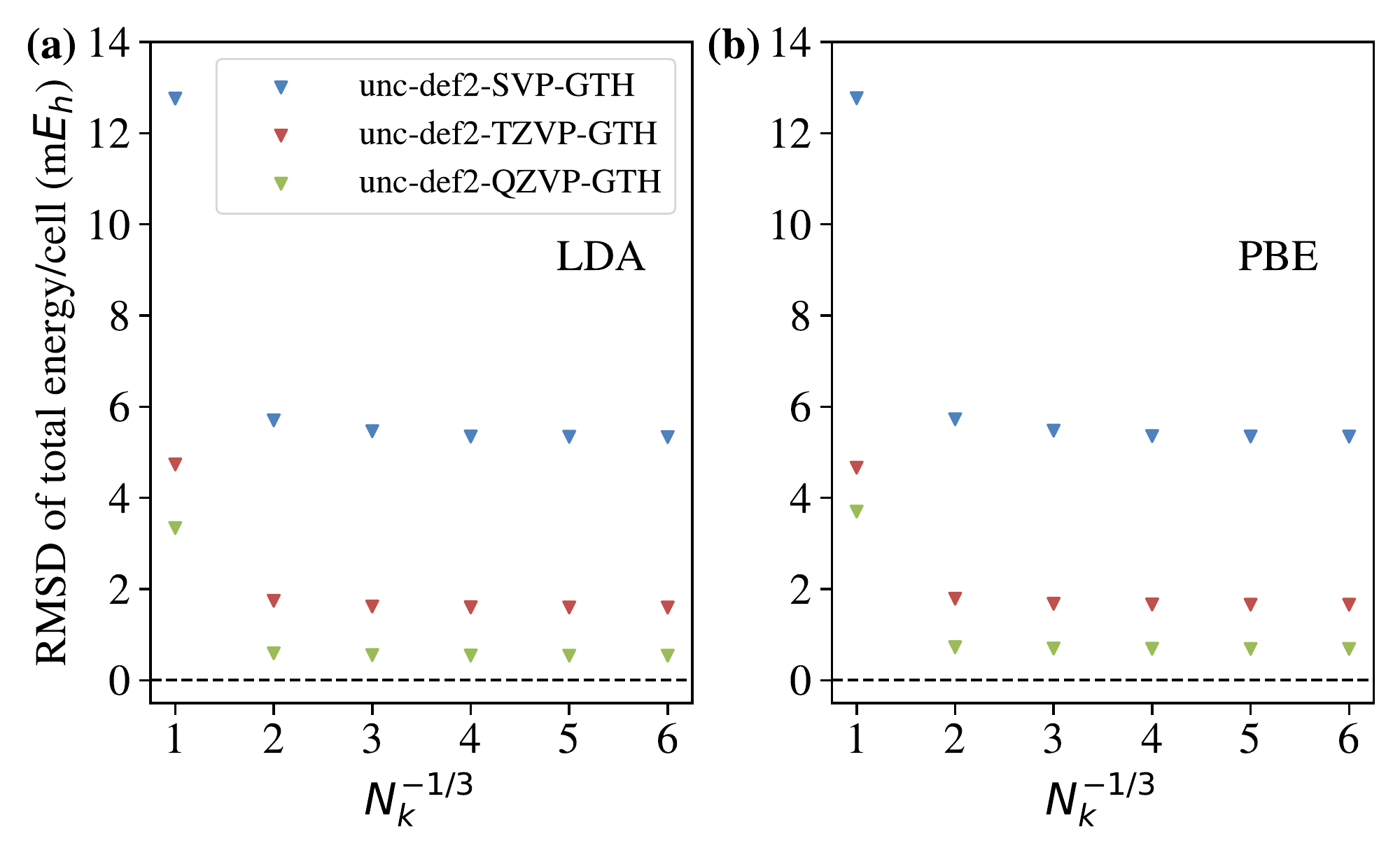}
    \caption{Root mean square deviation (RMSD) of DFT total energies (m$E_h$) per cell with respect to that of QE over 43 solids as a function of the number of $\mathbf k$-points for (a) LDA and (b) PBE functionals using unc-def2-GTH bases. }
    \label{fig:fig2}
\end{figure}
In \cref{fig:fig2}, we repeat the same analysis but over the entire benchmark set of 43 solids.
As before, unc-def2-GTH bases struggle for $N_k=1$ but work well for larger $\mathbf k$-meshes. RMSD systematically decreases as we increase the size of the basis set.
With the largest basis set, unc-def2-QZVP-GTH, we achieve better than 1 m$E_h$ accuracy in the TDL for the LDA and PBE functionals, as measured by the RMSD values.
Systems with the largest error in the TDL are SrSe (1.2 m$E_h$) in the case of LDA and GaP (1.4 m$E_h$) in the case of PBE.
As observed in the case of even-tempered bases, we expect that the result can be systematically made better by adding more exponents and increasing the maximum angular momentum.\cite{Bardo1973Dec,Feller1979Sep}
For instance, in the case of SrSe/LDA,
employing unc-def2-QZVPP-GTH (adding two additional f functions to both Sr and Se),
we observe an error
of 0.4 m$E_h$ which is three times smaller than that of unc-def2-QZVP-GTH.
While we can obtain overall better results by using unc-def2-QZVPP-GTH, we will mainly focus on the use of the unc-def2-QZVP-GTH basis set for the rest of the paper for simplicity.
In summary, these benchmark calculations suggest that unc-def2-GTH basis sets can achieve near basis set limit DFT total energies reliably towards the TDL. This result implies that 
the range of exponents and angular momenta in our bases is quite appropriate for solids.

\subsection{Basis set convergence of DFT band gaps}
In many materials applications,
DFT calculations are used not just to compute the ground state energy but to obtain spectral information through Kohn-Sham orbital energies.\cite{perdew2017understanding}
In doing so, one uses information from virtual orbitals in addition to that from occupied orbitals.
In the case of total energies presented in \cref{sec:total}, only occupied orbitals affect the results.
Here, we are assessing the quality of the difference between the lowest energy virtual orbital (i.e., the conduction band minimum) and the higher energy occupied orbital. It is possible that some BSIEs may cancel when taking energy differences.
\begin{figure*}[!ht]
    \centering
\includegraphics[scale=0.55]{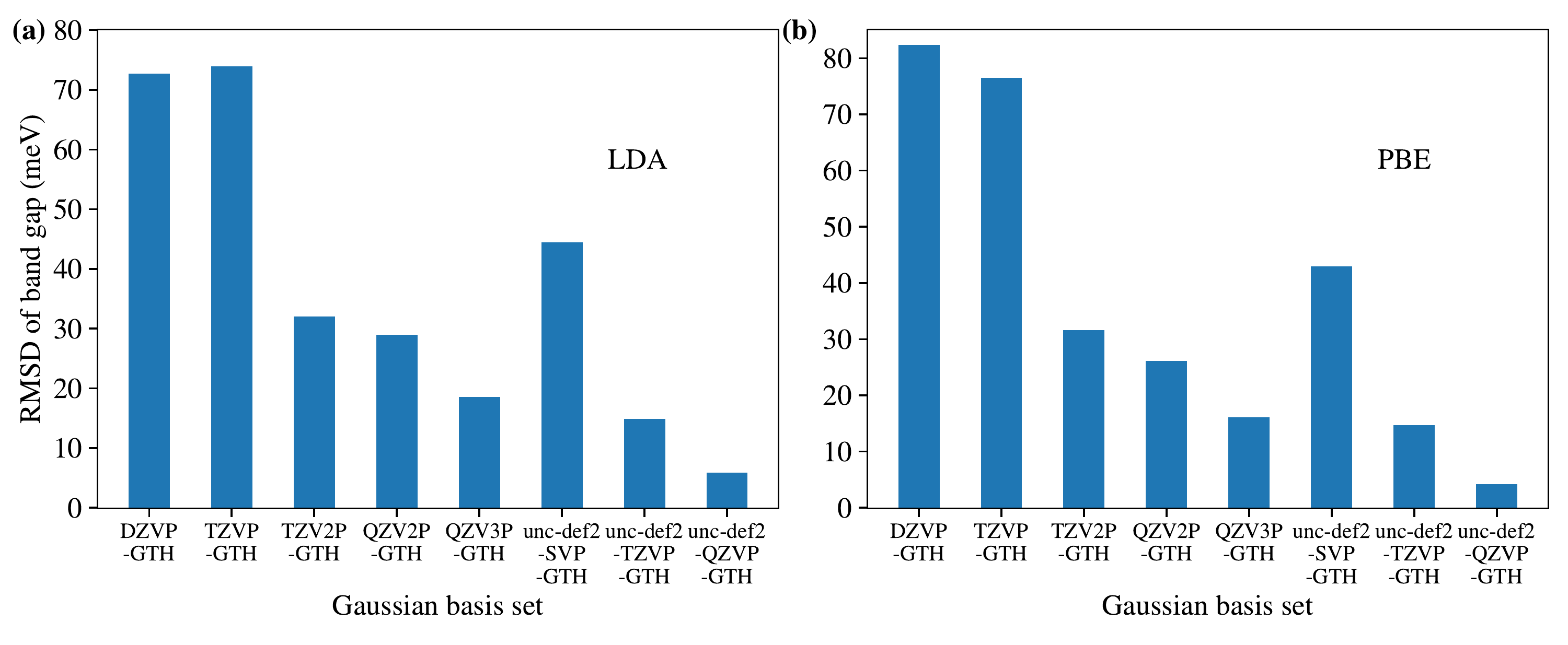}
    \caption{Root mean square deviation (RMSD) of DFT band gaps (meV) with respect to those of QE over 11 solids (a) LDA and (b) PBE functionals using GTH and unc-def2-GTH bases. }
    \label{fig:fig3}
\end{figure*}

In \cref{fig:fig3}, we present the RMSD of band gaps using GTO bases compared to the basis set limit results for LDA and PBE functionals. To compare unc-def2-GTH bases with GTH bases, we limit ourselves to the subset of 11 solids for the time being.
Somewhat surprisingly, we observe almost no improvement in the band gap when going from DZVP-GTH to TZVP-GTH.
By contrast, \cref{fig:fig1} shows a reduction in the RMSD of total energies of about 8 m$E_h$ when increasing the basis set size from DZVP-GTH to TZVP-GTH. However, this total energy improvement does not result in any band gap improvement. Nonetheless, past TZVP-GTH, the GTH bases do show systematic improvement in the band gap estimation. With the largest GTH basis set (QZV3P-GTH), RMSD in the band gap is 18-20 meV depending on the XC functional. 
Consistent with the total energy benchmark presented in \cref{sec:total}, 
unc-def2-GTH bases also exhibit systematic improvement. While the quality of unc-def2-SVP-GTH was on par with TZV2P-GTH in \cref{fig:fig1}, its band gap is clearly worse than that of TZV2P-GTH highlighting favorable error cancellation in TZV2P-GTH.
Nonetheless, unc-def2-TZVP-GTH is similar to QZV3P-GTH and unc-def2-QZVP-GTH has RMSD of 5.8 meV and 4.2 meV, respectively for LDA and PBE, showing its ability to converge band gaps to the basis set limit.

\begin{figure}[!ht]
    \centering
\includegraphics[scale=0.5]{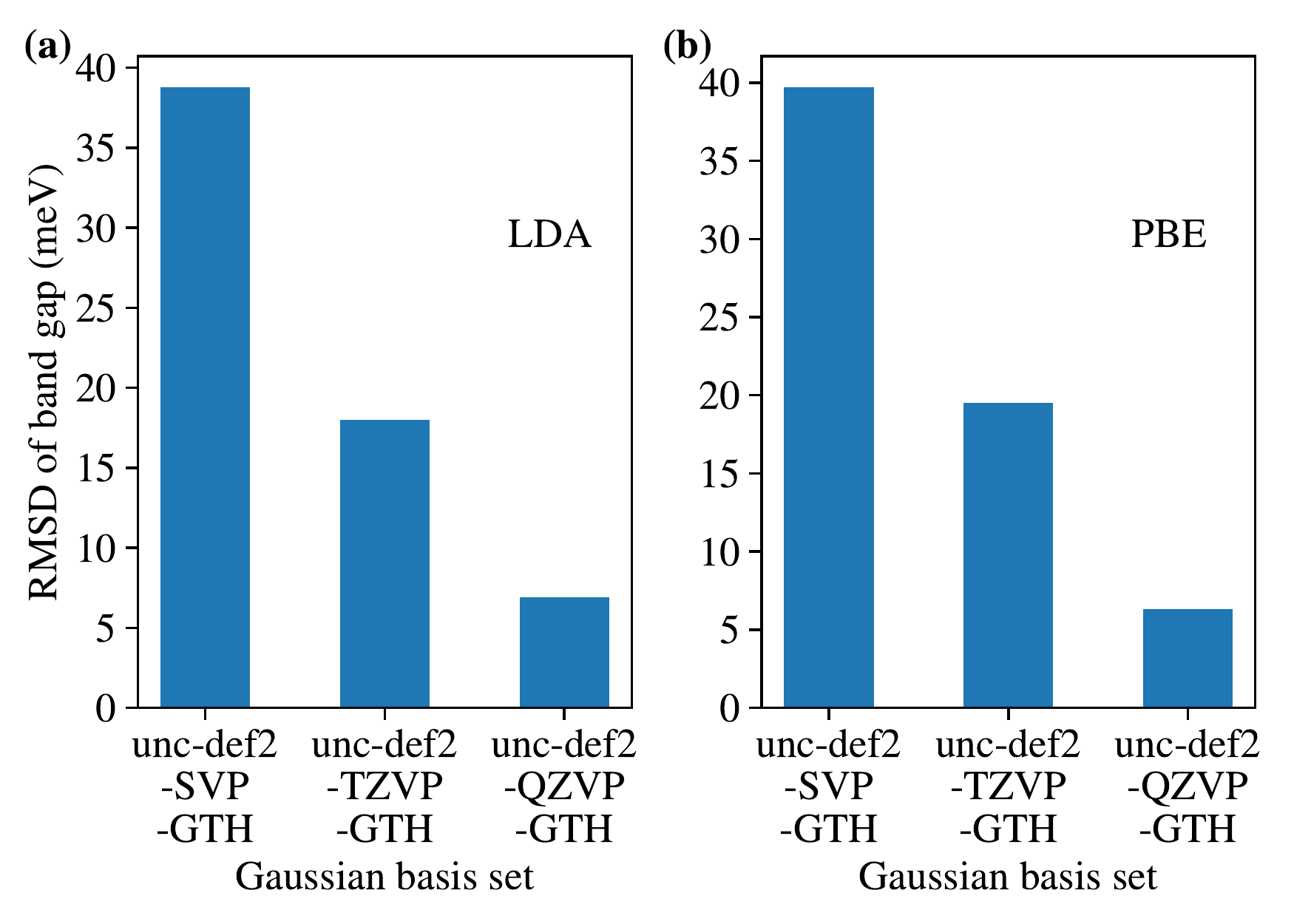}
    \caption{Root mean square deviation (RMSD) of DFT band gaps (meV) with respect to those of QE over 43 solids (a) LDA and (b) PBE functionals using unc-def2-GTH bases. }
    \label{fig:fig4}
\end{figure}

Encouraged by these results, we also analyzed the BSIEs in band gaps over all 43 solids using unc-def2-GTH bases as presented in \cref{fig:fig4}.
With unc-def2-SVP-GTH, the RMSD value is about 40 meV and it becomes less than 20 meV when using unc-def2-TZVP-GTH.
Lastly, with unc-def2-QZVP-GTH, the RMSD value becomes 6.9 meV and 6.3 meV, respectively, for LDA and PBE.
However, we note that the largest deviation is about 20 meV in both functionals, which corresponds to the band gap of SrSe. SrSe is the system with the largest total energy error for LDA as noted in the discussion of \cref{fig:fig2} in \cref{sec:total}.
Again, this remaining error can be further reduced by adding more GTOs to unc-def2-QZVP-GTH (e.g., using unc-def2-QZVPP-GTH), but we do not pursue this here.
The central message of this section is that the BSIE in the band gap reported in this paper using unc-def2-QZVP-GTH is smaller than 20 meV based on the numerical data. This is about 50 times smaller than the intrinsic errors in standard functionals for band gaps, so unc-def2-QZVP-GTH should be suitable for benchmarking purposes.

\subsection{Performance of pure DFT functionals}
\begin{table*}[!ht]
\begin{tabular}{lrrrrrrrrrrr}\hline
{} &   LDA &   PBE &  PBEsol &  revPBE &  BLYP &  B97-D &  SCAN &  M06-L &  MN15-L &  B97M-rV &  Exp. \\ \hline
C    &  4.12 &  4.33 &    4.16 &    4.38 &  4.60 &   4.57 &  4.64 &   4.84 &    4.24 &     4.67 &  5.48 \\
Si   &  0.49 &  0.66 &    0.52 &    0.72 &  0.94 &   0.91 &  0.93 &   1.12 &    0.96 &     0.92 &  1.17 \\
Ge   &  0.00 &  0.00 &    0.00 &    0.00 &  0.00 &   0.00 &  0.20 &   0.46 &    0.43 &     0.26 &  0.74 \\
SiC  &  1.33 &  1.50 &    1.35 &    1.54 &  1.85 &   1.95 &  1.81 &   1.83 &    1.92 &     2.07 &  2.42 \\
BN   &  4.36 &  4.64 &    4.42 &    4.71 &  5.03 &   5.07 &  5.03 &   4.94 &    4.98 &     5.28 &  6.22 \\
BP   &  1.20 &  1.38 &    1.23 &    1.42 &  1.66 &   1.62 &  1.67 &   1.95 &    1.61 &     1.70 &   2.4 \\
BAs  &  1.16 &  1.34 &    1.19 &    1.40 &  1.60 &   1.59 &  1.57 &   1.85 &    1.55 &     1.66 &  1.46 \\
BSb  &  0.76 &  0.91 &    0.78 &    0.96 &  1.15 &   1.17 &  1.06 &   1.21 &    1.08 &     1.15 &   N/A \\
AlP  &  1.47 &  1.67 &    1.50 &    1.75 &  1.98 &   2.04 &  1.99 &   2.20 &    2.12 &     2.16 &  2.51 \\
AlAs &  1.36 &  1.58 &    1.40 &    1.66 &  1.89 &   1.95 &  1.86 &   2.00 &    1.99 &     2.03 &  2.23 \\
AlSb &  1.17 &  1.36 &    1.20 &    1.45 &  1.57 &   1.58 &  1.56 &   1.78 &    1.63 &     1.63 &  1.68 \\
bGaN &  1.61 &  1.79 &    1.68 &    1.85 &  1.86 &   1.96 &  1.86 &   1.88 &    1.43 &     1.86 &   3.3 \\
GaP  &  1.44 &  1.66 &    1.50 &    1.75 &  1.70 &   1.72 &  1.85 &   1.89 &    1.84 &     2.05 &  2.35 \\
GaAs &  0.29 &  0.51 &    0.41 &    0.58 &  0.44 &   0.46 &  0.62 &   0.92 &    0.72 &     1.01 &  1.52 \\
GaSb &  0.00 &  0.17 &    0.08 &    0.23 &  0.08 &   0.08 &  0.21 &   0.50 &    0.36 &     0.64 &  0.73 \\
InP  &  0.42 &  0.61 &    0.52 &    0.68 &  0.57 &   0.56 &  0.57 &   0.86 &    0.36 &     0.88 &  1.42 \\
InAs &  0.00 &  0.00 &    0.00 &    0.00 &  0.00 &   0.00 &  0.00 &   0.00 &    0.00 &     0.03 &  0.41 \\
InSb &  0.00 &  0.00 &    0.00 &    0.00 &  0.00 &   0.00 &  0.00 &   0.00 &    0.00 &     0.13 &  0.23 \\
ZnS  &  1.80 &  2.12 &    1.94 &    2.24 &  2.15 &   2.24 &  2.38 &   2.56 &    2.00 &     2.36 &  3.66 \\
ZnSe &  1.01 &  1.33 &    1.15 &    1.45 &  1.35 &   1.43 &  1.60 &   1.80 &    1.31 &     1.67 &   2.7 \\
ZnTe &  1.07 &  1.38 &    1.22 &    1.49 &  1.38 &   1.43 &  1.60 &   1.83 &    1.40 &     1.79 &  2.38 \\
CdS  &  0.85 &  1.15 &    0.98 &    1.27 &  1.16 &   1.22 &  1.22 &   1.34 &    0.77 &     1.23 &  2.55 \\
CdSe &  0.34 &  0.64 &    0.48 &    0.76 &  0.66 &   0.70 &  0.72 &   0.88 &    0.34 &     0.82 &   1.9 \\
CdTe &  0.52 &  0.81 &    0.66 &    0.93 &  0.81 &   0.82 &  0.84 &   1.07 &    0.52 &     1.06 &  1.92 \\
MgS  &  3.27 &  3.57 &    3.39 &    3.73 &  3.71 &   3.88 &  4.14 &   4.03 &    4.06 &     4.17 &   5.4 \\
MgTe &  2.30 &  2.59 &    2.43 &    2.75 &  2.72 &   2.89 &  3.16 &   3.18 &    2.99 &     3.23 &   3.6 \\
MgO  &  4.68 &  4.93 &    4.79 &    5.06 &  5.17 &   5.49 &  5.59 &   5.01 &    5.69 &     5.55 &  7.22 \\
MgSe &  1.70 &  2.01 &    1.89 &    2.16 &  2.10 &   2.47 &  2.58 &   2.66 &    3.20 &     2.70 &  2.47 \\
CaS  &  2.17 &  2.41 &    2.27 &    2.51 &  2.52 &   2.64 &  2.92 &   2.56 &    3.03 &     3.05 &   N/A \\
CaSe &  1.90 &  2.14 &    2.00 &    2.24 &  2.27 &   2.38 &  2.63 &   2.28 &    2.73 &     2.78 &   N/A \\
CaTe &  1.42 &  1.65 &    1.51 &    1.74 &  1.80 &   1.90 &  2.08 &   1.74 &    2.19 &     2.27 &   N/A \\
SrS  &  2.22 &  2.49 &    2.33 &    2.61 &  2.62 &   2.74 &  2.92 &   2.57 &    2.86 &     2.95 &   N/A \\
SrSe &  2.01 &  2.28 &    2.12 &    2.40 &  2.43 &   2.54 &  2.69 &   2.35 &    2.63 &     2.74 &   N/A \\
SrTe &  1.57 &  1.83 &    1.66 &    1.94 &  2.00 &   2.10 &  2.20 &   1.87 &    2.17 &     2.29 &   N/A \\
BaS  &  2.01 &  2.26 &    2.11 &    2.38 &  2.36 &   2.44 &  2.58 &   2.25 &    2.38 &     2.51 &  3.88 \\
BaSe &  1.83 &  2.07 &    1.92 &    2.19 &  2.19 &   2.26 &  2.39 &   2.07 &    2.21 &     2.35 &  3.58 \\
BaTe &  1.49 &  1.74 &    1.58 &    1.85 &  1.87 &   1.93 &  2.03 &   1.74 &    1.90 &     2.03 &  3.08 \\
LiH  &  2.64 &  3.01 &    2.78 &    3.15 &  3.44 &   3.69 &  3.61 &   3.87 &    4.52 &     4.39 &   4.9 \\
LiF  &  8.92 &  9.33 &    9.11 &    9.56 &  9.49 &   9.92 & 10.08 &   9.64 &   10.27 &     9.77 &  14.2 \\
LiCl &  6.01 &  6.40 &    6.18 &    6.61 &  6.56 &   6.83 &  7.21 &   7.08 &    7.48 &     7.22 &   9.4 \\
AlN  &  4.25 &  4.38 &    4.25 &    4.43 &  4.66 &   4.78 &  4.87 &   4.75 &    4.94 &     5.24 &  6.13 \\
GaN  &  1.86 &  2.05 &    1.94 &    2.12 &  2.12 &   2.21 &  2.12 &   2.13 &    1.68 &     2.11 &   3.5 \\
InN  &  0.00 &  0.00 &    0.00 &    0.01 &  0.00 &   0.04 &  0.00 &   0.00 &    0.00 &     0.00 &  0.69 \\ \hline
RMSD &  1.72 &  1.50 &    1.64 &    1.41 &  1.39 &   1.28 &  1.20 &   1.26 &    1.27 &     1.17 &   N/A \\
MAD  & -1.46 & -1.23 &   -1.37 &   -1.14 & -1.10 &  -1.02 & -0.95 &  -0.90 &   -0.99 &    -0.84 &   N/A \\
MAX  &  5.28 &  4.87 &    5.09 &    4.64 &  4.71 &   4.28 &  4.12 &   4.56 &    3.93 &     4.43 &   N/A \\ \hline
\end{tabular}
  \caption{
Experimental and theoretical band gaps (or fundamental gaps) (eV) from various functionals over 43 solids.
N/A means ``not available''.
RMSD, MAD, and MAX denote, respectively,
root-mean-square-deviation,
mean-average-deviation,
and
maximum deviation
in reference to experimental values.
\rev{We took experimental references for three rocksalt solids
(LiH,LiF,LiCl)
from refs \citenum{vidal1986accurate,nolan2009calculation,hoang2013lih}, \citenum{matsushita2011comparative,garza2016predicting}, and \citenum{matsushita2011comparative,garza2016predicting}, respectively.
The rest of experimental values were taken from ref. \citenum{heyd2005energy}.}
\label{tab:gap1} 
  }
\end{table*}

\begin{figure*}[!ht]
    \centering
\includegraphics[scale=0.60]{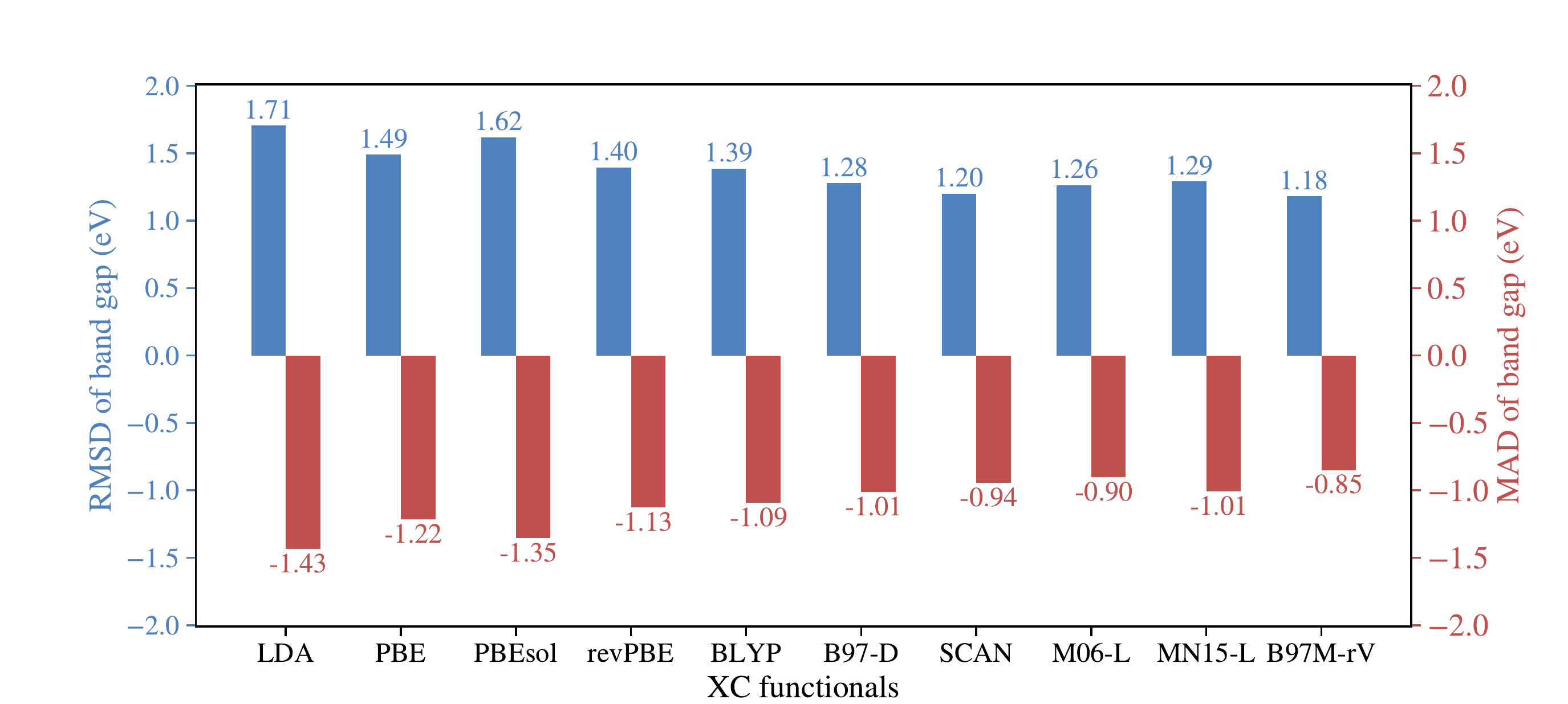}
    \caption{Band gap (eV) comparison over 36 solids between DFT (ten different functionals) and experiments: Blue: root-mean-square-deviation (RMSD) of DFT band gaps (eV) with respect to those of experiments and Red: mean-average-deviation (MAD) of DFT band gaps (eV) with respect to those of experiments}
    \label{fig:fig5}
\end{figure*}
Having established the accuracy of unc-def2-GTH bases, we assess the performance of pure DFT functionals over these simple solids.
Unfortunately, some of the 43 solids considered here do not have experimental band gaps. These solids are BSb, CaS, CaSe, CaTe, SrS, SrSe, and SrTe. Leaving aside these seven cases, we have a total of 36 experimental band gaps. 
unc-def2-QZVP-GTH is used with all XC functionals considered in this section.
The DFT band gaps over 43 solids along with the available experimental gaps are presented in \cref{tab:gap1}.
For an overall summary, it may be more instructuve to look at statistics of the band gap results as shown in \cref{fig:fig5}.
Looking at the mean-average-deviation (MAD), it is immediately evident that all pure functionals examined here exhibit the infamous
band gap underestimation problem of
pure functionals.\cite{perdew2017understanding}
Nonetheless, one can still find systematic improvement for going from the simplest functional, LDA, to more modern meta GGA functionals, SCAN, M06-L, MN15-L, and B97M-rV in
terms of the root-mean-square-deviation (RMSD) values. 
While the performance of B97M-rV is not great for those band gaps,
it still stays as one of the more accurate pure functionals for these problems.
This is encouraging because B97M-rV is statistically the most accurate pure XC functional in main group chemistry applications.\cite{mardirossian2017thirty}
\rev{Overall, all pure functionals perform poorly in this benchmark study and the inclusion of exact exchange seems necessary. 
This is not a new observation on its own and has been well-documented even for nearly the same benchmark set that we study here.\cite{Civalleri2012Oct,heyd2005energy}
In ref. \citenum{Civalleri2012Oct}, the authors consider many different ways to analyze the statistical data of LDA, PBE, and PBEsol along with other hybrid functionals
and revealed that hybrid functionals always perform the best in nine out of ten statistical analyses. 
It will be interesting to revisit this benchmark set with modern hybrid functionals in the future.}

For simplicity and due to the unavailability of functional-specific GTH pseudopotentials for most XC functionals considered here, 
we employed the GTH-LDA pseudopotential for all functionals in this section. 
Since this is not ideal, we checked the sensitivity of our conclusions with respect to the choice of the pseudopotential by testing GTH-LDA, GTH-PBE, and GTH-BLYP pseudopotentials with the BLYP functional. 
In all cases the RMSD and MAD are affected by less than 0.1 eV, which is a smaller energy scale than that of the band gap errors by roughly a factor of 10. 
For completeness, we provide the relevant numerical data in the Supplementary Material (see Table S1).
In the future, all-electron calculations could be done with all-electron basis sets generated via a similar protocol presented here. Alternatively, one could generate functional-specific GTH pseudopotentials for the modern XC functionals considered here.

\section{Outlook for future basis set design}
In this section, we would like to deliver cautionary notes on using our proposed bases and some discussion on future research in basis set design for solids.
\subsection{Our basis set is accurate but very large}\label{subsec:size}
\begin{table}[h]
\begin{tabular}{c|c|c|c|c}
& Si& C& O& Mg \\ \hline
SZV-GTH& 4& 4& 4& 5 \\ \hline
DZVP-GTH& 13& 13& 13& 14 \\ \hline
TZVP-GTH& 17& 17& 17& 18 \\ \hline
TZV2P-GTH& 22& 22& 22& 23 \\ \hline
QZV2P-GTH& 26& 26& 26& 27 \\ \hline
QZV3P-GTH& 31& 31& 31& 32 \\ \hline
unc-def2-SVP-GTH& 40& 41& 40& 53 \\ \hline
unc-def2-TZVP-GTH& 62& 58& 57& 68 \\ \hline
unc-def2-QZVP-GTH& 90& 83& 81& 86 \\ \hline
\end{tabular}
\caption {Number of basis functions in the basis sets used in this work for selected elements (Si, C, O, and Mg).}
\label{tab:basis}
\end{table}
While our proposed unc-def2-GTH bases are of high quality, these bases are very large due to the decontraction from the original contracted GTO bases. This large size carries a significant computational cost.
This is the major drawback of even-tempered and well-tempered bases, and it is one that our unc-def2-GTH bases also share.
To be more concrete, we provide the number of basis functions for selected elements (Si, C, O, Mg) in \cref{tab:basis}.
unc-def2-SVP-GTH is about three times bigger than DZVP-GTH while our unc-def2-TZVP-GTH is roughly three times bigger than TZV2P-GTH. Similarly, our largest basis set unc-def2-QZVP-GTH is about 2.5--3 times larger than QZV3P-GTH.

Because of compute cost and memory demand, there is a need to compress these bases for practical calculations.
Perhaps, the most difficult (but most effective if done correctly) way to compress them is to obtain transferable contraction coefficients. One could start by inspecting the molecular orbitals (or Bloch orbitals) that our calculations produce for those simple solids.
Another strategy is to take these mean-field molecular orbitals and compress the virtual space for subsequent correlation calculations, for instance using the random phase approximation (RPA). The use of natural orbitals to compress the virtual space was shown to be effective, and would be a good starting point for making our basis more compact \cite{ramberger2019rpa}. We note that it is also unclear whether our proposed bases exhibit any scaling properties which will allow for higher accuracy by using basis set extrapolation for correlation energy calculations, which could be further investigated in the future.

\subsection{Even low-lying virtual orbitals can be difficult to describe well}
\begin{figure}[!ht]
    \centering
\includegraphics[scale=0.45]{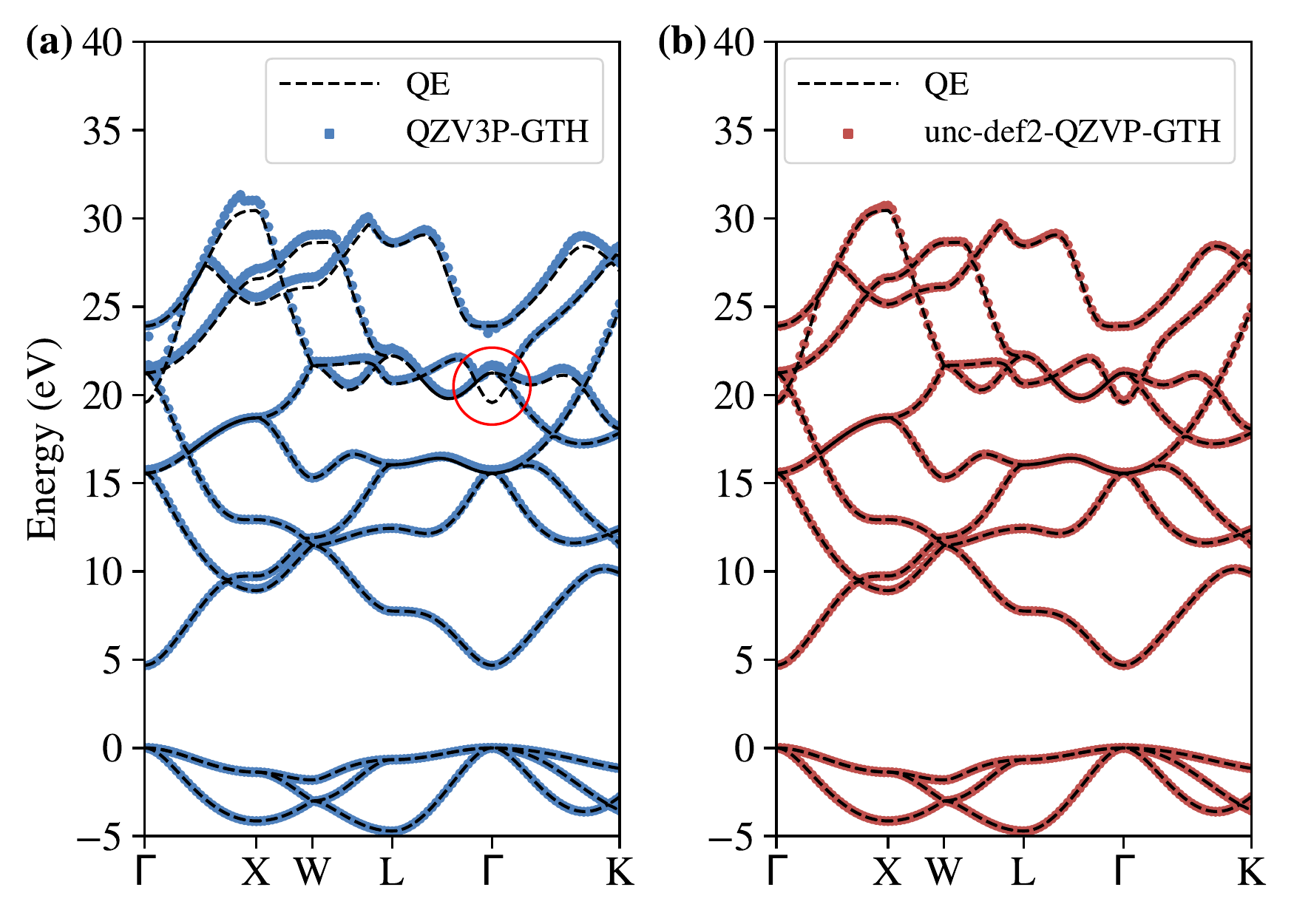}
    \caption{LDA band structure of MgO: (a) comparing QZV3P-GTH against QE and (b) comparing unc-def2-QZVP-GTH against QE.
    The band energies are shifted such that the highest valence band energy is located at zero.
    The red circle in (a) highlights the qualitative failure of QZV3P-GTH virtual orbitals.
     }
    \label{fig:fig6}
\end{figure}

Basis sets that are optimized for mean-field calculations such as GTH bases often behave erratically in correlated calculations.\cite{Morales2020Nov} 
Since these bases tend to yield good occupied orbitals, the poor performance of correlation calculations can be attributed to virtual orbitals.
Furthermore, low-lying virtual orbitals play important roles in describing optical properties and related excited states.
Therefore, high-quality basis sets should produce qualitatively accurate virtuals.
As an example, we present the band structure of MgO using QZV3P-GTH and unc-def2-QZVP-GTH and compare them against that of QE.
MgO has a total of 8 occupied orbitals and we computed up to the 16-th band in QE for comparison purposes. 
We note that the challenge of MgO conduction bands for GTOs was noted before in ref. \citenum{irmler2018robust}, but we focus on a wider range of conduction bands here.
The pertinent band structures are presented in \cref{fig:fig6}.

In both bases, the valence bands and the first few conduction bands are in an excellent agreement with those of QE. 
However,
the higher-lying virtuals of QZV3P-GTH (in \cref{fig:fig6}(a)) start to deviate significantly from those of QE.
The most striking failure is the lack of the 5-th virtual orbital highlighted under a red circle in \cref{fig:fig6} (a).
On the other hand, the virtuals from unc-def2-QZVP-GTH have visually indistinguishable energies when compared to QE
highlighting its potential utility for correlated calculations as well.
We also tried a smaller unc-def2-GTH basis set, namely unc-def2-TZVP-GTH.
It turns out that even unc-def2-TZVP-GTH misses the same virtual that QZV3P-GTH misses as well.
\rev{
With further investigations, we found that the 5-th virtual orbital is missed by
basis sets without any f function on the Mg atom.
To be more concrete, we added one f-function to Mg in the QZV3P-GTH basis where the exponent of 0.16 was taken from unc-def2-QZVP-GTH.
With this basis set, we recover the missing band at the $\Gamma$-point. This additional f-function introduces only a 0.2 m$E_h$ energy lowering in the ground state, but
it is essential to capture one of the low-lying conduction bands.}

This example emphasizes that more attention to the low-lying virtual orbitals should be paid when designing GTO basis sets for applications such as conduction band structure, time-dependent DFT and correlated methods such as RPA. Existing GTO bases designed primarily to describe the occupied space may likewise exhibit qualitative failures like this case.
\subsection{Transferability across different lattice constants is challenging}\label{sec:cold}
Cold curves of solids are often of great interest for experimentalists. 
Cold curves are analogous to potential energy curves (PECs) in molecular quantum chemistry. 
Similar to PECs, 
as one shrinks the lattice constant and brings atoms close to one another,
a larger number of near linear dependencies occur, and 
the quality of the underlying GTO basis degrades because of discarding such functions by canonical orthogonalization.
Furthermore, system-dependent optimization strategies can struggle for cold curves because
basis sets are usually optimized for one specific geometry (usually equilibrium geometries).\cite{Morales2020Nov,Daga2020Apr}
As a result of this, varying lattice constants can be challenging using GTO basis sets as the system approaches its high-pressure configuration (shorter lattice constants) or atomic limits (longer lattice constants).

\begin{figure}[!ht]
    \centering
\includegraphics[scale=0.43]{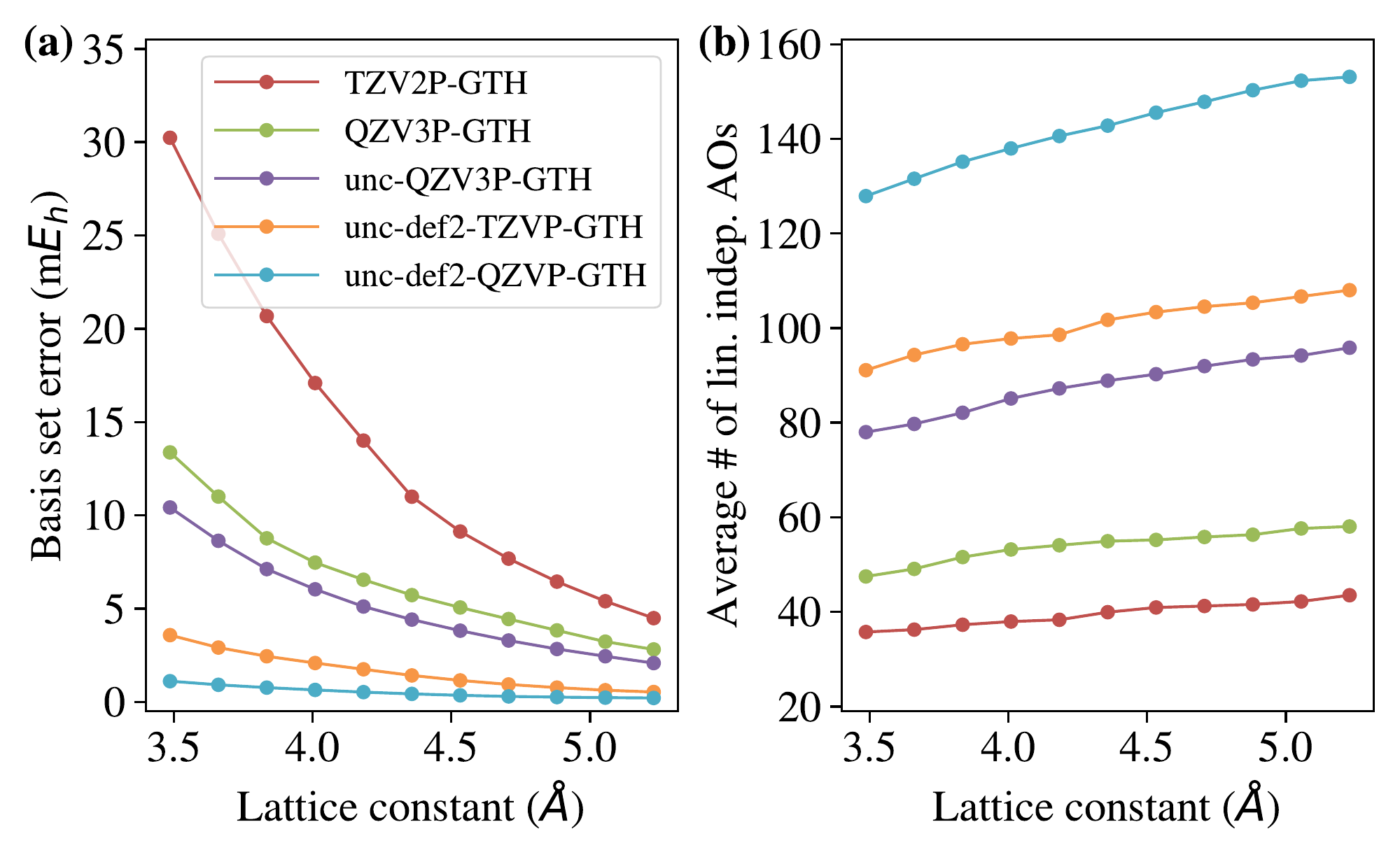}
    \caption{Investigation of a SiC cold curve using PBE: (a) Basis set error with respect to QE as a function of lattice constant for various basis sets. (b) Average number of linearly independent AOs as a function of the lattice constant for various basis sets.
     }
    \label{fig:fig7}
\end{figure}
As an example to illustrate this point, we computed a cold curve of SiC using PBE and the GTH-LDA pseudopotential with TZV2P-GTH, QZV3P-GTH, unc-QZV3P-GTH, unc-def2-TZVP, and unc-def2-QZVP. The Brillouin zone was sampled with $6\times 6 \times 6$ $\mathbf k$-points via the Monkhorst-Pack algorithm.
Here unc-QZV3P-GTH is the basis obtained from decontracting QZV3P-GTH. Using unc-QZV3P-GTH, we can quantify the error coming from the contraction coefficients of QZV3P-GTH.
As before, the QE results (the same functional and pseudopotential) serve as the basis set limit reference values. 
The pertinent results are presented in \cref{fig:fig7}.

\cref{fig:fig7} (a)
shows that TZV2P-GTH, QZV3P-GTH, and unc-QZV3P-GTH bases make a large error especially when compressing the lattice.
It is also instructive to quantify the nonparallelity error (NPE) over the range of lattice constants examined here as a means to measure error cancellation. 
The NPE is 25.7 m$E_h$, 10.6 m$E_h$, 8.3 m$E_h$, 3.0 m$E_h$, and 0.9 m$E_h$, respectively, for TZV2P-GTH, QZV3P-GTH, unc-QZV3P-GTH, unc-def2-TZVP-GTH, and unc-def2-QZVP-GTH. 
Interestingly, unc-QZV3P-GTH reduces the basis set error by only a small amount, which implies
that the contraction coefficients of QZV3P-GTH for those elements are transferable over a wide range of lattice constants. 
It also suggests that the range of exponents in QZV3P-GTH becomes inappropriate for smaller lattice constants.
Comparing the exponents of unc-QZV3P-GTH and unc-def2-TZVP-GTH, we find that unc-def2-TZVP-GTH has more compact exponents for spd shells and has an f shell for Si that is not present in unc-QZV3P-GTH. These more compact GTOs likely become more important at closer distances and hence explain the differences between two bases.

The main cause of these generally large NPEs is the fact that at closer distances the quality of those GTO bases degrades as shown in \cref{fig:fig7} (b) which quantifies the number of orthogonalized basis functions retained after canonical orthogonalization.
Since each $\mathbf k$-point has a slightly different number of linearly independent AOs, we present the average values over 216 $\mathbf k$-points as a function of lattice constant.
Evidently, the number of linearly independent AOs decreases as the lattice constant decreases, which in turn increases the basis set incompleteness error.
Nonetheless, the largest basis set, unc-def2-QZVP-GTH, is able to achieve a satisfying NPE in this case, which highlights the utility of this basis set for accurate cold curve simulations.

\section{Conclusions}
In this manuscript, we discussed 
strategies 
for generating 
large and accurate uncontracted Gaussian bases (unc-def2-GTH bases)
which do not resort to system- or method- specific optimizations.
Using a new implementation of the Gaussian atomic orbital plus planewave density fitting approach in Q-Chem, the basis set incompleteness error in our proposed bases were then
assessed over 43 prototypical semiconductors
by comparing the pure density functional theory total energies per cell and band gaps against those from fully converged planewave results.
We found that the basis set incompleteness error in total energy and band gap with our largest GTO basis (unc-def2-QZVP-GTH) is smaller than 0.7 m$E_h$ per atom in the unit cell and less than 20 meV, respectively, verifying the validity of the range of exponents and angular momenta in the proposed bases.

In the application of our bases, 
we focused on the assessment of ten pure density functionals for predicting the band gaps of 36 semiconductors whose experimental gaps are well documented.
Not surprisingly, we found that all examined pure functionals (LDA, PBE, PBEsol, revPBE, BLYP, B97-D,  SCAN, M06-L, MN15-L, B97M-rV) significantly underestimate the band gaps of these materials.
The combinatorially optimized mGGA functional, B97M-rV, performs as well as do other modern mGGA functionals. 
Our work suggests that combinatorially optimized range-separated hybrid functionals such as $\omega$B97X-rV and $\omega$B97M-rV will be highly interesting to study
since they may also exhibit better accuracy compared to other relatively older range-separated hybrid functionals or even short-range hybrid functionals.

We also made several cautionary remarks on our GTO bases as well as on the future research in GTO basis design for solids:
\begin{enumerate}
\item Our basis sets are accurate but large so there is a need for a way to compress our basis sets further for both mean-field and correlation calculations.
\item The widely used GTH bases may qualitatively fail for describing low-lying virtual orbitals which will affect the subsequent correlation and optical calculations. At much greater compute cost, our unc-def2-QZVP-GTH basis set was shown to accurately capture all of the low-lying virtual orbitals of MgO including the one missed by QZV3P-GTH.
\item Reducing the non-parallelity error of the basis set incompleteness error is challenging particularly due to the high pressure region of cold curves that exhibits a higher number of near linear dependencies.
\end{enumerate}
In the future, we will test several ways (e.g., finding universal contraction coefficients and frozen natural orbitals\cite{ramberger2019rpa}) to compress our unc-def2-GTH bases and
investigate the basis set convergence of correlation and optical methods with these bases in the future.
\rev{Furthermore, simple solids like LiH have many numerical data of the total Hartree-Fock energies
towards the basis set limit,\cite{Gillan2008Oct,Marsman2009May,Paier2009Nov,Civalleri2010Mar,Usvyat2011Jun} which could be a good testbed for our basis sets in the future.}
\section{Supplementary Material}
The supplementary material of this work is available, which contains the test of the impact of different GTH pseudopotentials in the band gaps. 
\section{Acknowledgement}
This work was supported by the National Institutes of Health SBIR program through Grant No. 2R44GM128480-02A1. We thank Eloy Ramos for initial engagement with this project in 2017
and
Kuan-Yu Liu for help with the implementation of the $\mathbf k$-point parser used in this work.
JL thanks David Reichman for support.
\section{Data Availability Statement}
The data that supports the findings of this study are available within the article and its supplementary material.

\section{Conflict of Interest}
E.E. and M.H.-G. are part-owners of Q-Chem, Inc.
\bibliography{refs}

\begin{thebibliography}{104}%
\makeatletter
\providecommand \@ifxundefined [1]{%
 \@ifx{#1\undefined}
}%
\providecommand \@ifnum [1]{%
 \ifnum #1\expandafter \@firstoftwo
 \else \expandafter \@secondoftwo
 \fi
}%
\providecommand \@ifx [1]{%
 \ifx #1\expandafter \@firstoftwo
 \else \expandafter \@secondoftwo
 \fi
}%
\providecommand \natexlab [1]{#1}%
\providecommand \enquote  [1]{``#1''}%
\providecommand \bibnamefont  [1]{#1}%
\providecommand \bibfnamefont [1]{#1}%
\providecommand \citenamefont [1]{#1}%
\providecommand \href@noop [0]{\@secondoftwo}%
\providecommand \href [0]{\begingroup \@sanitize@url \@href}%
\providecommand \@href[1]{\@@startlink{#1}\@@href}%
\providecommand \@@href[1]{\endgroup#1\@@endlink}%
\providecommand \@sanitize@url [0]{\catcode `\\12\catcode `\$12\catcode
  `\&12\catcode `\#12\catcode `\^12\catcode `\_12\catcode `\%12\relax}%
\providecommand \@@startlink[1]{}%
\providecommand \@@endlink[0]{}%
\providecommand \url  [0]{\begingroup\@sanitize@url \@url }%
\providecommand \@url [1]{\endgroup\@href {#1}{\urlprefix }}%
\providecommand \urlprefix  [0]{URL }%
\providecommand \Eprint [0]{\href }%
\providecommand \doibase [0]{http://dx.doi.org/}%
\providecommand \selectlanguage [0]{\@gobble}%
\providecommand \bibinfo  [0]{\@secondoftwo}%
\providecommand \bibfield  [0]{\@secondoftwo}%
\providecommand \translation [1]{[#1]}%
\providecommand \BibitemOpen [0]{}%
\providecommand \bibitemStop [0]{}%
\providecommand \bibitemNoStop [0]{.\EOS\space}%
\providecommand \EOS [0]{\spacefactor3000\relax}%
\providecommand \BibitemShut  [1]{\csname bibitem#1\endcsname}%
\let\auto@bib@innerbib\@empty
\bibitem [{\citenamefont {Ayala}\ \emph {et~al.}(2001)\citenamefont {Ayala},
  \citenamefont {Kudin},\ and\ \citenamefont {Scuseria}}]{Ayala2001Dec}%
  \BibitemOpen
  \bibfield  {author} {\bibinfo {author} {\bibfnamefont {Philippe~Y.}\
  \bibnamefont {Ayala}}, \bibinfo {author} {\bibfnamefont {Konstantin~N.}\
  \bibnamefont {Kudin}}, \ and\ \bibinfo {author} {\bibfnamefont {Gustavo~E.}\
  \bibnamefont {Scuseria}},\ }\bibfield  {title} {\enquote {\bibinfo {title}
  {{Atomic orbital Laplace-transformed second-order
  M{\o}ller{\textendash}Plesset theory for periodic systems}},}\ }\href
  {\doibase 10.1063/1.1414369} {\bibfield  {journal} {\bibinfo  {journal} {J.
  Chem. Phys.}\ }\textbf {\bibinfo {volume} {115}},\ \bibinfo {pages}
  {9698--9707} (\bibinfo {year} {2001})}\BibitemShut {NoStop}%
\bibitem [{\citenamefont {Katagiri}(2005)}]{Katagiri2005Jun}%
  \BibitemOpen
  \bibfield  {author} {\bibinfo {author} {\bibfnamefont {Hideki}\ \bibnamefont
  {Katagiri}},\ }\bibfield  {title} {\enquote {\bibinfo {title}
  {{Equation-of-motion coupled-cluster study on exciton states of polyethylene
  with periodic boundary condition}},}\ }\href {\doibase 10.1063/1.1929731}
  {\bibfield  {journal} {\bibinfo  {journal} {J. Chem. Phys.}\ }\textbf
  {\bibinfo {volume} {122}},\ \bibinfo {pages} {224901} (\bibinfo {year}
  {2005})}\BibitemShut {NoStop}%
\bibitem [{\citenamefont {Izmaylov}\ and\ \citenamefont
  {Scuseria}(2008)}]{Izmaylov2008}%
  \BibitemOpen
  \bibfield  {author} {\bibinfo {author} {\bibfnamefont {Artur~F.}\
  \bibnamefont {Izmaylov}}\ and\ \bibinfo {author} {\bibfnamefont {Gustavo~E.}\
  \bibnamefont {Scuseria}},\ }\bibfield  {title} {\enquote {\bibinfo {title}
  {{Resolution of the identity atomic orbital Laplace transformed second order
  M{\o}ller{\textendash}Plesset theory for nonconducting periodic systems}},}\
  }\href {\doibase 10.1039/B803274M} {\bibfield  {journal} {\bibinfo  {journal}
  {Phys. Chem. Chem. Phys.}\ }\textbf {\bibinfo {volume} {10}},\ \bibinfo
  {pages} {3421--3429} (\bibinfo {year} {2008})}\BibitemShut {NoStop}%
\bibitem [{\citenamefont {Pisani}\ \emph {et~al.}(2008)\citenamefont {Pisani},
  \citenamefont {Maschio}, \citenamefont {Casassa}, \citenamefont {Halo},
  \citenamefont {Sch{\ifmmode\ddot{u}\else\"{u}\fi}tz},\ and\ \citenamefont
  {Usvyat}}]{Pisani2008Oct}%
  \BibitemOpen
  \bibfield  {author} {\bibinfo {author} {\bibfnamefont {Cesare}\ \bibnamefont
  {Pisani}}, \bibinfo {author} {\bibfnamefont {Lorenzo}\ \bibnamefont
  {Maschio}}, \bibinfo {author} {\bibfnamefont {Silvia}\ \bibnamefont
  {Casassa}}, \bibinfo {author} {\bibfnamefont {Migen}\ \bibnamefont {Halo}},
  \bibinfo {author} {\bibfnamefont {Martin}\ \bibnamefont
  {Sch{\ifmmode\ddot{u}\else\"{u}\fi}tz}}, \ and\ \bibinfo {author}
  {\bibfnamefont {Denis}\ \bibnamefont {Usvyat}},\ }\bibfield  {title}
  {\enquote {\bibinfo {title} {{Periodic local MP2 method for the study of
  electronic correlation in crystals: Theory and preliminary applications}},}\
  }\href {\doibase 10.1002/jcc.20975} {\bibfield  {journal} {\bibinfo
  {journal} {J. Comput. Chem.}\ }\textbf {\bibinfo {volume} {29}},\ \bibinfo
  {pages} {2113--2124} (\bibinfo {year} {2008})}\BibitemShut {NoStop}%
\bibitem [{\citenamefont {Pisani}\ \emph {et~al.}(2012)\citenamefont {Pisani},
  \citenamefont {Sch{\ifmmode\ddot{u}\else\"{u}\fi}tz}, \citenamefont
  {Casassa}, \citenamefont {Usvyat}, \citenamefont {Maschio}, \citenamefont
  {Lorenz},\ and\ \citenamefont {Erba}}]{Pisani2012}%
  \BibitemOpen
  \bibfield  {author} {\bibinfo {author} {\bibfnamefont {Cesare}\ \bibnamefont
  {Pisani}}, \bibinfo {author} {\bibfnamefont {Martin}\ \bibnamefont
  {Sch{\ifmmode\ddot{u}\else\"{u}\fi}tz}}, \bibinfo {author} {\bibfnamefont
  {Silvia}\ \bibnamefont {Casassa}}, \bibinfo {author} {\bibfnamefont {Denis}\
  \bibnamefont {Usvyat}}, \bibinfo {author} {\bibfnamefont {Lorenzo}\
  \bibnamefont {Maschio}}, \bibinfo {author} {\bibfnamefont {Marco}\
  \bibnamefont {Lorenz}}, \ and\ \bibinfo {author} {\bibfnamefont {Alessandro}\
  \bibnamefont {Erba}},\ }\bibfield  {title} {\enquote {\bibinfo {title} {{C
  ryscor : a program for the post-Hartree{\textendash}Fock treatment of
  periodic systems}},}\ }\href {\doibase 10.1039/C2CP23927B} {\bibfield
  {journal} {\bibinfo  {journal} {Phys. Chem. Chem. Phys.}\ }\textbf {\bibinfo
  {volume} {14}},\ \bibinfo {pages} {7615--7628} (\bibinfo {year}
  {2012})}\BibitemShut {NoStop}%
\bibitem [{\citenamefont {Del~Ben}\ \emph {et~al.}(2012)\citenamefont
  {Del~Ben}, \citenamefont {Hutter},\ and\ \citenamefont
  {VandeVondele}}]{DelBen2012Nov}%
  \BibitemOpen
  \bibfield  {author} {\bibinfo {author} {\bibfnamefont {Mauro}\ \bibnamefont
  {Del~Ben}}, \bibinfo {author} {\bibfnamefont
  {J{\ifmmode\ddot{u}\else\"{u}\fi}rg}\ \bibnamefont {Hutter}}, \ and\ \bibinfo
  {author} {\bibfnamefont {Joost}\ \bibnamefont {VandeVondele}},\ }\bibfield
  {title} {\enquote {\bibinfo {title} {{Second-Order
  M{\o}ller{\textendash}Plesset Perturbation Theory in the Condensed Phase: An
  Efficient and Massively Parallel Gaussian and Plane Waves Approach}},}\
  }\href {\doibase 10.1021/ct300531w} {\bibfield  {journal} {\bibinfo
  {journal} {J. Chem. Theory Comput.}\ }\textbf {\bibinfo {volume} {8}},\
  \bibinfo {pages} {4177--4188} (\bibinfo {year} {2012})}\BibitemShut {NoStop}%
\bibitem [{\citenamefont {Usvyat}\ \emph {et~al.}(2015)\citenamefont {Usvyat},
  \citenamefont {Maschio},\ and\ \citenamefont
  {Sch{\ifmmode\ddot{u}\else\"{u}\fi}tz}}]{Usvyat2015Sep}%
  \BibitemOpen
  \bibfield  {author} {\bibinfo {author} {\bibfnamefont {Denis}\ \bibnamefont
  {Usvyat}}, \bibinfo {author} {\bibfnamefont {Lorenzo}\ \bibnamefont
  {Maschio}}, \ and\ \bibinfo {author} {\bibfnamefont {Martin}\ \bibnamefont
  {Sch{\ifmmode\ddot{u}\else\"{u}\fi}tz}},\ }\bibfield  {title} {\enquote
  {\bibinfo {title} {{Periodic local MP2 method employing orbital specific
  virtuals}},}\ }\href {\doibase 10.1063/1.4921301} {\bibfield  {journal}
  {\bibinfo  {journal} {J. Chem. Phys.}\ }\textbf {\bibinfo {volume} {143}},\
  \bibinfo {pages} {102805} (\bibinfo {year} {2015})}\BibitemShut {NoStop}%
\bibitem [{\citenamefont {McClain}\ \emph {et~al.}(2017)\citenamefont
  {McClain}, \citenamefont {Sun}, \citenamefont {Chan},\ and\ \citenamefont
  {Berkelbach}}]{McClain2017Mar}%
  \BibitemOpen
  \bibfield  {author} {\bibinfo {author} {\bibfnamefont {James}\ \bibnamefont
  {McClain}}, \bibinfo {author} {\bibfnamefont {Qiming}\ \bibnamefont {Sun}},
  \bibinfo {author} {\bibfnamefont {Garnet Kin-Lic}\ \bibnamefont {Chan}}, \
  and\ \bibinfo {author} {\bibfnamefont {Timothy~C.}\ \bibnamefont
  {Berkelbach}},\ }\bibfield  {title} {\enquote {\bibinfo {title}
  {{Gaussian-Based Coupled-Cluster Theory for the Ground-State and Band
  Structure of Solids}},}\ }\href {\doibase 10.1021/acs.jctc.7b00049}
  {\bibfield  {journal} {\bibinfo  {journal} {J. Chem. Theory Comput.}\
  }\textbf {\bibinfo {volume} {13}},\ \bibinfo {pages} {1209--1218} (\bibinfo
  {year} {2017})}\BibitemShut {NoStop}%
\bibitem [{\citenamefont {Shang}\ and\ \citenamefont
  {Yang}(2020)}]{Shang2020Nov}%
  \BibitemOpen
  \bibfield  {author} {\bibinfo {author} {\bibfnamefont {Honghui}\ \bibnamefont
  {Shang}}\ and\ \bibinfo {author} {\bibfnamefont {Jinlong}\ \bibnamefont
  {Yang}},\ }\bibfield  {title} {\enquote {\bibinfo {title} {{Implementation of
  Laplace Transformed MP2 for Periodic Systems With Numerical Atomic
  Orbitals}},}\ }\href {\doibase 10.3389/fchem.2020.589992} {\bibfield
  {journal} {\bibinfo  {journal} {Front. Chem.}\ }\textbf {\bibinfo {volume}
  {8}} (\bibinfo {year} {2020}),\ 10.3389/fchem.2020.589992}\BibitemShut
  {NoStop}%
\bibitem [{\citenamefont {Maurer}\ \emph {et~al.}(2019)\citenamefont {Maurer},
  \citenamefont {Freysoldt}, \citenamefont {Reilly}, \citenamefont
  {Brandenburg}, \citenamefont {Hofmann}, \citenamefont
  {Bj{\ifmmode\ddot{o}\else\"{o}\fi}rkman}, \citenamefont
  {Leb{\ifmmode\grave{e}\else\`{e}\fi}gue},\ and\ \citenamefont
  {Tkatchenko}}]{Maurer2019Jul}%
  \BibitemOpen
  \bibfield  {author} {\bibinfo {author} {\bibfnamefont {Reinhard~J.}\
  \bibnamefont {Maurer}}, \bibinfo {author} {\bibfnamefont {Christoph}\
  \bibnamefont {Freysoldt}}, \bibinfo {author} {\bibfnamefont {Anthony~M.}\
  \bibnamefont {Reilly}}, \bibinfo {author} {\bibfnamefont {Jan~Gerit}\
  \bibnamefont {Brandenburg}}, \bibinfo {author} {\bibfnamefont {Oliver~T.}\
  \bibnamefont {Hofmann}}, \bibinfo {author} {\bibfnamefont
  {Torbj{\ifmmode\ddot{o}\else\"{o}\fi}rn}\ \bibnamefont
  {Bj{\ifmmode\ddot{o}\else\"{o}\fi}rkman}}, \bibinfo {author} {\bibfnamefont
  {S{\ifmmode\acute{e}\else\'{e}\fi}bastien}\ \bibnamefont
  {Leb{\ifmmode\grave{e}\else\`{e}\fi}gue}}, \ and\ \bibinfo {author}
  {\bibfnamefont {Alexandre}\ \bibnamefont {Tkatchenko}},\ }\bibfield  {title}
  {\enquote {\bibinfo {title} {{Advances in Density-Functional Calculations for
  Materials Modeling}},}\ }\href {\doibase
  10.1146/annurev-matsci-070218-010143} {\bibfield  {journal} {\bibinfo
  {journal} {Annu. Rev. Mater. Res.}\ }\textbf {\bibinfo {volume} {49}},\
  \bibinfo {pages} {1--30} (\bibinfo {year} {2019})}\BibitemShut {NoStop}%
\bibitem [{\citenamefont {Hybertsen}\ and\ \citenamefont
  {Louie}(1986)}]{Hybertsen1986Oct}%
  \BibitemOpen
  \bibfield  {author} {\bibinfo {author} {\bibfnamefont {Mark~S.}\ \bibnamefont
  {Hybertsen}}\ and\ \bibinfo {author} {\bibfnamefont {Steven~G.}\ \bibnamefont
  {Louie}},\ }\bibfield  {title} {\enquote {\bibinfo {title} {{Electron
  correlation in semiconductors and insulators: Band gaps and quasiparticle
  energies}},}\ }\href {\doibase 10.1103/PhysRevB.34.5390} {\bibfield
  {journal} {\bibinfo  {journal} {Phys. Rev. B}\ }\textbf {\bibinfo {volume}
  {34}},\ \bibinfo {pages} {5390--5413} (\bibinfo {year} {1986})}\BibitemShut
  {NoStop}%
\bibitem [{\citenamefont {Aryasetiawan}\ and\ \citenamefont
  {Gunnarsson}(1998)}]{Aryasetiawan1998Mar}%
  \BibitemOpen
  \bibfield  {author} {\bibinfo {author} {\bibfnamefont {F.}~\bibnamefont
  {Aryasetiawan}}\ and\ \bibinfo {author} {\bibfnamefont {O.}~\bibnamefont
  {Gunnarsson}},\ }\bibfield  {title} {\enquote {\bibinfo {title} {{The GW
  method}},}\ }\href {\doibase 10.1088/0034-4885/61/3/002} {\bibfield
  {journal} {\bibinfo  {journal} {Rep. Prog. Phys.}\ }\textbf {\bibinfo
  {volume} {61}},\ \bibinfo {pages} {237--312} (\bibinfo {year}
  {1998})}\BibitemShut {NoStop}%
\bibitem [{\citenamefont
  {L{\ifmmode\ddot{o}\else\"{o}\fi}wdin}(1970)}]{Lowdin1970Jan}%
  \BibitemOpen
  \bibfield  {author} {\bibinfo {author} {\bibfnamefont {Per-Olov}\
  \bibnamefont {L{\ifmmode\ddot{o}\else\"{o}\fi}wdin}},\ }\bibfield  {title}
  {\enquote {\bibinfo {title} {{On the Nonorthogonality Problem}},}\ }in\ \href
  {\doibase 10.1016/S0065-3276(08)60339-1} {\emph {\bibinfo {booktitle}
  {{Advances in Quantum Chemistry}}}},\ Vol.~\bibinfo {volume} {5}\ (\bibinfo
  {publisher} {Academic Press},\ \bibinfo {address} {Cambridge, MA, USA},\
  \bibinfo {year} {1970})\ pp.\ \bibinfo {pages} {185--199}\BibitemShut
  {NoStop}%
\bibitem [{\citenamefont {Klahn}\ and\ \citenamefont
  {Bingel}(1977)}]{Klahn1977Jun}%
  \BibitemOpen
  \bibfield  {author} {\bibinfo {author} {\bibfnamefont {Bruno}\ \bibnamefont
  {Klahn}}\ and\ \bibinfo {author} {\bibfnamefont {Werner~A.}\ \bibnamefont
  {Bingel}},\ }\bibfield  {title} {\enquote {\bibinfo {title} {{Completeness
  and linear independence of basis sets used in quantum chemistry}},}\ }\href
  {\doibase 10.1002/qua.560110607} {\bibfield  {journal} {\bibinfo  {journal}
  {Int. J. Quantum Chem.}\ }\textbf {\bibinfo {volume} {11}},\ \bibinfo {pages}
  {943--957} (\bibinfo {year} {1977})}\BibitemShut {NoStop}%
\bibitem [{\citenamefont {VandeVondele}\ and\ \citenamefont
  {Hutter}(2007)}]{VandeVondele2007Sep}%
  \BibitemOpen
  \bibfield  {author} {\bibinfo {author} {\bibfnamefont {Joost}\ \bibnamefont
  {VandeVondele}}\ and\ \bibinfo {author} {\bibfnamefont
  {J{\ifmmode\ddot{u}\else\"{u}\fi}rg}\ \bibnamefont {Hutter}},\ }\bibfield
  {title} {\enquote {\bibinfo {title} {{Gaussian basis sets for accurate
  calculations on molecular systems in gas and condensed phases}},}\ }\href
  {\doibase 10.1063/1.2770708} {\bibfield  {journal} {\bibinfo  {journal} {J.
  Chem. Phys.}\ }\textbf {\bibinfo {volume} {127}},\ \bibinfo {pages} {114105}
  (\bibinfo {year} {2007})}\BibitemShut {NoStop}%
\bibitem [{\citenamefont {Peintinger}\ \emph {et~al.}(2013)\citenamefont
  {Peintinger}, \citenamefont {Oliveira},\ and\ \citenamefont
  {Bredow}}]{Peintinger2013Mar}%
  \BibitemOpen
  \bibfield  {author} {\bibinfo {author} {\bibfnamefont {Michael~F.}\
  \bibnamefont {Peintinger}}, \bibinfo {author} {\bibfnamefont {Daniel~Vilela}\
  \bibnamefont {Oliveira}}, \ and\ \bibinfo {author} {\bibfnamefont {Thomas}\
  \bibnamefont {Bredow}},\ }\bibfield  {title} {\enquote {\bibinfo {title}
  {{Consistent Gaussian basis sets of triple-zeta valence with polarization
  quality for solid-state calculations}},}\ }\href {\doibase 10.1002/jcc.23153}
  {\bibfield  {journal} {\bibinfo  {journal} {J. Comput. Chem.}\ }\textbf
  {\bibinfo {volume} {34}},\ \bibinfo {pages} {451--459} (\bibinfo {year}
  {2013})}\BibitemShut {NoStop}%
\bibitem [{\citenamefont {Laun}\ \emph {et~al.}(2018)\citenamefont {Laun},
  \citenamefont {Oliveira},\ and\ \citenamefont {Bredow}}]{Laun2018Jul}%
  \BibitemOpen
  \bibfield  {author} {\bibinfo {author} {\bibfnamefont {Joachim}\ \bibnamefont
  {Laun}}, \bibinfo {author} {\bibfnamefont {Daniel~Vilela}\ \bibnamefont
  {Oliveira}}, \ and\ \bibinfo {author} {\bibfnamefont {Thomas}\ \bibnamefont
  {Bredow}},\ }\bibfield  {title} {\enquote {\bibinfo {title} {{Consistent
  gaussian basis sets of double- and triple-zeta valence with polarization
  quality of the fifth period for solid-state calculations}},}\ }\href
  {\doibase 10.1002/jcc.25195} {\bibfield  {journal} {\bibinfo  {journal} {J.
  Comput. Chem.}\ }\textbf {\bibinfo {volume} {39}},\ \bibinfo {pages}
  {1285--1290} (\bibinfo {year} {2018})}\BibitemShut {NoStop}%
\bibitem [{\citenamefont {Oliveira}\ \emph {et~al.}(2019)\citenamefont
  {Oliveira}, \citenamefont {Laun}, \citenamefont {Peintinger},\ and\
  \citenamefont {Bredow}}]{Oliveira2019Oct}%
  \BibitemOpen
  \bibfield  {author} {\bibinfo {author} {\bibfnamefont {Daniel~Vilela}\
  \bibnamefont {Oliveira}}, \bibinfo {author} {\bibfnamefont {Joachim}\
  \bibnamefont {Laun}}, \bibinfo {author} {\bibfnamefont {Michael~F.}\
  \bibnamefont {Peintinger}}, \ and\ \bibinfo {author} {\bibfnamefont {Thomas}\
  \bibnamefont {Bredow}},\ }\bibfield  {title} {\enquote {\bibinfo {title}
  {{BSSE-correction scheme for consistent gaussian basis sets of double- and
  triple-zeta valence with polarization quality for solid-state
  calculations}},}\ }\href {\doibase 10.1002/jcc.26013} {\bibfield  {journal}
  {\bibinfo  {journal} {J. Comput. Chem.}\ }\textbf {\bibinfo {volume} {40}},\
  \bibinfo {pages} {2364--2376} (\bibinfo {year} {2019})}\BibitemShut {NoStop}%
\bibitem [{\citenamefont {Morales}\ and\ \citenamefont
  {Malone}(2020)}]{Morales2020Nov}%
  \BibitemOpen
  \bibfield  {author} {\bibinfo {author} {\bibfnamefont {Miguel~A.}\
  \bibnamefont {Morales}}\ and\ \bibinfo {author} {\bibfnamefont {Fionn~D.}\
  \bibnamefont {Malone}},\ }\bibfield  {title} {\enquote {\bibinfo {title}
  {{Accelerating the convergence of auxiliary-field quantum Monte Carlo in
  solids with optimized Gaussian basis sets}},}\ }\href {\doibase
  10.1063/5.0025390} {\bibfield  {journal} {\bibinfo  {journal} {J. Chem.
  Phys.}\ }\textbf {\bibinfo {volume} {153}},\ \bibinfo {pages} {194111}
  (\bibinfo {year} {2020})}\BibitemShut {NoStop}%
\bibitem [{\citenamefont {Daga}\ \emph {et~al.}(2020)\citenamefont {Daga},
  \citenamefont {Civalleri},\ and\ \citenamefont {Maschio}}]{Daga2020Apr}%
  \BibitemOpen
  \bibfield  {author} {\bibinfo {author} {\bibfnamefont {Loredana~Edith}\
  \bibnamefont {Daga}}, \bibinfo {author} {\bibfnamefont {Bartolomeo}\
  \bibnamefont {Civalleri}}, \ and\ \bibinfo {author} {\bibfnamefont {Lorenzo}\
  \bibnamefont {Maschio}},\ }\bibfield  {title} {\enquote {\bibinfo {title}
  {{Gaussian Basis Sets for Crystalline Solids: All-Purpose Basis Set Libraries
  vs System-Specific Optimizations}},}\ }\href {\doibase
  10.1021/acs.jctc.9b01004} {\bibfield  {journal} {\bibinfo  {journal} {J.
  Chem. Theory Comput.}\ }\textbf {\bibinfo {volume} {16}},\ \bibinfo {pages}
  {2192--2201} (\bibinfo {year} {2020})}\BibitemShut {NoStop}%
\bibitem [{\citenamefont {Zhou}\ \emph {et~al.}(2021)\citenamefont {Zhou},
  \citenamefont {Gull},\ and\ \citenamefont {Zgid}}]{Zhou2021May}%
  \BibitemOpen
  \bibfield  {author} {\bibinfo {author} {\bibfnamefont {Yanbing}\ \bibnamefont
  {Zhou}}, \bibinfo {author} {\bibfnamefont {Emanuel}\ \bibnamefont {Gull}}, \
  and\ \bibinfo {author} {\bibfnamefont {Dominika}\ \bibnamefont {Zgid}},\
  }\bibfield  {title} {\enquote {\bibinfo {title} {{Material-Specific
  Optimization of Gaussian Basis Sets against Plane Wave Data}},}\ }\href
  {\doibase 10.1021/acs.jctc.1c00491} {\bibfield  {journal} {\bibinfo
  {journal} {J. Chem. Theory Comput.}\ }\textbf {\bibinfo {volume} {17}},\
  \bibinfo {pages} {5611--5622} (\bibinfo {year} {2021})}\BibitemShut {NoStop}%
\bibitem [{\citenamefont {Li}\ \emph {et~al.}(1995)\citenamefont {Li},
  \citenamefont {Wrinn}, \citenamefont {Newsam},\ and\ \citenamefont
  {Sears}}]{Li1995Feb}%
  \BibitemOpen
  \bibfield  {author} {\bibinfo {author} {\bibfnamefont {Y.~S.}\ \bibnamefont
  {Li}}, \bibinfo {author} {\bibfnamefont {M.~C.}\ \bibnamefont {Wrinn}},
  \bibinfo {author} {\bibfnamefont {J.~M.}\ \bibnamefont {Newsam}}, \ and\
  \bibinfo {author} {\bibfnamefont {M.~P.}\ \bibnamefont {Sears}},\ }\bibfield
  {title} {\enquote {\bibinfo {title} {{Parallel implementation of a mesh-based
  density functional electronic structure code}},}\ }\href {\doibase
  10.1002/jcc.540160209} {\bibfield  {journal} {\bibinfo  {journal} {J. Comput.
  Chem.}\ }\textbf {\bibinfo {volume} {16}},\ \bibinfo {pages} {226--234}
  (\bibinfo {year} {1995})}\BibitemShut {NoStop}%
\bibitem [{\citenamefont {Soler}\ \emph {et~al.}(2002)\citenamefont {Soler},
  \citenamefont {Artacho}, \citenamefont {Gale}, \citenamefont
  {Garc{\ifmmode\acute{\imath}\else\'{\i}\fi}a}, \citenamefont {Junquera},
  \citenamefont {Ordej{\ifmmode\acute{o}\else\'{o}\fi}n},\ and\ \citenamefont
  {S{\ifmmode\acute{a}\else\'{a}\fi}nchez-Portal}}]{Soler2002Mar}%
  \BibitemOpen
  \bibfield  {author} {\bibinfo {author} {\bibfnamefont
  {Jos{\ifmmode\acute{e}\else\'{e}\fi}~M.}\ \bibnamefont {Soler}}, \bibinfo
  {author} {\bibfnamefont {Emilio}\ \bibnamefont {Artacho}}, \bibinfo {author}
  {\bibfnamefont {Julian~D.}\ \bibnamefont {Gale}}, \bibinfo {author}
  {\bibfnamefont {Alberto}\ \bibnamefont
  {Garc{\ifmmode\acute{\imath}\else\'{\i}\fi}a}}, \bibinfo {author}
  {\bibfnamefont {Javier}\ \bibnamefont {Junquera}}, \bibinfo {author}
  {\bibfnamefont {Pablo}\ \bibnamefont
  {Ordej{\ifmmode\acute{o}\else\'{o}\fi}n}}, \ and\ \bibinfo {author}
  {\bibfnamefont {Daniel}\ \bibnamefont
  {S{\ifmmode\acute{a}\else\'{a}\fi}nchez-Portal}},\ }\bibfield  {title}
  {\enquote {\bibinfo {title} {{The SIESTA method for ab initio order-N}},}\
  }\href {\doibase 10.1088/0953-8984/14/11/302} {\bibfield  {journal} {\bibinfo
   {journal} {J. Phys.: Condens. Matter}\ }\textbf {\bibinfo {volume} {14}},\
  \bibinfo {pages} {2745--2779} (\bibinfo {year} {2002})}\BibitemShut {NoStop}%
\bibitem [{\citenamefont {Artacho}\ \emph {et~al.}(2008)\citenamefont
  {Artacho}, \citenamefont {Anglada}, \citenamefont
  {Di{\ifmmode\acute{e}\else\'{e}\fi}guez}, \citenamefont {Gale}, \citenamefont
  {Garc{\ifmmode\acute{\imath}\else\'{\i}\fi}a}, \citenamefont {Junquera},
  \citenamefont {Martin}, \citenamefont
  {Ordej{\ifmmode\acute{o}\else\'{o}\fi}n}, \citenamefont {Pruneda},
  \citenamefont {S{\ifmmode\acute{a}\else\'{a}\fi}nchez-Portal},\ and\
  \citenamefont {Soler}}]{Artacho2008Jan}%
  \BibitemOpen
  \bibfield  {author} {\bibinfo {author} {\bibfnamefont {Emilio}\ \bibnamefont
  {Artacho}}, \bibinfo {author} {\bibfnamefont {E.}~\bibnamefont {Anglada}},
  \bibinfo {author} {\bibfnamefont {O.}~\bibnamefont
  {Di{\ifmmode\acute{e}\else\'{e}\fi}guez}}, \bibinfo {author} {\bibfnamefont
  {J.~D.}\ \bibnamefont {Gale}}, \bibinfo {author} {\bibfnamefont
  {A.}~\bibnamefont {Garc{\ifmmode\acute{\imath}\else\'{\i}\fi}a}}, \bibinfo
  {author} {\bibfnamefont {J.}~\bibnamefont {Junquera}}, \bibinfo {author}
  {\bibfnamefont {R.~M.}\ \bibnamefont {Martin}}, \bibinfo {author}
  {\bibfnamefont {P.}~\bibnamefont {Ordej{\ifmmode\acute{o}\else\'{o}\fi}n}},
  \bibinfo {author} {\bibfnamefont {J.~M.}\ \bibnamefont {Pruneda}}, \bibinfo
  {author} {\bibfnamefont {D.}~\bibnamefont
  {S{\ifmmode\acute{a}\else\'{a}\fi}nchez-Portal}}, \ and\ \bibinfo {author}
  {\bibfnamefont {J.~M.}\ \bibnamefont {Soler}},\ }\bibfield  {title} {\enquote
  {\bibinfo {title} {{The SIESTA method; developments and applicability}},}\
  }\href {\doibase 10.1088/0953-8984/20/6/064208} {\bibfield  {journal}
  {\bibinfo  {journal} {J. Phys.: Condens. Matter}\ }\textbf {\bibinfo {volume}
  {20}},\ \bibinfo {pages} {064208} (\bibinfo {year} {2008})}\BibitemShut
  {NoStop}%
\bibitem [{\citenamefont {Verzijl}\ and\ \citenamefont
  {Thijssen}(2012)}]{Verzijl2012Nov}%
  \BibitemOpen
  \bibfield  {author} {\bibinfo {author} {\bibfnamefont {C.~J.~O.}\
  \bibnamefont {Verzijl}}\ and\ \bibinfo {author} {\bibfnamefont {J.~M.}\
  \bibnamefont {Thijssen}},\ }\bibfield  {title} {\enquote {\bibinfo {title}
  {{DFT-Based Molecular Transport Implementation in ADF/BAND}},}\ }\href
  {\doibase 10.1021/jp3044225} {\bibfield  {journal} {\bibinfo  {journal} {J.
  Phys. Chem. C}\ }\textbf {\bibinfo {volume} {116}},\ \bibinfo {pages}
  {24393--24412} (\bibinfo {year} {2012})}\BibitemShut {NoStop}%
\bibitem [{\citenamefont {Kudin}\ and\ \citenamefont
  {Scuseria}(2000)}]{Kudin2000Jun}%
  \BibitemOpen
  \bibfield  {author} {\bibinfo {author} {\bibfnamefont {Konstantin~N.}\
  \bibnamefont {Kudin}}\ and\ \bibinfo {author} {\bibfnamefont {Gustavo~E.}\
  \bibnamefont {Scuseria}},\ }\bibfield  {title} {\enquote {\bibinfo {title}
  {{Linear-scaling density-functional theory with Gaussian orbitals and
  periodic boundary conditions: Efficient evaluation of energy and forces via
  the fast multipole method}},}\ }\href {\doibase 10.1103/PhysRevB.61.16440}
  {\bibfield  {journal} {\bibinfo  {journal} {Phys. Rev. B}\ }\textbf {\bibinfo
  {volume} {61}},\ \bibinfo {pages} {16440--16453} (\bibinfo {year}
  {2000})}\BibitemShut {NoStop}%
\bibitem [{\citenamefont {Balasubramani}\ \emph {et~al.}(2020)\citenamefont
  {Balasubramani}, \citenamefont {Chen}, \citenamefont {Coriani}, \citenamefont
  {Diedenhofen}, \citenamefont {Frank}, \citenamefont {Franzke}, \citenamefont
  {Furche}, \citenamefont {Grotjahn}, \citenamefont {Harding}, \citenamefont
  {H{\ifmmode\ddot{a}\else\"{a}\fi}ttig}, \citenamefont {Hellweg},
  \citenamefont {Helmich-Paris}, \citenamefont {Holzer}, \citenamefont
  {Huniar}, \citenamefont {Kaupp}, \citenamefont {Marefat~Khah}, \citenamefont
  {Karbalaei~Khani}, \citenamefont {M{\ifmmode\ddot{u}\else\"{u}\fi}ller},
  \citenamefont {Mack}, \citenamefont {Nguyen}, \citenamefont {Parker},
  \citenamefont {Perlt}, \citenamefont {Rappoport}, \citenamefont {Reiter},
  \citenamefont {Roy}, \citenamefont {R{\ifmmode\ddot{u}\else\"{u}\fi}ckert},
  \citenamefont {Schmitz}, \citenamefont {Sierka}, \citenamefont {Tapavicza},
  \citenamefont {Tew}, \citenamefont {van
  W{\ifmmode\ddot{u}\else\"{u}\fi}llen}, \citenamefont {Voora}, \citenamefont
  {Weigend}, \citenamefont {Wody{\ifmmode\acute{n}\else\'{n}\fi}ski},\ and\
  \citenamefont {Yu}}]{Balasubramani2020May}%
  \BibitemOpen
  \bibfield  {author} {\bibinfo {author} {\bibfnamefont {Sree~Ganesh}\
  \bibnamefont {Balasubramani}}, \bibinfo {author} {\bibfnamefont {Guo~P.}\
  \bibnamefont {Chen}}, \bibinfo {author} {\bibfnamefont {Sonia}\ \bibnamefont
  {Coriani}}, \bibinfo {author} {\bibfnamefont {Michael}\ \bibnamefont
  {Diedenhofen}}, \bibinfo {author} {\bibfnamefont {Marius~S.}\ \bibnamefont
  {Frank}}, \bibinfo {author} {\bibfnamefont {Yannick~J.}\ \bibnamefont
  {Franzke}}, \bibinfo {author} {\bibfnamefont {Filipp}\ \bibnamefont
  {Furche}}, \bibinfo {author} {\bibfnamefont {Robin}\ \bibnamefont
  {Grotjahn}}, \bibinfo {author} {\bibfnamefont {Michael~E.}\ \bibnamefont
  {Harding}}, \bibinfo {author} {\bibfnamefont {Christof}\ \bibnamefont
  {H{\ifmmode\ddot{a}\else\"{a}\fi}ttig}}, \bibinfo {author} {\bibfnamefont
  {Arnim}\ \bibnamefont {Hellweg}}, \bibinfo {author} {\bibfnamefont
  {Benjamin}\ \bibnamefont {Helmich-Paris}}, \bibinfo {author} {\bibfnamefont
  {Christof}\ \bibnamefont {Holzer}}, \bibinfo {author} {\bibfnamefont {Uwe}\
  \bibnamefont {Huniar}}, \bibinfo {author} {\bibfnamefont {Martin}\
  \bibnamefont {Kaupp}}, \bibinfo {author} {\bibfnamefont {Alireza}\
  \bibnamefont {Marefat~Khah}}, \bibinfo {author} {\bibfnamefont {Sarah}\
  \bibnamefont {Karbalaei~Khani}}, \bibinfo {author} {\bibfnamefont {Thomas}\
  \bibnamefont {M{\ifmmode\ddot{u}\else\"{u}\fi}ller}}, \bibinfo {author}
  {\bibfnamefont {Fabian}\ \bibnamefont {Mack}}, \bibinfo {author}
  {\bibfnamefont {Brian~D.}\ \bibnamefont {Nguyen}}, \bibinfo {author}
  {\bibfnamefont {Shane~M.}\ \bibnamefont {Parker}}, \bibinfo {author}
  {\bibfnamefont {Eva}\ \bibnamefont {Perlt}}, \bibinfo {author} {\bibfnamefont
  {Dmitrij}\ \bibnamefont {Rappoport}}, \bibinfo {author} {\bibfnamefont
  {Kevin}\ \bibnamefont {Reiter}}, \bibinfo {author} {\bibfnamefont {Saswata}\
  \bibnamefont {Roy}}, \bibinfo {author} {\bibfnamefont {Matthias}\
  \bibnamefont {R{\ifmmode\ddot{u}\else\"{u}\fi}ckert}}, \bibinfo {author}
  {\bibfnamefont {Gunnar}\ \bibnamefont {Schmitz}}, \bibinfo {author}
  {\bibfnamefont {Marek}\ \bibnamefont {Sierka}}, \bibinfo {author}
  {\bibfnamefont {Enrico}\ \bibnamefont {Tapavicza}}, \bibinfo {author}
  {\bibfnamefont {David~P.}\ \bibnamefont {Tew}}, \bibinfo {author}
  {\bibfnamefont {Christoph}\ \bibnamefont {van
  W{\ifmmode\ddot{u}\else\"{u}\fi}llen}}, \bibinfo {author} {\bibfnamefont
  {Vamsee~K.}\ \bibnamefont {Voora}}, \bibinfo {author} {\bibfnamefont
  {Florian}\ \bibnamefont {Weigend}}, \bibinfo {author} {\bibfnamefont {Artur}\
  \bibnamefont {Wody{\ifmmode\acute{n}\else\'{n}\fi}ski}}, \ and\ \bibinfo
  {author} {\bibfnamefont {Jason~M.}\ \bibnamefont {Yu}},\ }\bibfield  {title}
  {\enquote {\bibinfo {title} {{TURBOMOLE: Modular program suite for ab initio
  quantum-chemical and condensed-matter simulations}},}\ }\href {\doibase
  10.1063/5.0004635} {\bibfield  {journal} {\bibinfo  {journal} {J. Chem.
  Phys.}\ }\textbf {\bibinfo {volume} {152}},\ \bibinfo {pages} {184107}
  (\bibinfo {year} {2020})}\BibitemShut {NoStop}%
\bibitem [{\citenamefont {Dovesi}\ \emph {et~al.}(2020)\citenamefont {Dovesi},
  \citenamefont {Pascale}, \citenamefont {Civalleri}, \citenamefont {Doll},
  \citenamefont {Harrison}, \citenamefont {Bush}, \citenamefont {D{'}Arco},
  \citenamefont {No{\ifmmode\ddot{e}\else\"{e}\fi}l}, \citenamefont
  {R{\ifmmode\acute{e}\else\'{e}\fi}rat}, \citenamefont
  {Carbonni{\ifmmode\grave{e}\else\`{e}\fi}re}, \citenamefont
  {Caus{\ifmmode\grave{a}\else\`{a}\fi}}, \citenamefont {Salustro},
  \citenamefont {Lacivita}, \citenamefont {Kirtman}, \citenamefont {Ferrari},
  \citenamefont {Gentile}, \citenamefont {Baima}, \citenamefont {Ferrero},
  \citenamefont {Demichelis},\ and\ \citenamefont
  {De~La~Pierre}}]{Dovesi2020May}%
  \BibitemOpen
  \bibfield  {author} {\bibinfo {author} {\bibfnamefont {Roberto}\ \bibnamefont
  {Dovesi}}, \bibinfo {author} {\bibfnamefont {Fabien}\ \bibnamefont
  {Pascale}}, \bibinfo {author} {\bibfnamefont {Bartolomeo}\ \bibnamefont
  {Civalleri}}, \bibinfo {author} {\bibfnamefont {Klaus}\ \bibnamefont {Doll}},
  \bibinfo {author} {\bibfnamefont {Nicholas~M.}\ \bibnamefont {Harrison}},
  \bibinfo {author} {\bibfnamefont {Ian}\ \bibnamefont {Bush}}, \bibinfo
  {author} {\bibfnamefont {Philippe}\ \bibnamefont {D{'}Arco}}, \bibinfo
  {author} {\bibfnamefont {Yves}\ \bibnamefont
  {No{\ifmmode\ddot{e}\else\"{e}\fi}l}}, \bibinfo {author} {\bibfnamefont
  {Michel}\ \bibnamefont {R{\ifmmode\acute{e}\else\'{e}\fi}rat}}, \bibinfo
  {author} {\bibfnamefont {Philippe}\ \bibnamefont
  {Carbonni{\ifmmode\grave{e}\else\`{e}\fi}re}}, \bibinfo {author}
  {\bibfnamefont {Mauro}\ \bibnamefont {Caus{\ifmmode\grave{a}\else\`{a}\fi}}},
  \bibinfo {author} {\bibfnamefont {Simone}\ \bibnamefont {Salustro}}, \bibinfo
  {author} {\bibfnamefont {Valentina}\ \bibnamefont {Lacivita}}, \bibinfo
  {author} {\bibfnamefont {Bernard}\ \bibnamefont {Kirtman}}, \bibinfo {author}
  {\bibfnamefont {Anna~Maria}\ \bibnamefont {Ferrari}}, \bibinfo {author}
  {\bibfnamefont {Francesco~Silvio}\ \bibnamefont {Gentile}}, \bibinfo {author}
  {\bibfnamefont {Jacopo}\ \bibnamefont {Baima}}, \bibinfo {author}
  {\bibfnamefont {Mauro}\ \bibnamefont {Ferrero}}, \bibinfo {author}
  {\bibfnamefont {Raffaella}\ \bibnamefont {Demichelis}}, \ and\ \bibinfo
  {author} {\bibfnamefont {Marco}\ \bibnamefont {De~La~Pierre}},\ }\bibfield
  {title} {\enquote {\bibinfo {title} {{The CRYSTAL code, 1976{\textendash}2020
  and beyond, a long story}},}\ }\href {\doibase 10.1063/5.0004892} {\bibfield
  {journal} {\bibinfo  {journal} {J. Chem. Phys.}\ }\textbf {\bibinfo {volume}
  {152}},\ \bibinfo {pages} {204111} (\bibinfo {year} {2020})}\BibitemShut
  {NoStop}%
\bibitem [{\citenamefont {K{\ifmmode\ddot{u}\else\"{u}\fi}hne}\ \emph
  {et~al.}(2020)\citenamefont {K{\ifmmode\ddot{u}\else\"{u}\fi}hne},
  \citenamefont {Iannuzzi}, \citenamefont {Del~Ben}, \citenamefont {Rybkin},
  \citenamefont {Seewald}, \citenamefont {Stein}, \citenamefont {Laino},
  \citenamefont {Khaliullin}, \citenamefont
  {Sch{\ifmmode\ddot{u}\else\"{u}\fi}tt}, \citenamefont {Schiffmann},
  \citenamefont {Golze}, \citenamefont {Wilhelm}, \citenamefont {Chulkov},
  \citenamefont {Bani-Hashemian}, \citenamefont {Weber}, \citenamefont
  {Bor{\ifmmode\check{s}\else\v{s}\fi}tnik}, \citenamefont {Taillefumier},
  \citenamefont {Jakobovits}, \citenamefont {Lazzaro}, \citenamefont {Pabst},
  \citenamefont {M{\ifmmode\ddot{u}\else\"{u}\fi}ller}, \citenamefont {Schade},
  \citenamefont {Guidon}, \citenamefont {Andermatt}, \citenamefont {Holmberg},
  \citenamefont {Schenter}, \citenamefont {Hehn}, \citenamefont {Bussy},
  \citenamefont {Belleflamme}, \citenamefont {Tabacchi}, \citenamefont
  {Gl{\ifmmode\ddot{o}\else\"{o}\fi}{\ss}}, \citenamefont {Lass}, \citenamefont
  {Bethune}, \citenamefont {Mundy}, \citenamefont {Plessl}, \citenamefont
  {Watkins}, \citenamefont {VandeVondele}, \citenamefont {Krack},\ and\
  \citenamefont {Hutter}}]{Kuhne2020May}%
  \BibitemOpen
  \bibfield  {author} {\bibinfo {author} {\bibfnamefont {Thomas~D.}\
  \bibnamefont {K{\ifmmode\ddot{u}\else\"{u}\fi}hne}}, \bibinfo {author}
  {\bibfnamefont {Marcella}\ \bibnamefont {Iannuzzi}}, \bibinfo {author}
  {\bibfnamefont {Mauro}\ \bibnamefont {Del~Ben}}, \bibinfo {author}
  {\bibfnamefont {Vladimir~V.}\ \bibnamefont {Rybkin}}, \bibinfo {author}
  {\bibfnamefont {Patrick}\ \bibnamefont {Seewald}}, \bibinfo {author}
  {\bibfnamefont {Frederick}\ \bibnamefont {Stein}}, \bibinfo {author}
  {\bibfnamefont {Teodoro}\ \bibnamefont {Laino}}, \bibinfo {author}
  {\bibfnamefont {Rustam~Z.}\ \bibnamefont {Khaliullin}}, \bibinfo {author}
  {\bibfnamefont {Ole}\ \bibnamefont {Sch{\ifmmode\ddot{u}\else\"{u}\fi}tt}},
  \bibinfo {author} {\bibfnamefont {Florian}\ \bibnamefont {Schiffmann}},
  \bibinfo {author} {\bibfnamefont {Dorothea}\ \bibnamefont {Golze}}, \bibinfo
  {author} {\bibfnamefont {Jan}\ \bibnamefont {Wilhelm}}, \bibinfo {author}
  {\bibfnamefont {Sergey}\ \bibnamefont {Chulkov}}, \bibinfo {author}
  {\bibfnamefont {Mohammad~Hossein}\ \bibnamefont {Bani-Hashemian}}, \bibinfo
  {author} {\bibfnamefont {Val{\ifmmode\acute{e}\else\'{e}\fi}ry}\ \bibnamefont
  {Weber}}, \bibinfo {author} {\bibfnamefont {Urban}\ \bibnamefont
  {Bor{\ifmmode\check{s}\else\v{s}\fi}tnik}}, \bibinfo {author} {\bibfnamefont
  {Mathieu}\ \bibnamefont {Taillefumier}}, \bibinfo {author} {\bibfnamefont
  {Alice~Shoshana}\ \bibnamefont {Jakobovits}}, \bibinfo {author}
  {\bibfnamefont {Alfio}\ \bibnamefont {Lazzaro}}, \bibinfo {author}
  {\bibfnamefont {Hans}\ \bibnamefont {Pabst}}, \bibinfo {author}
  {\bibfnamefont {Tiziano}\ \bibnamefont
  {M{\ifmmode\ddot{u}\else\"{u}\fi}ller}}, \bibinfo {author} {\bibfnamefont
  {Robert}\ \bibnamefont {Schade}}, \bibinfo {author} {\bibfnamefont {Manuel}\
  \bibnamefont {Guidon}}, \bibinfo {author} {\bibfnamefont {Samuel}\
  \bibnamefont {Andermatt}}, \bibinfo {author} {\bibfnamefont {Nico}\
  \bibnamefont {Holmberg}}, \bibinfo {author} {\bibfnamefont {Gregory~K.}\
  \bibnamefont {Schenter}}, \bibinfo {author} {\bibfnamefont {Anna}\
  \bibnamefont {Hehn}}, \bibinfo {author} {\bibfnamefont {Augustin}\
  \bibnamefont {Bussy}}, \bibinfo {author} {\bibfnamefont {Fabian}\
  \bibnamefont {Belleflamme}}, \bibinfo {author} {\bibfnamefont {Gloria}\
  \bibnamefont {Tabacchi}}, \bibinfo {author} {\bibfnamefont {Andreas}\
  \bibnamefont {Gl{\ifmmode\ddot{o}\else\"{o}\fi}{\ss}}}, \bibinfo {author}
  {\bibfnamefont {Michael}\ \bibnamefont {Lass}}, \bibinfo {author}
  {\bibfnamefont {Iain}\ \bibnamefont {Bethune}}, \bibinfo {author}
  {\bibfnamefont {Christopher~J.}\ \bibnamefont {Mundy}}, \bibinfo {author}
  {\bibfnamefont {Christian}\ \bibnamefont {Plessl}}, \bibinfo {author}
  {\bibfnamefont {Matt}\ \bibnamefont {Watkins}}, \bibinfo {author}
  {\bibfnamefont {Joost}\ \bibnamefont {VandeVondele}}, \bibinfo {author}
  {\bibfnamefont {Matthias}\ \bibnamefont {Krack}}, \ and\ \bibinfo {author}
  {\bibfnamefont {J{\ifmmode\ddot{u}\else\"{u}\fi}rg}\ \bibnamefont {Hutter}},\
  }\bibfield  {title} {\enquote {\bibinfo {title} {{CP2K: An electronic
  structure and molecular dynamics software package - Quickstep: Efficient and
  accurate electronic structure calculations}},}\ }\href {\doibase
  10.1063/5.0007045} {\bibfield  {journal} {\bibinfo  {journal} {J. Chem.
  Phys.}\ }\textbf {\bibinfo {volume} {152}},\ \bibinfo {pages} {194103}
  (\bibinfo {year} {2020})}\BibitemShut {NoStop}%
\bibitem [{\citenamefont {Prentice}\ \emph {et~al.}(2020)\citenamefont
  {Prentice}, \citenamefont {Aarons}, \citenamefont {Womack}, \citenamefont
  {Allen}, \citenamefont {Andrinopoulos}, \citenamefont {Anton}, \citenamefont
  {Bell}, \citenamefont {Bhandari}, \citenamefont {Bramley}, \citenamefont
  {Charlton}, \citenamefont {Clements}, \citenamefont {Cole}, \citenamefont
  {Constantinescu}, \citenamefont {Corsetti}, \citenamefont {Dubois},
  \citenamefont {Duff}, \citenamefont
  {Escart{\ifmmode\acute{\imath}\else\'{\i}\fi}n}, \citenamefont {Greco},
  \citenamefont {Hill}, \citenamefont {Lee}, \citenamefont {Linscott},
  \citenamefont {O{'}Regan}, \citenamefont {Phipps}, \citenamefont {Ratcliff},
  \citenamefont {Serrano}, \citenamefont {Tait}, \citenamefont {Teobaldi},
  \citenamefont {Vitale}, \citenamefont {Yeung}, \citenamefont {Zuehlsdorff},
  \citenamefont {Dziedzic}, \citenamefont {Haynes}, \citenamefont {Hine},
  \citenamefont {Mostofi}, \citenamefont {Payne},\ and\ \citenamefont
  {Skylaris}}]{Prentice2020May}%
  \BibitemOpen
  \bibfield  {author} {\bibinfo {author} {\bibfnamefont {Joseph C.~A.}\
  \bibnamefont {Prentice}}, \bibinfo {author} {\bibfnamefont {Jolyon}\
  \bibnamefont {Aarons}}, \bibinfo {author} {\bibfnamefont {James~C.}\
  \bibnamefont {Womack}}, \bibinfo {author} {\bibfnamefont {Alice E.~A.}\
  \bibnamefont {Allen}}, \bibinfo {author} {\bibfnamefont {Lampros}\
  \bibnamefont {Andrinopoulos}}, \bibinfo {author} {\bibfnamefont {Lucian}\
  \bibnamefont {Anton}}, \bibinfo {author} {\bibfnamefont {Robert~A.}\
  \bibnamefont {Bell}}, \bibinfo {author} {\bibfnamefont {Arihant}\
  \bibnamefont {Bhandari}}, \bibinfo {author} {\bibfnamefont {Gabriel~A.}\
  \bibnamefont {Bramley}}, \bibinfo {author} {\bibfnamefont {Robert~J.}\
  \bibnamefont {Charlton}}, \bibinfo {author} {\bibfnamefont {Rebecca~J.}\
  \bibnamefont {Clements}}, \bibinfo {author} {\bibfnamefont {Daniel~J.}\
  \bibnamefont {Cole}}, \bibinfo {author} {\bibfnamefont {Gabriel}\
  \bibnamefont {Constantinescu}}, \bibinfo {author} {\bibfnamefont {Fabiano}\
  \bibnamefont {Corsetti}}, \bibinfo {author} {\bibfnamefont {Simon M.-M.}\
  \bibnamefont {Dubois}}, \bibinfo {author} {\bibfnamefont {Kevin K.~B.}\
  \bibnamefont {Duff}}, \bibinfo {author} {\bibfnamefont
  {Jos{\ifmmode\acute{e}\else\'{e}\fi}~Mar{\ifmmode\acute{\imath}\else\'{\i}\fi}a}\
  \bibnamefont {Escart{\ifmmode\acute{\imath}\else\'{\i}\fi}n}}, \bibinfo
  {author} {\bibfnamefont {Andrea}\ \bibnamefont {Greco}}, \bibinfo {author}
  {\bibfnamefont {Quintin}\ \bibnamefont {Hill}}, \bibinfo {author}
  {\bibfnamefont {Louis~P.}\ \bibnamefont {Lee}}, \bibinfo {author}
  {\bibfnamefont {Edward}\ \bibnamefont {Linscott}}, \bibinfo {author}
  {\bibfnamefont {David~D.}\ \bibnamefont {O{'}Regan}}, \bibinfo {author}
  {\bibfnamefont {Maximillian J.~S.}\ \bibnamefont {Phipps}}, \bibinfo {author}
  {\bibfnamefont {Laura~E.}\ \bibnamefont {Ratcliff}}, \bibinfo {author}
  {\bibfnamefont {{\ifmmode\acute{A}\else\'{A}\fi}lvaro~Ruiz}\ \bibnamefont
  {Serrano}}, \bibinfo {author} {\bibfnamefont {Edward~W.}\ \bibnamefont
  {Tait}}, \bibinfo {author} {\bibfnamefont {Gilberto}\ \bibnamefont
  {Teobaldi}}, \bibinfo {author} {\bibfnamefont {Valerio}\ \bibnamefont
  {Vitale}}, \bibinfo {author} {\bibfnamefont {Nelson}\ \bibnamefont {Yeung}},
  \bibinfo {author} {\bibfnamefont {Tim~J.}\ \bibnamefont {Zuehlsdorff}},
  \bibinfo {author} {\bibfnamefont {Jacek}\ \bibnamefont {Dziedzic}}, \bibinfo
  {author} {\bibfnamefont {Peter~D.}\ \bibnamefont {Haynes}}, \bibinfo {author}
  {\bibfnamefont {Nicholas D.~M.}\ \bibnamefont {Hine}}, \bibinfo {author}
  {\bibfnamefont {Arash~A.}\ \bibnamefont {Mostofi}}, \bibinfo {author}
  {\bibfnamefont {Mike~C.}\ \bibnamefont {Payne}}, \ and\ \bibinfo {author}
  {\bibfnamefont {Chris-Kriton}\ \bibnamefont {Skylaris}},\ }\bibfield  {title}
  {\enquote {\bibinfo {title} {{The ONETEP linear-scaling density functional
  theory program}},}\ }\href {\doibase 10.1063/5.0004445} {\bibfield  {journal}
  {\bibinfo  {journal} {J. Chem. Phys.}\ }\textbf {\bibinfo {volume} {152}},\
  \bibinfo {pages} {174111} (\bibinfo {year} {2020})}\BibitemShut {NoStop}%
\bibitem [{\citenamefont {Jensen}(2013)}]{jensen2013atomic}%
  \BibitemOpen
  \bibfield  {author} {\bibinfo {author} {\bibfnamefont {Frank}\ \bibnamefont
  {Jensen}},\ }\bibfield  {title} {\enquote {\bibinfo {title} {Atomic orbital
  basis sets},}\ }\href@noop {} {\bibfield  {journal} {\bibinfo  {journal}
  {WIRES: Comput. Mol. Sci.}\ }\textbf {\bibinfo {volume} {3}},\ \bibinfo
  {pages} {273--295} (\bibinfo {year} {2013})}\BibitemShut {NoStop}%
\bibitem [{\citenamefont {Nagy}\ and\ \citenamefont
  {Jensen}(2017)}]{nagy2017basis}%
  \BibitemOpen
  \bibfield  {author} {\bibinfo {author} {\bibfnamefont {Balazs}\ \bibnamefont
  {Nagy}}\ and\ \bibinfo {author} {\bibfnamefont {Frank}\ \bibnamefont
  {Jensen}},\ }\bibfield  {title} {\enquote {\bibinfo {title} {Basis sets in
  quantum chemistry},}\ }\href@noop {} {\bibfield  {journal} {\bibinfo
  {journal} {Rev. Computat. Chem.}\ }\textbf {\bibinfo {volume} {30}},\
  \bibinfo {pages} {93--149} (\bibinfo {year} {2017})}\BibitemShut {NoStop}%
\bibitem [{\citenamefont {Shavitt}(1993)}]{Shavitt1993Jan}%
  \BibitemOpen
  \bibfield  {author} {\bibinfo {author} {\bibfnamefont {Isaiah}\ \bibnamefont
  {Shavitt}},\ }\bibfield  {title} {\enquote {\bibinfo {title} {{The History
  and Evolution of Gaussian Basis Sets}},}\ }\href {\doibase
  10.1002/ijch.199300044} {\bibfield  {journal} {\bibinfo  {journal} {Isr. J.
  Chem.}\ }\textbf {\bibinfo {volume} {33}},\ \bibinfo {pages} {357--367}
  (\bibinfo {year} {1993})}\BibitemShut {NoStop}%
\bibitem [{\citenamefont {Pritchard}\ \emph {et~al.}(2019)\citenamefont
  {Pritchard}, \citenamefont {Altarawy}, \citenamefont {Didier}, \citenamefont
  {Gibson},\ and\ \citenamefont {Windus}}]{Pritchard2019Nov}%
  \BibitemOpen
  \bibfield  {author} {\bibinfo {author} {\bibfnamefont {Benjamin~P.}\
  \bibnamefont {Pritchard}}, \bibinfo {author} {\bibfnamefont {Doaa}\
  \bibnamefont {Altarawy}}, \bibinfo {author} {\bibfnamefont {Brett}\
  \bibnamefont {Didier}}, \bibinfo {author} {\bibfnamefont {Tara~D.}\
  \bibnamefont {Gibson}}, \ and\ \bibinfo {author} {\bibfnamefont {Theresa~L.}\
  \bibnamefont {Windus}},\ }\bibfield  {title} {\enquote {\bibinfo {title}
  {{New Basis Set Exchange: An Open, Up-to-Date Resource for the Molecular
  Sciences Community}},}\ }\href {\doibase 10.1021/acs.jcim.9b00725} {\bibfield
   {journal} {\bibinfo  {journal} {J. Chem. Inf. Model.}\ }\textbf {\bibinfo
  {volume} {59}},\ \bibinfo {pages} {4814--4820} (\bibinfo {year}
  {2019})}\BibitemShut {NoStop}%
\bibitem [{\citenamefont {Mcweeny}(1950)}]{Mcweeny1950Jul}%
  \BibitemOpen
  \bibfield  {author} {\bibinfo {author} {\bibfnamefont {R.}~\bibnamefont
  {Mcweeny}},\ }\bibfield  {title} {\enquote {\bibinfo {title} {{Gaussian
  Approximations, to Wave Functions - Nature}},}\ }\href {\doibase
  10.1038/166021a0} {\bibfield  {journal} {\bibinfo  {journal} {Nature}\
  }\textbf {\bibinfo {volume} {166}},\ \bibinfo {pages} {21--22} (\bibinfo
  {year} {1950})}\BibitemShut {NoStop}%
\bibitem [{\citenamefont {Boys~S.}(1950)}]{Boys1950Feb}%
  \BibitemOpen
  \bibfield  {author} {\bibinfo {author} {\bibfnamefont {F.}~\bibnamefont
  {Boys~S.}},\ }\bibfield  {title} {\enquote {\bibinfo {title} {{Electronic
  wave functions - I. A general method of calculation for the stationary states
  of any molecular system}},}\ }\href {\doibase 10.1098/rspa.1950.0036}
  {\bibfield  {journal} {\bibinfo  {journal} {Proc. R. Soc. London A - Math.
  Phys. Sci.}\ }\textbf {\bibinfo {volume} {200}},\ \bibinfo {pages} {542--554}
  (\bibinfo {year} {1950})}\BibitemShut {NoStop}%
\bibitem [{\citenamefont {Alml{\ifmmode\ddot{o}\else\"{o}\fi}f}\ and\
  \citenamefont {Taylor}(1987)}]{Almlof1987Apr}%
  \BibitemOpen
  \bibfield  {author} {\bibinfo {author} {\bibfnamefont {Jan}\ \bibnamefont
  {Alml{\ifmmode\ddot{o}\else\"{o}\fi}f}}\ and\ \bibinfo {author}
  {\bibfnamefont {Peter~R.}\ \bibnamefont {Taylor}},\ }\bibfield  {title}
  {\enquote {\bibinfo {title} {{General contraction of Gaussian basis sets. I.
  Atomic natural orbitals for first{-} and second{-}row atoms}},}\ }\href
  {\doibase 10.1063/1.451917} {\bibfield  {journal} {\bibinfo  {journal} {J.
  Chem. Phys.}\ }\textbf {\bibinfo {volume} {86}},\ \bibinfo {pages}
  {4070--4077} (\bibinfo {year} {1987})}\BibitemShut {NoStop}%
\bibitem [{\citenamefont {Dunning}(1989)}]{Dunning1989Jan}%
  \BibitemOpen
  \bibfield  {author} {\bibinfo {author} {\bibfnamefont {Thom~H.}\ \bibnamefont
  {Dunning}},\ }\bibfield  {title} {\enquote {\bibinfo {title} {{Gaussian basis
  sets for use in correlated molecular calculations. I. The atoms boron through
  neon and hydrogen}},}\ }\href {\doibase 10.1063/1.456153} {\bibfield
  {journal} {\bibinfo  {journal} {J. Chem. Phys.}\ }\textbf {\bibinfo {volume}
  {90}},\ \bibinfo {pages} {1007--1023} (\bibinfo {year} {1989})}\BibitemShut
  {NoStop}%
\bibitem [{\citenamefont {Jensen}(2001)}]{jensen2001polarization}%
  \BibitemOpen
  \bibfield  {author} {\bibinfo {author} {\bibfnamefont {Frank}\ \bibnamefont
  {Jensen}},\ }\bibfield  {title} {\enquote {\bibinfo {title} {Polarization
  consistent basis sets: Principles},}\ }\href@noop {} {\bibfield  {journal}
  {\bibinfo  {journal} {J. Chem. Phys.}\ }\textbf {\bibinfo {volume} {115}},\
  \bibinfo {pages} {9113--9125} (\bibinfo {year} {2001})}\BibitemShut {NoStop}%
\bibitem [{\citenamefont {Jensen}(2002)}]{jensen2002polarization}%
  \BibitemOpen
  \bibfield  {author} {\bibinfo {author} {\bibfnamefont {Frank}\ \bibnamefont
  {Jensen}},\ }\bibfield  {title} {\enquote {\bibinfo {title} {Polarization
  consistent basis sets. ii. estimating the kohn--sham basis set limit},}\
  }\href@noop {} {\bibfield  {journal} {\bibinfo  {journal} {J. Chem. Phys.}\
  }\textbf {\bibinfo {volume} {116}},\ \bibinfo {pages} {7372--7379} (\bibinfo
  {year} {2002})}\BibitemShut {NoStop}%
\bibitem [{\citenamefont {Bardo}\ and\ \citenamefont
  {Ruedenberg}(1973)}]{Bardo1973Dec}%
  \BibitemOpen
  \bibfield  {author} {\bibinfo {author} {\bibfnamefont {Richard~D.}\
  \bibnamefont {Bardo}}\ and\ \bibinfo {author} {\bibfnamefont {Klaus}\
  \bibnamefont {Ruedenberg}},\ }\bibfield  {title} {\enquote {\bibinfo {title}
  {{Even{-}tempered atomic orbitals. III. Economic deployment of Gaussian
  primitives in expanding atomic SCF orbitals}},}\ }\href {\doibase
  10.1063/1.1679964} {\bibfield  {journal} {\bibinfo  {journal} {J. Chem.
  Phys.}\ }\textbf {\bibinfo {volume} {59}},\ \bibinfo {pages} {5956--5965}
  (\bibinfo {year} {1973})}\BibitemShut {NoStop}%
\bibitem [{\citenamefont {Feller}\ and\ \citenamefont
  {Ruedenberg}(1979)}]{Feller1979Sep}%
  \BibitemOpen
  \bibfield  {author} {\bibinfo {author} {\bibfnamefont {David~F.}\
  \bibnamefont {Feller}}\ and\ \bibinfo {author} {\bibfnamefont {Klaus}\
  \bibnamefont {Ruedenberg}},\ }\bibfield  {title} {\enquote {\bibinfo {title}
  {{Systematic approach to extended even-tempered orbital bases for atomic and
  molecular calculations}},}\ }\href {\doibase 10.1007/BF00547681} {\bibfield
  {journal} {\bibinfo  {journal} {Theor. Chim. Acta}\ }\textbf {\bibinfo
  {volume} {52}},\ \bibinfo {pages} {231--251} (\bibinfo {year}
  {1979})}\BibitemShut {NoStop}%
\bibitem [{\citenamefont {Huzinaga}\ \emph {et~al.}(1985)\citenamefont
  {Huzinaga}, \citenamefont {Klobukowski},\ and\ \citenamefont
  {Tatewaki}}]{Huzinaga2011Feb}%
  \BibitemOpen
  \bibfield  {author} {\bibinfo {author} {\bibfnamefont {S.}~\bibnamefont
  {Huzinaga}}, \bibinfo {author} {\bibfnamefont {M.}~\bibnamefont
  {Klobukowski}}, \ and\ \bibinfo {author} {\bibfnamefont {H.}~\bibnamefont
  {Tatewaki}},\ }\bibfield  {title} {\enquote {\bibinfo {title} {{The
  well-tempered GTF basis sets and their applications in the SCF calculations
  on N2, CO, Na2, and P2}},}\ }\href {\doibase 10.1139/v85-302} {\bibfield
  {journal} {\bibinfo  {journal} {Can. J. Chem.}\ }\textbf {\bibinfo {volume}
  {63}},\ \bibinfo {pages} {1812} (\bibinfo {year} {1985})}\BibitemShut
  {NoStop}%
\bibitem [{\citenamefont {Weigend}\ and\ \citenamefont
  {Ahlrichs}(2005)}]{Weigend2005Aug}%
  \BibitemOpen
  \bibfield  {author} {\bibinfo {author} {\bibfnamefont {Florian}\ \bibnamefont
  {Weigend}}\ and\ \bibinfo {author} {\bibfnamefont {Reinhart}\ \bibnamefont
  {Ahlrichs}},\ }\bibfield  {title} {\enquote {\bibinfo {title} {{Balanced
  basis sets of split valence, triple zeta valence and quadruple zeta valence
  quality for H to Rn: Design and assessment of accuracy}},}\ }\href {\doibase
  10.1039/B508541A} {\bibfield  {journal} {\bibinfo  {journal} {Phys. Chem.
  Chem. Phys.}\ }\textbf {\bibinfo {volume} {7}},\ \bibinfo {pages}
  {3297--3305} (\bibinfo {year} {2005})}\BibitemShut {NoStop}%
\bibitem [{\citenamefont {Goedecker}\ \emph {et~al.}(1996)\citenamefont
  {Goedecker}, \citenamefont {Teter},\ and\ \citenamefont
  {Hutter}}]{Goedecker1996Jul}%
  \BibitemOpen
  \bibfield  {author} {\bibinfo {author} {\bibfnamefont {S.}~\bibnamefont
  {Goedecker}}, \bibinfo {author} {\bibfnamefont {M.}~\bibnamefont {Teter}}, \
  and\ \bibinfo {author} {\bibfnamefont {J.}~\bibnamefont {Hutter}},\
  }\bibfield  {title} {\enquote {\bibinfo {title} {{Separable dual-space
  Gaussian pseudopotentials}},}\ }\href {\doibase 10.1103/PhysRevB.54.1703}
  {\bibfield  {journal} {\bibinfo  {journal} {Phys. Rev. B}\ }\textbf {\bibinfo
  {volume} {54}},\ \bibinfo {pages} {1703--1710} (\bibinfo {year}
  {1996})}\BibitemShut {NoStop}%
\bibitem [{\citenamefont {Hartwigsen}\ \emph {et~al.}(1998)\citenamefont
  {Hartwigsen}, \citenamefont {Goedecker},\ and\ \citenamefont
  {Hutter}}]{Hartwigsen1998Aug}%
  \BibitemOpen
  \bibfield  {author} {\bibinfo {author} {\bibfnamefont {C.}~\bibnamefont
  {Hartwigsen}}, \bibinfo {author} {\bibfnamefont {S.}~\bibnamefont
  {Goedecker}}, \ and\ \bibinfo {author} {\bibfnamefont {J.}~\bibnamefont
  {Hutter}},\ }\bibfield  {title} {\enquote {\bibinfo {title} {{Relativistic
  separable dual-space Gaussian pseudopotentials from H to Rn}},}\ }\href
  {\doibase 10.1103/PhysRevB.58.3641} {\bibfield  {journal} {\bibinfo
  {journal} {Phys. Rev. B}\ }\textbf {\bibinfo {volume} {58}},\ \bibinfo
  {pages} {3641--3662} (\bibinfo {year} {1998})}\BibitemShut {NoStop}%
\bibitem [{\citenamefont {Giannozzi}\ \emph {et~al.}(2020)\citenamefont
  {Giannozzi}, \citenamefont {Baseggio}, \citenamefont {Bonf{\`a}},
  \citenamefont {Brunato}, \citenamefont {Car}, \citenamefont {Carnimeo},
  \citenamefont {Cavazzoni}, \citenamefont {De~Gironcoli}, \citenamefont
  {Delugas}, \citenamefont {Ferrari~Ruffino} \emph
  {et~al.}}]{giannozzi2020quantum}%
  \BibitemOpen
  \bibfield  {author} {\bibinfo {author} {\bibfnamefont {Paolo}\ \bibnamefont
  {Giannozzi}}, \bibinfo {author} {\bibfnamefont {Oscar}\ \bibnamefont
  {Baseggio}}, \bibinfo {author} {\bibfnamefont {Pietro}\ \bibnamefont
  {Bonf{\`a}}}, \bibinfo {author} {\bibfnamefont {Davide}\ \bibnamefont
  {Brunato}}, \bibinfo {author} {\bibfnamefont {Roberto}\ \bibnamefont {Car}},
  \bibinfo {author} {\bibfnamefont {Ivan}\ \bibnamefont {Carnimeo}}, \bibinfo
  {author} {\bibfnamefont {Carlo}\ \bibnamefont {Cavazzoni}}, \bibinfo {author}
  {\bibfnamefont {Stefano}\ \bibnamefont {De~Gironcoli}}, \bibinfo {author}
  {\bibfnamefont {Pietro}\ \bibnamefont {Delugas}}, \bibinfo {author}
  {\bibfnamefont {Fabrizio}\ \bibnamefont {Ferrari~Ruffino}},  \emph {et~al.},\
  }\bibfield  {title} {\enquote {\bibinfo {title} {Quantum espresso toward the
  exascale},}\ }\href@noop {} {\bibfield  {journal} {\bibinfo  {journal} {J.
  Chem. Phys.}\ }\textbf {\bibinfo {volume} {152}},\ \bibinfo {pages} {154105}
  (\bibinfo {year} {2020})}\BibitemShut {NoStop}%
\bibitem [{\citenamefont {Slater}(1951)}]{slater1951simplification}%
  \BibitemOpen
  \bibfield  {author} {\bibinfo {author} {\bibfnamefont {John~C}\ \bibnamefont
  {Slater}},\ }\bibfield  {title} {\enquote {\bibinfo {title} {A simplification
  of the hartree-fock method},}\ }\href@noop {} {\bibfield  {journal} {\bibinfo
   {journal} {Phys. Rev.}\ }\textbf {\bibinfo {volume} {81}},\ \bibinfo {pages}
  {385} (\bibinfo {year} {1951})}\BibitemShut {NoStop}%
\bibitem [{\citenamefont {Perdew}\ and\ \citenamefont
  {Zunger}(1981)}]{perdew1981self}%
  \BibitemOpen
  \bibfield  {author} {\bibinfo {author} {\bibfnamefont {John~P}\ \bibnamefont
  {Perdew}}\ and\ \bibinfo {author} {\bibfnamefont {Alex}\ \bibnamefont
  {Zunger}},\ }\bibfield  {title} {\enquote {\bibinfo {title} {Self-interaction
  correction to density-functional approximations for many-electron systems},}\
  }\href@noop {} {\bibfield  {journal} {\bibinfo  {journal} {Phys. Rev. B}\
  }\textbf {\bibinfo {volume} {23}},\ \bibinfo {pages} {5048} (\bibinfo {year}
  {1981})}\BibitemShut {NoStop}%
\bibitem [{\citenamefont {Perdew}\ \emph {et~al.}(1996)\citenamefont {Perdew},
  \citenamefont {Burke},\ and\ \citenamefont
  {Ernzerhof}}]{perdew1996generalized}%
  \BibitemOpen
  \bibfield  {author} {\bibinfo {author} {\bibfnamefont {John~P}\ \bibnamefont
  {Perdew}}, \bibinfo {author} {\bibfnamefont {Kieron}\ \bibnamefont {Burke}},
  \ and\ \bibinfo {author} {\bibfnamefont {Matthias}\ \bibnamefont
  {Ernzerhof}},\ }\bibfield  {title} {\enquote {\bibinfo {title} {Generalized
  gradient approximation made simple},}\ }\href@noop {} {\bibfield  {journal}
  {\bibinfo  {journal} {Phys. Rev. Lett.}\ }\textbf {\bibinfo {volume} {77}},\
  \bibinfo {pages} {3865} (\bibinfo {year} {1996})}\BibitemShut {NoStop}%
\bibitem [{\citenamefont {Perdew}\ \emph {et~al.}(2008)\citenamefont {Perdew},
  \citenamefont {Ruzsinszky}, \citenamefont {Csonka}, \citenamefont {Vydrov},
  \citenamefont {Scuseria}, \citenamefont {Constantin}, \citenamefont {Zhou},\
  and\ \citenamefont {Burke}}]{Perdew2008Apr}%
  \BibitemOpen
  \bibfield  {author} {\bibinfo {author} {\bibfnamefont {John~P.}\ \bibnamefont
  {Perdew}}, \bibinfo {author} {\bibfnamefont {Adrienn}\ \bibnamefont
  {Ruzsinszky}}, \bibinfo {author} {\bibfnamefont
  {G{\ifmmode\acute{a}\else\'{a}\fi}bor~I.}\ \bibnamefont {Csonka}}, \bibinfo
  {author} {\bibfnamefont {Oleg~A.}\ \bibnamefont {Vydrov}}, \bibinfo {author}
  {\bibfnamefont {Gustavo~E.}\ \bibnamefont {Scuseria}}, \bibinfo {author}
  {\bibfnamefont {Lucian~A.}\ \bibnamefont {Constantin}}, \bibinfo {author}
  {\bibfnamefont {Xiaolan}\ \bibnamefont {Zhou}}, \ and\ \bibinfo {author}
  {\bibfnamefont {Kieron}\ \bibnamefont {Burke}},\ }\bibfield  {title}
  {\enquote {\bibinfo {title} {{Restoring the Density-Gradient Expansion for
  Exchange in Solids and Surfaces}},}\ }\href {\doibase
  10.1103/PhysRevLett.100.136406} {\bibfield  {journal} {\bibinfo  {journal}
  {Phys. Rev. Lett.}\ }\textbf {\bibinfo {volume} {100}},\ \bibinfo {pages}
  {136406} (\bibinfo {year} {2008})}\BibitemShut {NoStop}%
\bibitem [{\citenamefont {Zhang}\ and\ \citenamefont
  {Yang}(1998)}]{Zhang1998Jan}%
  \BibitemOpen
  \bibfield  {author} {\bibinfo {author} {\bibfnamefont {Yingkai}\ \bibnamefont
  {Zhang}}\ and\ \bibinfo {author} {\bibfnamefont {Weitao}\ \bibnamefont
  {Yang}},\ }\bibfield  {title} {\enquote {\bibinfo {title} {{Comment on
  ``Generalized Gradient Approximation Made Simple''}},}\ }\href {\doibase
  10.1103/PhysRevLett.80.890} {\bibfield  {journal} {\bibinfo  {journal} {Phys.
  Rev. Lett.}\ }\textbf {\bibinfo {volume} {80}},\ \bibinfo {pages} {890}
  (\bibinfo {year} {1998})}\BibitemShut {NoStop}%
\bibitem [{\citenamefont {Becke}(1988)}]{becke1988density}%
  \BibitemOpen
  \bibfield  {author} {\bibinfo {author} {\bibfnamefont {Axel~D}\ \bibnamefont
  {Becke}},\ }\bibfield  {title} {\enquote {\bibinfo {title}
  {Density-functional exchange-energy approximation with correct asymptotic
  behavior},}\ }\href@noop {} {\bibfield  {journal} {\bibinfo  {journal} {Phys.
  Rev. A}\ }\textbf {\bibinfo {volume} {38}},\ \bibinfo {pages} {3098}
  (\bibinfo {year} {1988})}\BibitemShut {NoStop}%
\bibitem [{\citenamefont {Lee}\ \emph {et~al.}(1988)\citenamefont {Lee},
  \citenamefont {Yang},\ and\ \citenamefont {Parr}}]{lee1988development}%
  \BibitemOpen
  \bibfield  {author} {\bibinfo {author} {\bibfnamefont {Chengteh}\
  \bibnamefont {Lee}}, \bibinfo {author} {\bibfnamefont {Weitao}\ \bibnamefont
  {Yang}}, \ and\ \bibinfo {author} {\bibfnamefont {Robert~G}\ \bibnamefont
  {Parr}},\ }\bibfield  {title} {\enquote {\bibinfo {title} {Development of the
  colle-salvetti correlation-energy formula into a functional of the electron
  density},}\ }\href@noop {} {\bibfield  {journal} {\bibinfo  {journal} {Phys.
  Rev. B}\ }\textbf {\bibinfo {volume} {37}},\ \bibinfo {pages} {785} (\bibinfo
  {year} {1988})}\BibitemShut {NoStop}%
\bibitem [{\citenamefont {Grimme}(2006)}]{Grimme2006Nov}%
  \BibitemOpen
  \bibfield  {author} {\bibinfo {author} {\bibfnamefont {Stefan}\ \bibnamefont
  {Grimme}},\ }\bibfield  {title} {\enquote {\bibinfo {title} {{Semiempirical
  GGA-type density functional constructed with a long-range dispersion
  correction}},}\ }\href {\doibase 10.1002/jcc.20495} {\bibfield  {journal}
  {\bibinfo  {journal} {J. Comput. Chem.}\ }\textbf {\bibinfo {volume} {27}},\
  \bibinfo {pages} {1787--1799} (\bibinfo {year} {2006})}\BibitemShut {NoStop}%
\bibitem [{\citenamefont {Sun}\ \emph {et~al.}(2015)\citenamefont {Sun},
  \citenamefont {Ruzsinszky},\ and\ \citenamefont {Perdew}}]{sun2015strongly}%
  \BibitemOpen
  \bibfield  {author} {\bibinfo {author} {\bibfnamefont {Jianwei}\ \bibnamefont
  {Sun}}, \bibinfo {author} {\bibfnamefont {Adrienn}\ \bibnamefont
  {Ruzsinszky}}, \ and\ \bibinfo {author} {\bibfnamefont {John~P}\ \bibnamefont
  {Perdew}},\ }\bibfield  {title} {\enquote {\bibinfo {title} {Strongly
  constrained and appropriately normed semilocal density functional},}\
  }\href@noop {} {\bibfield  {journal} {\bibinfo  {journal} {Phys. Rev. Lett.}\
  }\textbf {\bibinfo {volume} {115}},\ \bibinfo {pages} {036402} (\bibinfo
  {year} {2015})}\BibitemShut {NoStop}%
\bibitem [{\citenamefont {Zhao}\ and\ \citenamefont
  {Truhlar}(2006)}]{Zhao2006Nov}%
  \BibitemOpen
  \bibfield  {author} {\bibinfo {author} {\bibfnamefont {Yan}\ \bibnamefont
  {Zhao}}\ and\ \bibinfo {author} {\bibfnamefont {Donald~G.}\ \bibnamefont
  {Truhlar}},\ }\bibfield  {title} {\enquote {\bibinfo {title} {{A new local
  density functional for main-group thermochemistry, transition metal bonding,
  thermochemical kinetics, and noncovalent interactions}},}\ }\href {\doibase
  10.1063/1.2370993} {\bibfield  {journal} {\bibinfo  {journal} {J. Chem.
  Phys.}\ }\textbf {\bibinfo {volume} {125}},\ \bibinfo {pages} {194101}
  (\bibinfo {year} {2006})}\BibitemShut {NoStop}%
\bibitem [{\citenamefont {Yu}\ \emph {et~al.}(2016)\citenamefont {Yu},
  \citenamefont {He},\ and\ \citenamefont {Truhlar}}]{Yu2016Mar}%
  \BibitemOpen
  \bibfield  {author} {\bibinfo {author} {\bibfnamefont {Haoyu~S.}\
  \bibnamefont {Yu}}, \bibinfo {author} {\bibfnamefont {Xiao}\ \bibnamefont
  {He}}, \ and\ \bibinfo {author} {\bibfnamefont {Donald~G.}\ \bibnamefont
  {Truhlar}},\ }\bibfield  {title} {\enquote {\bibinfo {title} {{MN15-L: A New
  Local Exchange-Correlation Functional for Kohn{\textendash}Sham Density
  Functional Theory with Broad Accuracy for Atoms, Molecules, and Solids}},}\
  }\href {\doibase 10.1021/acs.jctc.5b01082} {\bibfield  {journal} {\bibinfo
  {journal} {J. Chem. Theory Comput.}\ }\textbf {\bibinfo {volume} {12}},\
  \bibinfo {pages} {1280--1293} (\bibinfo {year} {2016})}\BibitemShut {NoStop}%
\bibitem [{\citenamefont {Mardirossian}\ and\ \citenamefont
  {Head-Gordon}(2015)}]{mardirossian2015mapping}%
  \BibitemOpen
  \bibfield  {author} {\bibinfo {author} {\bibfnamefont {Narbe}\ \bibnamefont
  {Mardirossian}}\ and\ \bibinfo {author} {\bibfnamefont {Martin}\ \bibnamefont
  {Head-Gordon}},\ }\bibfield  {title} {\enquote {\bibinfo {title} {Mapping the
  genome of meta-generalized gradient approximation density functionals: The
  search for b97m-v},}\ }\href@noop {} {\bibfield  {journal} {\bibinfo
  {journal} {J. Chem. Phys.}\ }\textbf {\bibinfo {volume} {142}},\ \bibinfo
  {pages} {074111} (\bibinfo {year} {2015})}\BibitemShut {NoStop}%
\bibitem [{\citenamefont {Mardirossian}\ \emph {et~al.}(2017)\citenamefont
  {Mardirossian}, \citenamefont {Ruiz~Pestana}, \citenamefont {Womack},
  \citenamefont {Skylaris}, \citenamefont {Head-Gordon},\ and\ \citenamefont
  {Head-Gordon}}]{mardirossian2017use}%
  \BibitemOpen
  \bibfield  {author} {\bibinfo {author} {\bibfnamefont {Narbe}\ \bibnamefont
  {Mardirossian}}, \bibinfo {author} {\bibfnamefont {Luis}\ \bibnamefont
  {Ruiz~Pestana}}, \bibinfo {author} {\bibfnamefont {James~C}\ \bibnamefont
  {Womack}}, \bibinfo {author} {\bibfnamefont {Chris-Kriton}\ \bibnamefont
  {Skylaris}}, \bibinfo {author} {\bibfnamefont {Teresa}\ \bibnamefont
  {Head-Gordon}}, \ and\ \bibinfo {author} {\bibfnamefont {Martin}\
  \bibnamefont {Head-Gordon}},\ }\bibfield  {title} {\enquote {\bibinfo {title}
  {Use of the rvv10 nonlocal correlation functional in the b97m-v density
  functional: Defining b97m-rv and related functionals},}\ }\href@noop {}
  {\bibfield  {journal} {\bibinfo  {journal} {J. Phys. Chem. Lett.}\ }\textbf
  {\bibinfo {volume} {8}},\ \bibinfo {pages} {35--40} (\bibinfo {year}
  {2017})}\BibitemShut {NoStop}%
\bibitem [{\citenamefont {Heyd}\ \emph {et~al.}(2005)\citenamefont {Heyd},
  \citenamefont {Peralta}, \citenamefont {Scuseria},\ and\ \citenamefont
  {Martin}}]{heyd2005energy}%
  \BibitemOpen
  \bibfield  {author} {\bibinfo {author} {\bibfnamefont {Jochen}\ \bibnamefont
  {Heyd}}, \bibinfo {author} {\bibfnamefont {Juan~E}\ \bibnamefont {Peralta}},
  \bibinfo {author} {\bibfnamefont {Gustavo~E}\ \bibnamefont {Scuseria}}, \
  and\ \bibinfo {author} {\bibfnamefont {Richard~L}\ \bibnamefont {Martin}},\
  }\bibfield  {title} {\enquote {\bibinfo {title} {Energy band gaps and lattice
  parameters evaluated with the heyd-scuseria-ernzerhof screened hybrid
  functional},}\ }\href@noop {} {\bibfield  {journal} {\bibinfo  {journal} {J.
  Chem. Phys.}\ }\textbf {\bibinfo {volume} {123}},\ \bibinfo {pages} {174101}
  (\bibinfo {year} {2005})}\BibitemShut {NoStop}%
\bibitem [{\citenamefont {Vidal}\ and\ \citenamefont
  {Vidal-Valat}(1986)}]{vidal1986accurate}%
  \BibitemOpen
  \bibfield  {author} {\bibinfo {author} {\bibfnamefont {JEAN~PIERRE}\
  \bibnamefont {Vidal}}\ and\ \bibinfo {author} {\bibfnamefont {G}~\bibnamefont
  {Vidal-Valat}},\ }\bibfield  {title} {\enquote {\bibinfo {title} {Accurate
  debye--waller factors of 7lih and 7lid by neutron diffraction at three
  temperatures},}\ }\href@noop {} {\bibfield  {journal} {\bibinfo  {journal}
  {Acta Crystallogr., Sect. B: Struct. Sci.}\ }\textbf {\bibinfo {volume}
  {42}},\ \bibinfo {pages} {131--137} (\bibinfo {year} {1986})}\BibitemShut
  {NoStop}%
\bibitem [{\citenamefont {Nolan}\ \emph {et~al.}(2009)\citenamefont {Nolan},
  \citenamefont {Gillan}, \citenamefont {Alf{\`e}}, \citenamefont {Allan},\
  and\ \citenamefont {Manby}}]{nolan2009calculation}%
  \BibitemOpen
  \bibfield  {author} {\bibinfo {author} {\bibfnamefont {SJ}~\bibnamefont
  {Nolan}}, \bibinfo {author} {\bibfnamefont {MJ}~\bibnamefont {Gillan}},
  \bibinfo {author} {\bibfnamefont {D}~\bibnamefont {Alf{\`e}}}, \bibinfo
  {author} {\bibfnamefont {NL}~\bibnamefont {Allan}}, \ and\ \bibinfo {author}
  {\bibfnamefont {FR}~\bibnamefont {Manby}},\ }\bibfield  {title} {\enquote
  {\bibinfo {title} {Calculation of properties of crystalline lithium hydride
  using correlated wave function theory},}\ }\href@noop {} {\bibfield
  {journal} {\bibinfo  {journal} {Phys. Rev. B}\ }\textbf {\bibinfo {volume}
  {80}},\ \bibinfo {pages} {165109} (\bibinfo {year} {2009})}\BibitemShut
  {NoStop}%
\bibitem [{\citenamefont {Hoang}\ and\ \citenamefont {Van~de
  Walle}(2013)}]{hoang2013lih}%
  \BibitemOpen
  \bibfield  {author} {\bibinfo {author} {\bibfnamefont {Khang}\ \bibnamefont
  {Hoang}}\ and\ \bibinfo {author} {\bibfnamefont {Chris~G}\ \bibnamefont
  {Van~de Walle}},\ }\bibfield  {title} {\enquote {\bibinfo {title} {Lih as a
  li+ and h- ion provider},}\ }\href@noop {} {\bibfield  {journal} {\bibinfo
  {journal} {Solid State Ion.}\ }\textbf {\bibinfo {volume} {253}},\ \bibinfo
  {pages} {53--56} (\bibinfo {year} {2013})}\BibitemShut {NoStop}%
\bibitem [{\citenamefont {Matsushita}\ \emph {et~al.}(2011)\citenamefont
  {Matsushita}, \citenamefont {Nakamura},\ and\ \citenamefont
  {Oshiyama}}]{matsushita2011comparative}%
  \BibitemOpen
  \bibfield  {author} {\bibinfo {author} {\bibfnamefont {Yu-ichiro}\
  \bibnamefont {Matsushita}}, \bibinfo {author} {\bibfnamefont {Kazuma}\
  \bibnamefont {Nakamura}}, \ and\ \bibinfo {author} {\bibfnamefont {Atsushi}\
  \bibnamefont {Oshiyama}},\ }\bibfield  {title} {\enquote {\bibinfo {title}
  {Comparative study of hybrid functionals applied to structural and electronic
  properties of semiconductors and insulators},}\ }\href@noop {} {\bibfield
  {journal} {\bibinfo  {journal} {Phys. Rev. B}\ }\textbf {\bibinfo {volume}
  {84}},\ \bibinfo {pages} {075205} (\bibinfo {year} {2011})}\BibitemShut
  {NoStop}%
\bibitem [{\citenamefont {Garza}\ and\ \citenamefont
  {Scuseria}(2016)}]{garza2016predicting}%
  \BibitemOpen
  \bibfield  {author} {\bibinfo {author} {\bibfnamefont {Alejandro~J}\
  \bibnamefont {Garza}}\ and\ \bibinfo {author} {\bibfnamefont {Gustavo~E}\
  \bibnamefont {Scuseria}},\ }\bibfield  {title} {\enquote {\bibinfo {title}
  {Predicting band gaps with hybrid density functionals},}\ }\href@noop {}
  {\bibfield  {journal} {\bibinfo  {journal} {J. Phys. Chem. Lett.}\ }\textbf
  {\bibinfo {volume} {7}},\ \bibinfo {pages} {4165--4170} (\bibinfo {year}
  {2016})}\BibitemShut {NoStop}%
\bibitem [{\citenamefont {Perdew}\ \emph {et~al.}(2017)\citenamefont {Perdew},
  \citenamefont {Yang}, \citenamefont {Burke}, \citenamefont {Yang},
  \citenamefont {Gross}, \citenamefont {Scheffler}, \citenamefont {Scuseria},
  \citenamefont {Henderson}, \citenamefont {Zhang}, \citenamefont {Ruzsinszky}
  \emph {et~al.}}]{perdew2017understanding}%
  \BibitemOpen
  \bibfield  {author} {\bibinfo {author} {\bibfnamefont {John~P}\ \bibnamefont
  {Perdew}}, \bibinfo {author} {\bibfnamefont {Weitao}\ \bibnamefont {Yang}},
  \bibinfo {author} {\bibfnamefont {Kieron}\ \bibnamefont {Burke}}, \bibinfo
  {author} {\bibfnamefont {Zenghui}\ \bibnamefont {Yang}}, \bibinfo {author}
  {\bibfnamefont {Eberhard~KU}\ \bibnamefont {Gross}}, \bibinfo {author}
  {\bibfnamefont {Matthias}\ \bibnamefont {Scheffler}}, \bibinfo {author}
  {\bibfnamefont {Gustavo~E}\ \bibnamefont {Scuseria}}, \bibinfo {author}
  {\bibfnamefont {Thomas~M}\ \bibnamefont {Henderson}}, \bibinfo {author}
  {\bibfnamefont {Igor~Ying}\ \bibnamefont {Zhang}}, \bibinfo {author}
  {\bibfnamefont {Adrienn}\ \bibnamefont {Ruzsinszky}},  \emph {et~al.},\
  }\bibfield  {title} {\enquote {\bibinfo {title} {Understanding band gaps of
  solids in generalized kohn--sham theory},}\ }\href@noop {} {\bibfield
  {journal} {\bibinfo  {journal} {Proc. Natl. Acad. Sci. U.S.A}\ }\textbf
  {\bibinfo {volume} {114}},\ \bibinfo {pages} {2801--2806} (\bibinfo {year}
  {2017})}\BibitemShut {NoStop}%
\bibitem [{\citenamefont {Yang}\ \emph {et~al.}(2016)\citenamefont {Yang},
  \citenamefont {Peng}, \citenamefont {Sun},\ and\ \citenamefont
  {Perdew}}]{yang2016more}%
  \BibitemOpen
  \bibfield  {author} {\bibinfo {author} {\bibfnamefont {Zeng-hui}\
  \bibnamefont {Yang}}, \bibinfo {author} {\bibfnamefont {Haowei}\ \bibnamefont
  {Peng}}, \bibinfo {author} {\bibfnamefont {Jianwei}\ \bibnamefont {Sun}}, \
  and\ \bibinfo {author} {\bibfnamefont {John~P}\ \bibnamefont {Perdew}},\
  }\bibfield  {title} {\enquote {\bibinfo {title} {More realistic band gaps
  from meta-generalized gradient approximations: Only in a generalized
  kohn-sham scheme},}\ }\href@noop {} {\bibfield  {journal} {\bibinfo
  {journal} {Phys. Rev. B}\ }\textbf {\bibinfo {volume} {93}},\ \bibinfo
  {pages} {205205} (\bibinfo {year} {2016})}\BibitemShut {NoStop}%
\bibitem [{\citenamefont {Mardirossian}\ and\ \citenamefont
  {Head-Gordon}(2014)}]{mardirossian2014omegab97x}%
  \BibitemOpen
  \bibfield  {author} {\bibinfo {author} {\bibfnamefont {Narbe}\ \bibnamefont
  {Mardirossian}}\ and\ \bibinfo {author} {\bibfnamefont {Martin}\ \bibnamefont
  {Head-Gordon}},\ }\bibfield  {title} {\enquote {\bibinfo {title}
  {$\omega$b97x-v: A 10-parameter, range-separated hybrid, generalized gradient
  approximation density functional with nonlocal correlation, designed by a
  survival-of-the-fittest strategy},}\ }\href@noop {} {\bibfield  {journal}
  {\bibinfo  {journal} {Phys. Chem. Chem. Phys.}\ }\textbf {\bibinfo {volume}
  {16}},\ \bibinfo {pages} {9904--9924} (\bibinfo {year} {2014})}\BibitemShut
  {NoStop}%
\bibitem [{\citenamefont {Mardirossian}\ and\ \citenamefont
  {Head-Gordon}(2016)}]{mardirossian2016omega}%
  \BibitemOpen
  \bibfield  {author} {\bibinfo {author} {\bibfnamefont {Narbe}\ \bibnamefont
  {Mardirossian}}\ and\ \bibinfo {author} {\bibfnamefont {Martin}\ \bibnamefont
  {Head-Gordon}},\ }\bibfield  {title} {\enquote {\bibinfo {title} {$\omega$
  b97m-v: A combinatorially optimized, range-separated hybrid, meta-gga density
  functional with vv10 nonlocal correlation},}\ }\href@noop {} {\bibfield
  {journal} {\bibinfo  {journal} {J. Chem. Phys.}\ }\textbf {\bibinfo {volume}
  {144}},\ \bibinfo {pages} {214110} (\bibinfo {year} {2016})}\BibitemShut
  {NoStop}%
\bibitem [{\citenamefont {Pestana}\ \emph {et~al.}(2017)\citenamefont
  {Pestana}, \citenamefont {Mardirossian}, \citenamefont {Head-Gordon},\ and\
  \citenamefont {Head-Gordon}}]{pestana2017ab}%
  \BibitemOpen
  \bibfield  {author} {\bibinfo {author} {\bibfnamefont {Luis~Ruiz}\
  \bibnamefont {Pestana}}, \bibinfo {author} {\bibfnamefont {Narbe}\
  \bibnamefont {Mardirossian}}, \bibinfo {author} {\bibfnamefont {Martin}\
  \bibnamefont {Head-Gordon}}, \ and\ \bibinfo {author} {\bibfnamefont
  {Teresa}\ \bibnamefont {Head-Gordon}},\ }\bibfield  {title} {\enquote
  {\bibinfo {title} {Ab initio molecular dynamics simulations of liquid water
  using high quality meta-gga functionals},}\ }\href@noop {} {\bibfield
  {journal} {\bibinfo  {journal} {Chem. Sci.}\ }\textbf {\bibinfo {volume}
  {8}},\ \bibinfo {pages} {3554--3565} (\bibinfo {year} {2017})}\BibitemShut
  {NoStop}%
\bibitem [{\citenamefont {Ruiz~Pestana}\ \emph {et~al.}(2018)\citenamefont
  {Ruiz~Pestana}, \citenamefont {Marsalek}, \citenamefont {Markland},\ and\
  \citenamefont {Head-Gordon}}]{ruiz2018quest}%
  \BibitemOpen
  \bibfield  {author} {\bibinfo {author} {\bibfnamefont {Luis}\ \bibnamefont
  {Ruiz~Pestana}}, \bibinfo {author} {\bibfnamefont {Ondrej}\ \bibnamefont
  {Marsalek}}, \bibinfo {author} {\bibfnamefont {Thomas~E}\ \bibnamefont
  {Markland}}, \ and\ \bibinfo {author} {\bibfnamefont {Teresa}\ \bibnamefont
  {Head-Gordon}},\ }\bibfield  {title} {\enquote {\bibinfo {title} {The quest
  for accurate liquid water properties from first principles},}\ }\href@noop {}
  {\bibfield  {journal} {\bibinfo  {journal} {J. Phys. Chem. Lett.}\ }\textbf
  {\bibinfo {volume} {9}},\ \bibinfo {pages} {5009--5016} (\bibinfo {year}
  {2018})}\BibitemShut {NoStop}%
\bibitem [{\citenamefont {Pestana}\ \emph {et~al.}(2019)\citenamefont
  {Pestana}, \citenamefont {Hao},\ and\ \citenamefont
  {Head-Gordon}}]{pestana2019diels}%
  \BibitemOpen
  \bibfield  {author} {\bibinfo {author} {\bibfnamefont {Luis~Ruiz}\
  \bibnamefont {Pestana}}, \bibinfo {author} {\bibfnamefont {Hongxia}\
  \bibnamefont {Hao}}, \ and\ \bibinfo {author} {\bibfnamefont {Teresa}\
  \bibnamefont {Head-Gordon}},\ }\bibfield  {title} {\enquote {\bibinfo {title}
  {Diels--alder reactions in water are determined by microsolvation},}\
  }\href@noop {} {\bibfield  {journal} {\bibinfo  {journal} {Nano Lett.}\
  }\textbf {\bibinfo {volume} {20}},\ \bibinfo {pages} {606--611} (\bibinfo
  {year} {2019})}\BibitemShut {NoStop}%
\bibitem [{\citenamefont {Lininger}\ \emph {et~al.}(2021)\citenamefont
  {Lininger}, \citenamefont {Gauthier}, \citenamefont {Li}, \citenamefont
  {Rossomme}, \citenamefont {Welborn}, \citenamefont {Lin}, \citenamefont
  {Head-Gordon}, \citenamefont {Head-Gordon},\ and\ \citenamefont
  {Bell}}]{lininger2021challenges}%
  \BibitemOpen
  \bibfield  {author} {\bibinfo {author} {\bibfnamefont {Christianna~N}\
  \bibnamefont {Lininger}}, \bibinfo {author} {\bibfnamefont {Joseph~A}\
  \bibnamefont {Gauthier}}, \bibinfo {author} {\bibfnamefont {Wan-Lu}\
  \bibnamefont {Li}}, \bibinfo {author} {\bibfnamefont {Elliot}\ \bibnamefont
  {Rossomme}}, \bibinfo {author} {\bibfnamefont {Valerie~Vaissier}\
  \bibnamefont {Welborn}}, \bibinfo {author} {\bibfnamefont {Zhou}\
  \bibnamefont {Lin}}, \bibinfo {author} {\bibfnamefont {Teresa}\ \bibnamefont
  {Head-Gordon}}, \bibinfo {author} {\bibfnamefont {Martin}\ \bibnamefont
  {Head-Gordon}}, \ and\ \bibinfo {author} {\bibfnamefont {Alexis~T}\
  \bibnamefont {Bell}},\ }\bibfield  {title} {\enquote {\bibinfo {title}
  {Challenges for density functional theory: calculation of co adsorption on
  electrocatalytically relevant metals},}\ }\href@noop {} {\bibfield  {journal}
  {\bibinfo  {journal} {Phys. Chem. Chem. Phys.}\ }\textbf {\bibinfo {volume}
  {23}},\ \bibinfo {pages} {9394--9406} (\bibinfo {year} {2021})}\BibitemShut
  {NoStop}%
\bibitem [{\citenamefont {Li}\ \emph {et~al.}(2021)\citenamefont {Li},
  \citenamefont {Lininger}, \citenamefont {Chen}, \citenamefont
  {Vaissier~Welborn}, \citenamefont {Rossomme}, \citenamefont {Bell},
  \citenamefont {Head-Gordon},\ and\ \citenamefont
  {Head-Gordon}}]{li2021critical}%
  \BibitemOpen
  \bibfield  {author} {\bibinfo {author} {\bibfnamefont {Wan-Lu}\ \bibnamefont
  {Li}}, \bibinfo {author} {\bibfnamefont {Christianna~N}\ \bibnamefont
  {Lininger}}, \bibinfo {author} {\bibfnamefont {Kaixuan}\ \bibnamefont
  {Chen}}, \bibinfo {author} {\bibfnamefont {Valerie}\ \bibnamefont
  {Vaissier~Welborn}}, \bibinfo {author} {\bibfnamefont {Elliot}\ \bibnamefont
  {Rossomme}}, \bibinfo {author} {\bibfnamefont {Alexis~T}\ \bibnamefont
  {Bell}}, \bibinfo {author} {\bibfnamefont {Martin}\ \bibnamefont
  {Head-Gordon}}, \ and\ \bibinfo {author} {\bibfnamefont {Teresa}\
  \bibnamefont {Head-Gordon}},\ }\bibfield  {title} {\enquote {\bibinfo {title}
  {Critical role of thermal fluctuations for co binding on electrocatalytic
  metal surfaces},}\ }\href@noop {} {\bibfield  {journal} {\bibinfo  {journal}
  {JACS Au}\ ,\ \bibinfo {pages} {(in press)}} (\bibinfo {year}
  {2021})}\BibitemShut {NoStop}%
\bibitem [{\citenamefont {Sabatini}\ \emph {et~al.}(2013)\citenamefont
  {Sabatini}, \citenamefont {Gorni},\ and\ \citenamefont
  {De~Gironcoli}}]{sabatini2013nonlocal}%
  \BibitemOpen
  \bibfield  {author} {\bibinfo {author} {\bibfnamefont {Riccardo}\
  \bibnamefont {Sabatini}}, \bibinfo {author} {\bibfnamefont {Tommaso}\
  \bibnamefont {Gorni}}, \ and\ \bibinfo {author} {\bibfnamefont {Stefano}\
  \bibnamefont {De~Gironcoli}},\ }\bibfield  {title} {\enquote {\bibinfo
  {title} {Nonlocal van der waals density functional made simple and
  efficient},}\ }\href@noop {} {\bibfield  {journal} {\bibinfo  {journal}
  {Phys. Rev. B}\ }\textbf {\bibinfo {volume} {87}},\ \bibinfo {pages} {041108}
  (\bibinfo {year} {2013})}\BibitemShut {NoStop}%
\bibitem [{\citenamefont {Vydrov}\ and\ \citenamefont
  {Van~Voorhis}(2010)}]{vydrov2010nonlocal}%
  \BibitemOpen
  \bibfield  {author} {\bibinfo {author} {\bibfnamefont {Oleg~A}\ \bibnamefont
  {Vydrov}}\ and\ \bibinfo {author} {\bibfnamefont {Troy}\ \bibnamefont
  {Van~Voorhis}},\ }\bibfield  {title} {\enquote {\bibinfo {title} {Nonlocal
  van der waals density functional: The simpler the better},}\ }\href@noop {}
  {\bibfield  {journal} {\bibinfo  {journal} {J. Chem. Phys.}\ }\textbf
  {\bibinfo {volume} {133}},\ \bibinfo {pages} {244103} (\bibinfo {year}
  {2010})}\BibitemShut {NoStop}%
\bibitem [{\citenamefont {Mardirossian}\ and\ \citenamefont
  {Head-Gordon}(2017)}]{mardirossian2017thirty}%
  \BibitemOpen
  \bibfield  {author} {\bibinfo {author} {\bibfnamefont {Narbe}\ \bibnamefont
  {Mardirossian}}\ and\ \bibinfo {author} {\bibfnamefont {Martin}\ \bibnamefont
  {Head-Gordon}},\ }\bibfield  {title} {\enquote {\bibinfo {title} {Thirty
  years of density functional theory in computational chemistry: an overview
  and extensive assessment of 200 density functionals},}\ }\href@noop {}
  {\bibfield  {journal} {\bibinfo  {journal} {Mol. Phys.}\ }\textbf {\bibinfo
  {volume} {115}},\ \bibinfo {pages} {2315--2372} (\bibinfo {year}
  {2017})}\BibitemShut {NoStop}%
\bibitem [{\citenamefont {Goerigk}\ \emph {et~al.}(2017)\citenamefont
  {Goerigk}, \citenamefont {Hansen}, \citenamefont {Bauer}, \citenamefont
  {Ehrlich}, \citenamefont {Najibi},\ and\ \citenamefont
  {Grimme}}]{Goerigk:2017}%
  \BibitemOpen
  \bibfield  {author} {\bibinfo {author} {\bibfnamefont {L.}~\bibnamefont
  {Goerigk}}, \bibinfo {author} {\bibfnamefont {A.}~\bibnamefont {Hansen}},
  \bibinfo {author} {\bibfnamefont {C.}~\bibnamefont {Bauer}}, \bibinfo
  {author} {\bibfnamefont {S.}~\bibnamefont {Ehrlich}}, \bibinfo {author}
  {\bibfnamefont {A.}~\bibnamefont {Najibi}}, \ and\ \bibinfo {author}
  {\bibfnamefont {S.}~\bibnamefont {Grimme}},\ }\bibfield  {title} {\enquote
  {\bibinfo {title} {A look at the density functional theory zoo with the
  advanced {GMTKN55} database for general main group thermochemistry, kinetics
  and noncovalent interactions},}\ }\href {\doibase 10.1039/C7CP04913G}
  {\bibfield  {journal} {\bibinfo  {journal} {Phys. Chem. Chem. Phys.}\
  }\textbf {\bibinfo {volume} {19}},\ \bibinfo {pages} {32184--32215} (\bibinfo
  {year} {2017})}\BibitemShut {NoStop}%
\bibitem [{\citenamefont {Najibi}\ and\ \citenamefont
  {Goerigk}(2018)}]{Najibi:2018b}%
  \BibitemOpen
  \bibfield  {author} {\bibinfo {author} {\bibfnamefont {A.}~\bibnamefont
  {Najibi}}\ and\ \bibinfo {author} {\bibfnamefont {L.}~\bibnamefont
  {Goerigk}},\ }\bibfield  {title} {\enquote {\bibinfo {title} {The nonlocal
  kernel in van der {Waals} density functionals as an additive correction: {An}
  extensive analysis with special emphasis on the {B97M}-{V} and
  $\omega${B97M}-{V} approches},}\ }\href {\doibase 10.1021/acs.jctc.8b00842}
  {\bibfield  {journal} {\bibinfo  {journal} {J. Chem. Theory Comput.}\
  }\textbf {\bibinfo {volume} {14}},\ \bibinfo {pages} {5725--5738} (\bibinfo
  {year} {2018})}\BibitemShut {NoStop}%
\bibitem [{\citenamefont {Evarestov}(2007)}]{Evarestov2007}%
  \BibitemOpen
  \bibfield  {author} {\bibinfo {author} {\bibfnamefont {Robert}\ \bibnamefont
  {Evarestov}},\ }\href {\doibase 10.1007/978-3-540-48748-7} {\emph {\bibinfo
  {title} {{Quantum Chemistry of Solids}}}}\ (\bibinfo  {publisher}
  {Springer-Verlag},\ \bibinfo {address} {Berlin, Germany},\ \bibinfo {year}
  {2007})\BibitemShut {NoStop}%
\bibitem [{\citenamefont {Evarestov}(2012)}]{Evarestov2012}%
  \BibitemOpen
  \bibfield  {author} {\bibinfo {author} {\bibfnamefont {R.~A.}\ \bibnamefont
  {Evarestov}},\ }\href {\doibase 10.1007/978-3-642-30356-2} {\emph {\bibinfo
  {title} {{Quantum Chemistry of Solids}}}}\ (\bibinfo  {publisher}
  {Springer-Verlag},\ \bibinfo {address} {Berlin, Germany},\ \bibinfo {year}
  {2012})\BibitemShut {NoStop}%
\bibitem [{\citenamefont {Kudin}\ and\ \citenamefont
  {Scuseria}(1998)}]{kudin1998fast}%
  \BibitemOpen
  \bibfield  {author} {\bibinfo {author} {\bibfnamefont {Konstantin~N}\
  \bibnamefont {Kudin}}\ and\ \bibinfo {author} {\bibfnamefont {Gustavo~E}\
  \bibnamefont {Scuseria}},\ }\bibfield  {title} {\enquote {\bibinfo {title} {A
  fast multipole algorithm for the efficient treatment of the coulomb problem
  in electronic structure calculations of periodic systems with gaussian
  orbitals},}\ }\href@noop {} {\bibfield  {journal} {\bibinfo  {journal} {Chem.
  Phys. Lett.}\ }\textbf {\bibinfo {volume} {289}},\ \bibinfo {pages}
  {611--616} (\bibinfo {year} {1998})}\BibitemShut {NoStop}%
\bibitem [{\citenamefont {Szabo}\ and\ \citenamefont
  {Ostlund}(1996)}]{Szabo1996Jul}%
  \BibitemOpen
  \bibfield  {author} {\bibinfo {author} {\bibfnamefont {Attila}\ \bibnamefont
  {Szabo}}\ and\ \bibinfo {author} {\bibfnamefont {Neil~S.}\ \bibnamefont
  {Ostlund}},\ }\href
  {https://books.google.com/books/about/Modern_Quantum_Chemistry.html?id=6mV9gYzEkgIC}
  {\emph {\bibinfo {title} {{Modern Quantum Chemistry: Introduction to Advanced
  Electronic Structure Theory}}}}\ (\bibinfo  {publisher} {Courier
  Corporation},\ \bibinfo {year} {1996})\BibitemShut {NoStop}%
\bibitem [{\citenamefont {Shao}\ \emph {et~al.}(2015)\citenamefont {Shao},
  \citenamefont {Gan}, \citenamefont {Epifanovsky}, \citenamefont {Gilbert},
  \citenamefont {Wormit}, \citenamefont {Kussmann}, \citenamefont {Lange},
  \citenamefont {Behn}, \citenamefont {Deng}, \citenamefont {Feng},
  \citenamefont {Ghosh}, \citenamefont {Goldey}, \citenamefont {Horn},
  \citenamefont {Jacobson}, \citenamefont {Kaliman}, \citenamefont
  {Khaliullin}, \citenamefont {Ku{\ifmmode\acute{s}\else\'{s}\fi}},
  \citenamefont {Landau}, \citenamefont {Liu}, \citenamefont {Proynov},
  \citenamefont {Rhee}, \citenamefont {Richard}, \citenamefont {Rohrdanz},
  \citenamefont {Steele}, \citenamefont {Sundstrom}, \citenamefont {Woodcock},
  \citenamefont {Zimmerman}, \citenamefont {Zuev}, \citenamefont {Albrecht},
  \citenamefont {Alguire}, \citenamefont {Austin}, \citenamefont {Beran},
  \citenamefont {Bernard}, \citenamefont {Berquist}, \citenamefont
  {Brandhorst}, \citenamefont {Bravaya}, \citenamefont {Brown}, \citenamefont
  {Casanova}, \citenamefont {Chang}, \citenamefont {Chen}, \citenamefont
  {Chien}, \citenamefont {Closser}, \citenamefont {Crittenden}, \citenamefont
  {Diedenhofen}, \citenamefont {DiStasio}, \citenamefont {Do}, \citenamefont
  {Dutoi}, \citenamefont {Edgar}, \citenamefont {Fatehi}, \citenamefont
  {Fusti-Molnar}, \citenamefont {Ghysels}, \citenamefont
  {Golubeva-Zadorozhnaya}, \citenamefont {Gomes}, \citenamefont {Hanson-Heine},
  \citenamefont {Harbach}, \citenamefont {Hauser}, \citenamefont {Hohenstein},
  \citenamefont {Holden}, \citenamefont {Jagau}, \citenamefont {Ji},
  \citenamefont {Kaduk}, \citenamefont {Khistyaev}, \citenamefont {Kim},
  \citenamefont {Kim}, \citenamefont {King}, \citenamefont {Klunzinger},
  \citenamefont {Kosenkov}, \citenamefont {Kowalczyk}, \citenamefont {Krauter},
  \citenamefont {Lao}, \citenamefont {Laurent}, \citenamefont {Lawler},
  \citenamefont {Levchenko}, \citenamefont {Lin}, \citenamefont {Liu},
  \citenamefont {Livshits}, \citenamefont {Lochan}, \citenamefont {Luenser},
  \citenamefont {Manohar}, \citenamefont {Manzer}, \citenamefont {Mao},
  \citenamefont {Mardirossian}, \citenamefont {Marenich}, \citenamefont
  {Maurer}, \citenamefont {Mayhall}, \citenamefont {Neuscamman}, \citenamefont
  {Oana}, \citenamefont {Olivares-Amaya}, \citenamefont {O{'}Neill},
  \citenamefont {Parkhill}, \citenamefont {Perrine}, \citenamefont {Peverati},
  \citenamefont {Prociuk}, \citenamefont {Rehn}, \citenamefont {Rosta},
  \citenamefont {Russ}, \citenamefont {Sharada}, \citenamefont {Sharma},
  \citenamefont {Small}, \citenamefont {Sodt}, \citenamefont {Stein},
  \citenamefont {St{\ifmmode\ddot{u}\else\"{u}\fi}ck}, \citenamefont {Su},
  \citenamefont {Thom}, \citenamefont {Tsuchimochi}, \citenamefont {Vanovschi},
  \citenamefont {Vogt}, \citenamefont {Vydrov}, \citenamefont {Wang},
  \citenamefont {Watson}, \citenamefont {Wenzel}, \citenamefont {White},
  \citenamefont {Williams}, \citenamefont {Yang}, \citenamefont {Yeganeh},
  \citenamefont {Yost}, \citenamefont {You}, \citenamefont {Zhang},
  \citenamefont {Zhang}, \citenamefont {Zhao}, \citenamefont {Brooks},
  \citenamefont {Chan}, \citenamefont {Chipman}, \citenamefont {Cramer},
  \citenamefont {Goddard}, \citenamefont {Gordon}, \citenamefont {Hehre},
  \citenamefont {Klamt}, \citenamefont {Schaefer}, \citenamefont {Schmidt},
  \citenamefont {Sherrill}, \citenamefont {Truhlar}, \citenamefont {Warshel},
  \citenamefont {Xu}, \citenamefont {Aspuru-Guzik}, \citenamefont {Baer},
  \citenamefont {Bell}, \citenamefont {Besley}, \citenamefont {Chai},
  \citenamefont {Dreuw}, \citenamefont {Dunietz}, \citenamefont {Furlani},
  \citenamefont {Gwaltney}, \citenamefont {Hsu}, \citenamefont {Jung},
  \citenamefont {Kong}, \citenamefont {Lambrecht}, \citenamefont {Liang},
  \citenamefont {Ochsenfeld}, \citenamefont {Rassolov}, \citenamefont
  {Slipchenko}, \citenamefont {Subotnik}, \citenamefont {Van~Voorhis},
  \citenamefont {Herbert}, \citenamefont {Krylov}, \citenamefont {Gill},\ and\
  \citenamefont {Head-Gordon}}]{Shao2015Jan}%
  \BibitemOpen
  \bibfield  {author} {\bibinfo {author} {\bibfnamefont {Yihan}\ \bibnamefont
  {Shao}}, \bibinfo {author} {\bibfnamefont {Zhengting}\ \bibnamefont {Gan}},
  \bibinfo {author} {\bibfnamefont {Evgeny}\ \bibnamefont {Epifanovsky}},
  \bibinfo {author} {\bibfnamefont {Andrew T.~B.}\ \bibnamefont {Gilbert}},
  \bibinfo {author} {\bibfnamefont {Michael}\ \bibnamefont {Wormit}}, \bibinfo
  {author} {\bibfnamefont {Joerg}\ \bibnamefont {Kussmann}}, \bibinfo {author}
  {\bibfnamefont {Adrian~W.}\ \bibnamefont {Lange}}, \bibinfo {author}
  {\bibfnamefont {Andrew}\ \bibnamefont {Behn}}, \bibinfo {author}
  {\bibfnamefont {Jia}\ \bibnamefont {Deng}}, \bibinfo {author} {\bibfnamefont
  {Xintian}\ \bibnamefont {Feng}}, \bibinfo {author} {\bibfnamefont
  {Debashree}\ \bibnamefont {Ghosh}}, \bibinfo {author} {\bibfnamefont
  {Matthew}\ \bibnamefont {Goldey}}, \bibinfo {author} {\bibfnamefont
  {Paul~R.}\ \bibnamefont {Horn}}, \bibinfo {author} {\bibfnamefont {Leif~D.}\
  \bibnamefont {Jacobson}}, \bibinfo {author} {\bibfnamefont {Ilya}\
  \bibnamefont {Kaliman}}, \bibinfo {author} {\bibfnamefont {Rustam~Z.}\
  \bibnamefont {Khaliullin}}, \bibinfo {author} {\bibfnamefont {Tomasz}\
  \bibnamefont {Ku{\ifmmode\acute{s}\else\'{s}\fi}}}, \bibinfo {author}
  {\bibfnamefont {Arie}\ \bibnamefont {Landau}}, \bibinfo {author}
  {\bibfnamefont {Jie}\ \bibnamefont {Liu}}, \bibinfo {author} {\bibfnamefont
  {Emil~I.}\ \bibnamefont {Proynov}}, \bibinfo {author} {\bibfnamefont
  {Young~Min}\ \bibnamefont {Rhee}}, \bibinfo {author} {\bibfnamefont
  {Ryan~M.}\ \bibnamefont {Richard}}, \bibinfo {author} {\bibfnamefont
  {Mary~A.}\ \bibnamefont {Rohrdanz}}, \bibinfo {author} {\bibfnamefont
  {Ryan~P.}\ \bibnamefont {Steele}}, \bibinfo {author} {\bibfnamefont
  {Eric~J.}\ \bibnamefont {Sundstrom}}, \bibinfo {author} {\bibfnamefont
  {H.~Lee}\ \bibnamefont {Woodcock}}, \bibinfo {author} {\bibfnamefont
  {Paul~M.}\ \bibnamefont {Zimmerman}}, \bibinfo {author} {\bibfnamefont
  {Dmitry}\ \bibnamefont {Zuev}}, \bibinfo {author} {\bibfnamefont {Ben}\
  \bibnamefont {Albrecht}}, \bibinfo {author} {\bibfnamefont {Ethan}\
  \bibnamefont {Alguire}}, \bibinfo {author} {\bibfnamefont {Brian}\
  \bibnamefont {Austin}}, \bibinfo {author} {\bibfnamefont {Gregory J.~O.}\
  \bibnamefont {Beran}}, \bibinfo {author} {\bibfnamefont {Yves~A.}\
  \bibnamefont {Bernard}}, \bibinfo {author} {\bibfnamefont {Eric}\
  \bibnamefont {Berquist}}, \bibinfo {author} {\bibfnamefont {Kai}\
  \bibnamefont {Brandhorst}}, \bibinfo {author} {\bibfnamefont {Ksenia~B.}\
  \bibnamefont {Bravaya}}, \bibinfo {author} {\bibfnamefont {Shawn~T.}\
  \bibnamefont {Brown}}, \bibinfo {author} {\bibfnamefont {David}\ \bibnamefont
  {Casanova}}, \bibinfo {author} {\bibfnamefont {Chun-Min}\ \bibnamefont
  {Chang}}, \bibinfo {author} {\bibfnamefont {Yunqing}\ \bibnamefont {Chen}},
  \bibinfo {author} {\bibfnamefont {Siu~Hung}\ \bibnamefont {Chien}}, \bibinfo
  {author} {\bibfnamefont {Kristina~D.}\ \bibnamefont {Closser}}, \bibinfo
  {author} {\bibfnamefont {Deborah~L.}\ \bibnamefont {Crittenden}}, \bibinfo
  {author} {\bibfnamefont {Michael}\ \bibnamefont {Diedenhofen}}, \bibinfo
  {author} {\bibfnamefont {Robert~A.}\ \bibnamefont {DiStasio}}, \bibinfo
  {author} {\bibfnamefont {Hainam}\ \bibnamefont {Do}}, \bibinfo {author}
  {\bibfnamefont {Anthony~D.}\ \bibnamefont {Dutoi}}, \bibinfo {author}
  {\bibfnamefont {Richard~G.}\ \bibnamefont {Edgar}}, \bibinfo {author}
  {\bibfnamefont {Shervin}\ \bibnamefont {Fatehi}}, \bibinfo {author}
  {\bibfnamefont {Laszlo}\ \bibnamefont {Fusti-Molnar}}, \bibinfo {author}
  {\bibfnamefont {An}~\bibnamefont {Ghysels}}, \bibinfo {author} {\bibfnamefont
  {Anna}\ \bibnamefont {Golubeva-Zadorozhnaya}}, \bibinfo {author}
  {\bibfnamefont {Joseph}\ \bibnamefont {Gomes}}, \bibinfo {author}
  {\bibfnamefont {Magnus W.~D.}\ \bibnamefont {Hanson-Heine}}, \bibinfo
  {author} {\bibfnamefont {Philipp H.~P.}\ \bibnamefont {Harbach}}, \bibinfo
  {author} {\bibfnamefont {Andreas~W.}\ \bibnamefont {Hauser}}, \bibinfo
  {author} {\bibfnamefont {Edward~G.}\ \bibnamefont {Hohenstein}}, \bibinfo
  {author} {\bibfnamefont {Zachary~C.}\ \bibnamefont {Holden}}, \bibinfo
  {author} {\bibfnamefont {Thomas-C.}\ \bibnamefont {Jagau}}, \bibinfo {author}
  {\bibfnamefont {Hyunjun}\ \bibnamefont {Ji}}, \bibinfo {author}
  {\bibfnamefont {Benjamin}\ \bibnamefont {Kaduk}}, \bibinfo {author}
  {\bibfnamefont {Kirill}\ \bibnamefont {Khistyaev}}, \bibinfo {author}
  {\bibfnamefont {Jaehoon}\ \bibnamefont {Kim}}, \bibinfo {author}
  {\bibfnamefont {Jihan}\ \bibnamefont {Kim}}, \bibinfo {author} {\bibfnamefont
  {Rollin~A.}\ \bibnamefont {King}}, \bibinfo {author} {\bibfnamefont {Phil}\
  \bibnamefont {Klunzinger}}, \bibinfo {author} {\bibfnamefont {Dmytro}\
  \bibnamefont {Kosenkov}}, \bibinfo {author} {\bibfnamefont {Tim}\
  \bibnamefont {Kowalczyk}}, \bibinfo {author} {\bibfnamefont {Caroline~M.}\
  \bibnamefont {Krauter}}, \bibinfo {author} {\bibfnamefont {Ka~Un}\
  \bibnamefont {Lao}}, \bibinfo {author} {\bibfnamefont
  {Ad{\ifmmode\grave{e}\else\`{e}\fi}le~D.}\ \bibnamefont {Laurent}}, \bibinfo
  {author} {\bibfnamefont {Keith~V.}\ \bibnamefont {Lawler}}, \bibinfo {author}
  {\bibfnamefont {Sergey~V.}\ \bibnamefont {Levchenko}}, \bibinfo {author}
  {\bibfnamefont {Ching~Yeh}\ \bibnamefont {Lin}}, \bibinfo {author}
  {\bibfnamefont {Fenglai}\ \bibnamefont {Liu}}, \bibinfo {author}
  {\bibfnamefont {Ester}\ \bibnamefont {Livshits}}, \bibinfo {author}
  {\bibfnamefont {Rohini~C.}\ \bibnamefont {Lochan}}, \bibinfo {author}
  {\bibfnamefont {Arne}\ \bibnamefont {Luenser}}, \bibinfo {author}
  {\bibfnamefont {Prashant}\ \bibnamefont {Manohar}}, \bibinfo {author}
  {\bibfnamefont {Samuel~F.}\ \bibnamefont {Manzer}}, \bibinfo {author}
  {\bibfnamefont {Shan-Ping}\ \bibnamefont {Mao}}, \bibinfo {author}
  {\bibfnamefont {Narbe}\ \bibnamefont {Mardirossian}}, \bibinfo {author}
  {\bibfnamefont {Aleksandr~V.}\ \bibnamefont {Marenich}}, \bibinfo {author}
  {\bibfnamefont {Simon~A.}\ \bibnamefont {Maurer}}, \bibinfo {author}
  {\bibfnamefont {Nicholas~J.}\ \bibnamefont {Mayhall}}, \bibinfo {author}
  {\bibfnamefont {Eric}\ \bibnamefont {Neuscamman}}, \bibinfo {author}
  {\bibfnamefont {C.~Melania}\ \bibnamefont {Oana}}, \bibinfo {author}
  {\bibfnamefont {Roberto}\ \bibnamefont {Olivares-Amaya}}, \bibinfo {author}
  {\bibfnamefont {Darragh~P.}\ \bibnamefont {O{'}Neill}}, \bibinfo {author}
  {\bibfnamefont {John~A.}\ \bibnamefont {Parkhill}}, \bibinfo {author}
  {\bibfnamefont {Trilisa~M.}\ \bibnamefont {Perrine}}, \bibinfo {author}
  {\bibfnamefont {Roberto}\ \bibnamefont {Peverati}}, \bibinfo {author}
  {\bibfnamefont {Alexander}\ \bibnamefont {Prociuk}}, \bibinfo {author}
  {\bibfnamefont {Dirk~R.}\ \bibnamefont {Rehn}}, \bibinfo {author}
  {\bibfnamefont {Edina}\ \bibnamefont {Rosta}}, \bibinfo {author}
  {\bibfnamefont {Nicholas~J.}\ \bibnamefont {Russ}}, \bibinfo {author}
  {\bibfnamefont {Shaama~M.}\ \bibnamefont {Sharada}}, \bibinfo {author}
  {\bibfnamefont {Sandeep}\ \bibnamefont {Sharma}}, \bibinfo {author}
  {\bibfnamefont {David~W.}\ \bibnamefont {Small}}, \bibinfo {author}
  {\bibfnamefont {Alexander}\ \bibnamefont {Sodt}}, \bibinfo {author}
  {\bibfnamefont {Tamar}\ \bibnamefont {Stein}}, \bibinfo {author}
  {\bibfnamefont {David}\ \bibnamefont {St{\ifmmode\ddot{u}\else\"{u}\fi}ck}},
  \bibinfo {author} {\bibfnamefont {Yu-Chuan}\ \bibnamefont {Su}}, \bibinfo
  {author} {\bibfnamefont {Alex J.~W.}\ \bibnamefont {Thom}}, \bibinfo {author}
  {\bibfnamefont {Takashi}\ \bibnamefont {Tsuchimochi}}, \bibinfo {author}
  {\bibfnamefont {Vitalii}\ \bibnamefont {Vanovschi}}, \bibinfo {author}
  {\bibfnamefont {Leslie}\ \bibnamefont {Vogt}}, \bibinfo {author}
  {\bibfnamefont {Oleg}\ \bibnamefont {Vydrov}}, \bibinfo {author}
  {\bibfnamefont {Tao}\ \bibnamefont {Wang}}, \bibinfo {author} {\bibfnamefont
  {Mark~A.}\ \bibnamefont {Watson}}, \bibinfo {author} {\bibfnamefont {Jan}\
  \bibnamefont {Wenzel}}, \bibinfo {author} {\bibfnamefont {Alec}\ \bibnamefont
  {White}}, \bibinfo {author} {\bibfnamefont {Christopher~F.}\ \bibnamefont
  {Williams}}, \bibinfo {author} {\bibfnamefont {Jun}\ \bibnamefont {Yang}},
  \bibinfo {author} {\bibfnamefont {Sina}\ \bibnamefont {Yeganeh}}, \bibinfo
  {author} {\bibfnamefont {Shane~R.}\ \bibnamefont {Yost}}, \bibinfo {author}
  {\bibfnamefont {Zhi-Qiang}\ \bibnamefont {You}}, \bibinfo {author}
  {\bibfnamefont {Igor~Ying}\ \bibnamefont {Zhang}}, \bibinfo {author}
  {\bibfnamefont {Xing}\ \bibnamefont {Zhang}}, \bibinfo {author}
  {\bibfnamefont {Yan}\ \bibnamefont {Zhao}}, \bibinfo {author} {\bibfnamefont
  {Bernard~R.}\ \bibnamefont {Brooks}}, \bibinfo {author} {\bibfnamefont
  {Garnet K.~L.}\ \bibnamefont {Chan}}, \bibinfo {author} {\bibfnamefont
  {Daniel~M.}\ \bibnamefont {Chipman}}, \bibinfo {author} {\bibfnamefont
  {Christopher~J.}\ \bibnamefont {Cramer}}, \bibinfo {author} {\bibfnamefont
  {William~A.}\ \bibnamefont {Goddard}}, \bibinfo {author} {\bibfnamefont
  {Mark~S.}\ \bibnamefont {Gordon}}, \bibinfo {author} {\bibfnamefont
  {Warren~J.}\ \bibnamefont {Hehre}}, \bibinfo {author} {\bibfnamefont
  {Andreas}\ \bibnamefont {Klamt}}, \bibinfo {author} {\bibfnamefont
  {Henry~F.}\ \bibnamefont {Schaefer}}, \bibinfo {author} {\bibfnamefont
  {Michael~W.}\ \bibnamefont {Schmidt}}, \bibinfo {author} {\bibfnamefont
  {C.~David}\ \bibnamefont {Sherrill}}, \bibinfo {author} {\bibfnamefont
  {Donald~G.}\ \bibnamefont {Truhlar}}, \bibinfo {author} {\bibfnamefont
  {Arieh}\ \bibnamefont {Warshel}}, \bibinfo {author} {\bibfnamefont {Xin}\
  \bibnamefont {Xu}}, \bibinfo {author} {\bibfnamefont
  {Al{\ifmmode\acute{a}\else\'{a}\fi}n}\ \bibnamefont {Aspuru-Guzik}}, \bibinfo
  {author} {\bibfnamefont {Roi}\ \bibnamefont {Baer}}, \bibinfo {author}
  {\bibfnamefont {Alexis~T.}\ \bibnamefont {Bell}}, \bibinfo {author}
  {\bibfnamefont {Nicholas~A.}\ \bibnamefont {Besley}}, \bibinfo {author}
  {\bibfnamefont {Jeng-Da}\ \bibnamefont {Chai}}, \bibinfo {author}
  {\bibfnamefont {Andreas}\ \bibnamefont {Dreuw}}, \bibinfo {author}
  {\bibfnamefont {Barry~D.}\ \bibnamefont {Dunietz}}, \bibinfo {author}
  {\bibfnamefont {Thomas~R.}\ \bibnamefont {Furlani}}, \bibinfo {author}
  {\bibfnamefont {Steven~R.}\ \bibnamefont {Gwaltney}}, \bibinfo {author}
  {\bibfnamefont {Chao-Ping}\ \bibnamefont {Hsu}}, \bibinfo {author}
  {\bibfnamefont {Yousung}\ \bibnamefont {Jung}}, \bibinfo {author}
  {\bibfnamefont {Jing}\ \bibnamefont {Kong}}, \bibinfo {author} {\bibfnamefont
  {Daniel~S.}\ \bibnamefont {Lambrecht}}, \bibinfo {author} {\bibfnamefont
  {WanZhen}\ \bibnamefont {Liang}}, \bibinfo {author} {\bibfnamefont
  {Christian}\ \bibnamefont {Ochsenfeld}}, \bibinfo {author} {\bibfnamefont
  {Vitaly~A.}\ \bibnamefont {Rassolov}}, \bibinfo {author} {\bibfnamefont
  {Lyudmila~V.}\ \bibnamefont {Slipchenko}}, \bibinfo {author} {\bibfnamefont
  {Joseph~E.}\ \bibnamefont {Subotnik}}, \bibinfo {author} {\bibfnamefont
  {Troy}\ \bibnamefont {Van~Voorhis}}, \bibinfo {author} {\bibfnamefont
  {John~M.}\ \bibnamefont {Herbert}}, \bibinfo {author} {\bibfnamefont
  {Anna~I.}\ \bibnamefont {Krylov}}, \bibinfo {author} {\bibfnamefont {Peter
  M.~W.}\ \bibnamefont {Gill}}, \ and\ \bibinfo {author} {\bibfnamefont
  {Martin}\ \bibnamefont {Head-Gordon}},\ }\bibfield  {title} {\enquote
  {\bibinfo {title} {{Advances in molecular quantum chemistry contained in the
  Q-Chem 4 program package}},}\ }\href {\doibase 10.1080/00268976.2014.952696}
  {\bibfield  {journal} {\bibinfo  {journal} {Mol. Phys.}\ }\textbf {\bibinfo
  {volume} {113}},\ \bibinfo {pages} {184--215} (\bibinfo {year}
  {2015})}\BibitemShut {NoStop}%
\bibitem [{\citenamefont {Epifanovsky}\ \emph {et~al.}(2021)\citenamefont
  {Epifanovsky}, \citenamefont {Gilbert}, \citenamefont {Feng}, \citenamefont
  {Lee}, \citenamefont {Mao}, \citenamefont {Mardirossian}, \citenamefont
  {Pokhilko}, \citenamefont {White}, \citenamefont {Coons}, \citenamefont
  {Dempwolff}, \citenamefont {Gan}, \citenamefont {Hait}, \citenamefont {Horn},
  \citenamefont {Jacobson}, \citenamefont {Kaliman}, \citenamefont {Kussmann},
  \citenamefont {Lange}, \citenamefont {Lao}, \citenamefont {Levine},
  \citenamefont {Liu}, \citenamefont {McKenzie}, \citenamefont {Morrison},
  \citenamefont {Nanda}, \citenamefont {Plasser}, \citenamefont {Rehn},
  \citenamefont {Vidal}, \citenamefont {You}, \citenamefont {Zhu},
  \citenamefont {Alam}, \citenamefont {Albrecht}, \citenamefont {Aldossary},
  \citenamefont {Alguire}, \citenamefont {Andersen}, \citenamefont {Athavale},
  \citenamefont {Barton}, \citenamefont {Begam}, \citenamefont {Behn},
  \citenamefont {Bellonzi}, \citenamefont {Bernard}, \citenamefont {Berquist},
  \citenamefont {Burton}, \citenamefont {Carreras}, \citenamefont
  {Carter-Fenk}, \citenamefont {Chakraborty}, \citenamefont {Chien},
  \citenamefont {Closser}, \citenamefont {Cofer-Shabica}, \citenamefont
  {Dasgupta}, \citenamefont {de~Wergifosse}, \citenamefont {Deng},
  \citenamefont {Diedenhofen}, \citenamefont {Do}, \citenamefont {Ehlert},
  \citenamefont {Fang}, \citenamefont {Fatehi}, \citenamefont {Feng},
  \citenamefont {Friedhoff}, \citenamefont {Gayvert}, \citenamefont {Ge},
  \citenamefont {Gidofalvi}, \citenamefont {Goldey}, \citenamefont {Gomes},
  \citenamefont {Gonz{\ifmmode\acute{a}\else\'{a}\fi}lez-Espinoza},
  \citenamefont {Gulania}, \citenamefont {Gunina}, \citenamefont
  {Hanson-Heine}, \citenamefont {Harbach}, \citenamefont {Hauser},
  \citenamefont {Herbst}, \citenamefont
  {Hern{\ifmmode\acute{a}\else\'{a}\fi}ndez~Vera}, \citenamefont {Hodecker},
  \citenamefont {Holden}, \citenamefont {Houck}, \citenamefont {Huang},
  \citenamefont {Hui}, \citenamefont {Huynh}, \citenamefont {Ivanov},
  \citenamefont {J{\ifmmode\acute{a}\else\'{a}\fi}sz}, \citenamefont {Ji},
  \citenamefont {Jiang}, \citenamefont {Kaduk}, \citenamefont
  {K{\ifmmode\ddot{a}\else\"{a}\fi}hler}, \citenamefont {Khistyaev},
  \citenamefont {Kim}, \citenamefont {Kis}, \citenamefont {Klunzinger},
  \citenamefont {Koczor-Benda}, \citenamefont {Koh}, \citenamefont {Kosenkov},
  \citenamefont {Koulias}, \citenamefont {Kowalczyk}, \citenamefont {Krauter},
  \citenamefont {Kue}, \citenamefont {Kunitsa}, \citenamefont {Kus},
  \citenamefont {Ladj{\ifmmode\acute{a}\else\'{a}\fi}nszki}, \citenamefont
  {Landau}, \citenamefont {Lawler}, \citenamefont {Lefrancois}, \citenamefont
  {Lehtola}, \citenamefont {Li}, \citenamefont {Li}, \citenamefont {Liang},
  \citenamefont {Liebenthal}, \citenamefont {Lin}, \citenamefont {Lin},
  \citenamefont {Liu}, \citenamefont {Liu}, \citenamefont {Loipersberger},
  \citenamefont {Luenser}, \citenamefont {Manjanath}, \citenamefont {Manohar},
  \citenamefont {Mansoor}, \citenamefont {Manzer}, \citenamefont {Mao},
  \citenamefont {Marenich}, \citenamefont {Markovich}, \citenamefont {Mason},
  \citenamefont {Maurer}, \citenamefont {McLaughlin}, \citenamefont {Menger},
  \citenamefont {Mewes}, \citenamefont {Mewes}, \citenamefont {Morgante},
  \citenamefont {Mullinax}, \citenamefont {Oosterbaan}, \citenamefont {Paran},
  \citenamefont {Paul}, \citenamefont {Paul}, \citenamefont
  {Pavo{\ifmmode\check{s}\else\v{s}\fi}evi{\ifmmode\acute{c}\else\'{c}\fi}},
  \citenamefont {Pei}, \citenamefont {Prager}, \citenamefont {Proynov},
  \citenamefont {R{\ifmmode\acute{a}\else\'{a}\fi}k}, \citenamefont
  {Ramos-Cordoba}, \citenamefont {Rana}, \citenamefont {Rask}, \citenamefont
  {Rettig}, \citenamefont {Richard}, \citenamefont {Rob}, \citenamefont
  {Rossomme}, \citenamefont {Scheele}, \citenamefont {Scheurer}, \citenamefont
  {Schneider}, \citenamefont {Sergueev}, \citenamefont {Sharada}, \citenamefont
  {Skomorowski}, \citenamefont {Small}, \citenamefont {Stein}, \citenamefont
  {Su}, \citenamefont {Sundstrom}, \citenamefont {Tao}, \citenamefont
  {Thirman}, \citenamefont {Tornai}, \citenamefont {Tsuchimochi}, \citenamefont
  {Tubman}, \citenamefont {Veccham}, \citenamefont {Vydrov}, \citenamefont
  {Wenzel}, \citenamefont {Witte}, \citenamefont {Yamada}, \citenamefont {Yao},
  \citenamefont {Yeganeh}, \citenamefont {Yost}, \citenamefont {Zech},
  \citenamefont {Zhang}, \citenamefont {Zhang}, \citenamefont {Zhang},
  \citenamefont {Zuev}, \citenamefont {Aspuru-Guzik}, \citenamefont {Bell},
  \citenamefont {Besley}, \citenamefont {Bravaya}, \citenamefont {Brooks},
  \citenamefont {Casanova}, \citenamefont {Chai}, \citenamefont {Coriani},
  \citenamefont {Cramer}, \citenamefont {Cserey}, \citenamefont {DePrince},
  \citenamefont {DiStasio}, \citenamefont {Dreuw}, \citenamefont {Dunietz},
  \citenamefont {Furlani}, \citenamefont {Goddard}, \citenamefont
  {Hammes-Schiffer}, \citenamefont {Head-Gordon}, \citenamefont {Hehre},
  \citenamefont {Hsu}, \citenamefont {Jagau}, \citenamefont {Jung},
  \citenamefont {Klamt}, \citenamefont {Kong}, \citenamefont {Lambrecht},
  \citenamefont {Liang}, \citenamefont {Mayhall}, \citenamefont {McCurdy},
  \citenamefont {Neaton}, \citenamefont {Ochsenfeld}, \citenamefont {Parkhill},
  \citenamefont {Peverati}, \citenamefont {Rassolov}, \citenamefont {Shao},
  \citenamefont {Slipchenko}, \citenamefont {Stauch}, \citenamefont {Steele},
  \citenamefont {Subotnik}, \citenamefont {Thom}, \citenamefont {Tkatchenko},
  \citenamefont {Truhlar}, \citenamefont {Van~Voorhis}, \citenamefont
  {Wesolowski}, \citenamefont {Whaley}, \citenamefont {Woodcock}, \citenamefont
  {Zimmerman}, \citenamefont {Faraji}, \citenamefont {Gill}, \citenamefont
  {Head-Gordon}, \citenamefont {Herbert},\ and\ \citenamefont
  {Krylov}}]{Epifanovsky2021Aug}%
  \BibitemOpen
  \bibfield  {author} {\bibinfo {author} {\bibfnamefont {Evgeny}\ \bibnamefont
  {Epifanovsky}}, \bibinfo {author} {\bibfnamefont {Andrew T.~B.}\ \bibnamefont
  {Gilbert}}, \bibinfo {author} {\bibfnamefont {Xintian}\ \bibnamefont {Feng}},
  \bibinfo {author} {\bibfnamefont {Joonho}\ \bibnamefont {Lee}}, \bibinfo
  {author} {\bibfnamefont {Yuezhi}\ \bibnamefont {Mao}}, \bibinfo {author}
  {\bibfnamefont {Narbe}\ \bibnamefont {Mardirossian}}, \bibinfo {author}
  {\bibfnamefont {Pavel}\ \bibnamefont {Pokhilko}}, \bibinfo {author}
  {\bibfnamefont {Alec~F.}\ \bibnamefont {White}}, \bibinfo {author}
  {\bibfnamefont {Marc~P.}\ \bibnamefont {Coons}}, \bibinfo {author}
  {\bibfnamefont {Adrian~L.}\ \bibnamefont {Dempwolff}}, \bibinfo {author}
  {\bibfnamefont {Zhengting}\ \bibnamefont {Gan}}, \bibinfo {author}
  {\bibfnamefont {Diptarka}\ \bibnamefont {Hait}}, \bibinfo {author}
  {\bibfnamefont {Paul~R.}\ \bibnamefont {Horn}}, \bibinfo {author}
  {\bibfnamefont {Leif~D.}\ \bibnamefont {Jacobson}}, \bibinfo {author}
  {\bibfnamefont {Ilya}\ \bibnamefont {Kaliman}}, \bibinfo {author}
  {\bibfnamefont {J{\ifmmode\ddot{o}\else\"{o}\fi}rg}\ \bibnamefont
  {Kussmann}}, \bibinfo {author} {\bibfnamefont {Adrian~W.}\ \bibnamefont
  {Lange}}, \bibinfo {author} {\bibfnamefont {Ka~Un}\ \bibnamefont {Lao}},
  \bibinfo {author} {\bibfnamefont {Daniel~S.}\ \bibnamefont {Levine}},
  \bibinfo {author} {\bibfnamefont {Jie}\ \bibnamefont {Liu}}, \bibinfo
  {author} {\bibfnamefont {Simon~C.}\ \bibnamefont {McKenzie}}, \bibinfo
  {author} {\bibfnamefont {Adrian~F.}\ \bibnamefont {Morrison}}, \bibinfo
  {author} {\bibfnamefont {Kaushik~D.}\ \bibnamefont {Nanda}}, \bibinfo
  {author} {\bibfnamefont {Felix}\ \bibnamefont {Plasser}}, \bibinfo {author}
  {\bibfnamefont {Dirk~R.}\ \bibnamefont {Rehn}}, \bibinfo {author}
  {\bibfnamefont {Marta~L.}\ \bibnamefont {Vidal}}, \bibinfo {author}
  {\bibfnamefont {Zhi-Qiang}\ \bibnamefont {You}}, \bibinfo {author}
  {\bibfnamefont {Ying}\ \bibnamefont {Zhu}}, \bibinfo {author} {\bibfnamefont
  {Bushra}\ \bibnamefont {Alam}}, \bibinfo {author} {\bibfnamefont
  {Benjamin~J.}\ \bibnamefont {Albrecht}}, \bibinfo {author} {\bibfnamefont
  {Abdulrahman}\ \bibnamefont {Aldossary}}, \bibinfo {author} {\bibfnamefont
  {Ethan}\ \bibnamefont {Alguire}}, \bibinfo {author} {\bibfnamefont
  {Josefine~H.}\ \bibnamefont {Andersen}}, \bibinfo {author} {\bibfnamefont
  {Vishikh}\ \bibnamefont {Athavale}}, \bibinfo {author} {\bibfnamefont
  {Dennis}\ \bibnamefont {Barton}}, \bibinfo {author} {\bibfnamefont {Khadiza}\
  \bibnamefont {Begam}}, \bibinfo {author} {\bibfnamefont {Andrew}\
  \bibnamefont {Behn}}, \bibinfo {author} {\bibfnamefont {Nicole}\ \bibnamefont
  {Bellonzi}}, \bibinfo {author} {\bibfnamefont {Yves~A.}\ \bibnamefont
  {Bernard}}, \bibinfo {author} {\bibfnamefont {Eric~J.}\ \bibnamefont
  {Berquist}}, \bibinfo {author} {\bibfnamefont {Hugh G.~A.}\ \bibnamefont
  {Burton}}, \bibinfo {author} {\bibfnamefont {Abel}\ \bibnamefont {Carreras}},
  \bibinfo {author} {\bibfnamefont {Kevin}\ \bibnamefont {Carter-Fenk}},
  \bibinfo {author} {\bibfnamefont {Romit}\ \bibnamefont {Chakraborty}},
  \bibinfo {author} {\bibfnamefont {Alan~D.}\ \bibnamefont {Chien}}, \bibinfo
  {author} {\bibfnamefont {Kristina~D.}\ \bibnamefont {Closser}}, \bibinfo
  {author} {\bibfnamefont {Vale}\ \bibnamefont {Cofer-Shabica}}, \bibinfo
  {author} {\bibfnamefont {Saswata}\ \bibnamefont {Dasgupta}}, \bibinfo
  {author} {\bibfnamefont {Marc}\ \bibnamefont {de~Wergifosse}}, \bibinfo
  {author} {\bibfnamefont {Jia}\ \bibnamefont {Deng}}, \bibinfo {author}
  {\bibfnamefont {Michael}\ \bibnamefont {Diedenhofen}}, \bibinfo {author}
  {\bibfnamefont {Hainam}\ \bibnamefont {Do}}, \bibinfo {author} {\bibfnamefont
  {Sebastian}\ \bibnamefont {Ehlert}}, \bibinfo {author} {\bibfnamefont
  {Po-Tung}\ \bibnamefont {Fang}}, \bibinfo {author} {\bibfnamefont {Shervin}\
  \bibnamefont {Fatehi}}, \bibinfo {author} {\bibfnamefont {Qingguo}\
  \bibnamefont {Feng}}, \bibinfo {author} {\bibfnamefont {Triet}\ \bibnamefont
  {Friedhoff}}, \bibinfo {author} {\bibfnamefont {James}\ \bibnamefont
  {Gayvert}}, \bibinfo {author} {\bibfnamefont {Qinghui}\ \bibnamefont {Ge}},
  \bibinfo {author} {\bibfnamefont {Gergely}\ \bibnamefont {Gidofalvi}},
  \bibinfo {author} {\bibfnamefont {Matthew}\ \bibnamefont {Goldey}}, \bibinfo
  {author} {\bibfnamefont {Joe}\ \bibnamefont {Gomes}}, \bibinfo {author}
  {\bibfnamefont {Cristina~E.}\ \bibnamefont
  {Gonz{\ifmmode\acute{a}\else\'{a}\fi}lez-Espinoza}}, \bibinfo {author}
  {\bibfnamefont {Sahil}\ \bibnamefont {Gulania}}, \bibinfo {author}
  {\bibfnamefont {Anastasia~O.}\ \bibnamefont {Gunina}}, \bibinfo {author}
  {\bibfnamefont {Magnus W.~D.}\ \bibnamefont {Hanson-Heine}}, \bibinfo
  {author} {\bibfnamefont {Phillip H.~P.}\ \bibnamefont {Harbach}}, \bibinfo
  {author} {\bibfnamefont {Andreas}\ \bibnamefont {Hauser}}, \bibinfo {author}
  {\bibfnamefont {Michael~F.}\ \bibnamefont {Herbst}}, \bibinfo {author}
  {\bibfnamefont {Mario}\ \bibnamefont
  {Hern{\ifmmode\acute{a}\else\'{a}\fi}ndez~Vera}}, \bibinfo {author}
  {\bibfnamefont {Manuel}\ \bibnamefont {Hodecker}}, \bibinfo {author}
  {\bibfnamefont {Zachary~C.}\ \bibnamefont {Holden}}, \bibinfo {author}
  {\bibfnamefont {Shannon}\ \bibnamefont {Houck}}, \bibinfo {author}
  {\bibfnamefont {Xunkun}\ \bibnamefont {Huang}}, \bibinfo {author}
  {\bibfnamefont {Kerwin}\ \bibnamefont {Hui}}, \bibinfo {author}
  {\bibfnamefont {Bang~C.}\ \bibnamefont {Huynh}}, \bibinfo {author}
  {\bibfnamefont {Maxim}\ \bibnamefont {Ivanov}}, \bibinfo {author}
  {\bibfnamefont
  {{\ifmmode\acute{A}\else\'{A}\fi}d{\ifmmode\acute{a}\else\'{a}\fi}m}\
  \bibnamefont {J{\ifmmode\acute{a}\else\'{a}\fi}sz}}, \bibinfo {author}
  {\bibfnamefont {Hyunjun}\ \bibnamefont {Ji}}, \bibinfo {author}
  {\bibfnamefont {Hanjie}\ \bibnamefont {Jiang}}, \bibinfo {author}
  {\bibfnamefont {Benjamin}\ \bibnamefont {Kaduk}}, \bibinfo {author}
  {\bibfnamefont {Sven}\ \bibnamefont {K{\ifmmode\ddot{a}\else\"{a}\fi}hler}},
  \bibinfo {author} {\bibfnamefont {Kirill}\ \bibnamefont {Khistyaev}},
  \bibinfo {author} {\bibfnamefont {Jaehoon}\ \bibnamefont {Kim}}, \bibinfo
  {author} {\bibfnamefont {Gergely}\ \bibnamefont {Kis}}, \bibinfo {author}
  {\bibfnamefont {Phil}\ \bibnamefont {Klunzinger}}, \bibinfo {author}
  {\bibfnamefont {Zsuzsanna}\ \bibnamefont {Koczor-Benda}}, \bibinfo {author}
  {\bibfnamefont {Joong~Hoon}\ \bibnamefont {Koh}}, \bibinfo {author}
  {\bibfnamefont {Dimitri}\ \bibnamefont {Kosenkov}}, \bibinfo {author}
  {\bibfnamefont {Laura}\ \bibnamefont {Koulias}}, \bibinfo {author}
  {\bibfnamefont {Tim}\ \bibnamefont {Kowalczyk}}, \bibinfo {author}
  {\bibfnamefont {Caroline~M.}\ \bibnamefont {Krauter}}, \bibinfo {author}
  {\bibfnamefont {Karl}\ \bibnamefont {Kue}}, \bibinfo {author} {\bibfnamefont
  {Alexander}\ \bibnamefont {Kunitsa}}, \bibinfo {author} {\bibfnamefont
  {Thomas}\ \bibnamefont {Kus}}, \bibinfo {author} {\bibfnamefont
  {Istv{\ifmmode\acute{a}\else\'{a}\fi}n}\ \bibnamefont
  {Ladj{\ifmmode\acute{a}\else\'{a}\fi}nszki}}, \bibinfo {author}
  {\bibfnamefont {Arie}\ \bibnamefont {Landau}}, \bibinfo {author}
  {\bibfnamefont {Keith~V.}\ \bibnamefont {Lawler}}, \bibinfo {author}
  {\bibfnamefont {Daniel}\ \bibnamefont {Lefrancois}}, \bibinfo {author}
  {\bibfnamefont {Susi}\ \bibnamefont {Lehtola}}, \bibinfo {author}
  {\bibfnamefont {Run~R.}\ \bibnamefont {Li}}, \bibinfo {author} {\bibfnamefont
  {Yi-Pei}\ \bibnamefont {Li}}, \bibinfo {author} {\bibfnamefont {Jiashu}\
  \bibnamefont {Liang}}, \bibinfo {author} {\bibfnamefont {Marcus}\
  \bibnamefont {Liebenthal}}, \bibinfo {author} {\bibfnamefont {Hung-Hsuan}\
  \bibnamefont {Lin}}, \bibinfo {author} {\bibfnamefont {You-Sheng}\
  \bibnamefont {Lin}}, \bibinfo {author} {\bibfnamefont {Fenglai}\ \bibnamefont
  {Liu}}, \bibinfo {author} {\bibfnamefont {Kuan-Yu}\ \bibnamefont {Liu}},
  \bibinfo {author} {\bibfnamefont {Matthias}\ \bibnamefont {Loipersberger}},
  \bibinfo {author} {\bibfnamefont {Arne}\ \bibnamefont {Luenser}}, \bibinfo
  {author} {\bibfnamefont {Aaditya}\ \bibnamefont {Manjanath}}, \bibinfo
  {author} {\bibfnamefont {Prashant}\ \bibnamefont {Manohar}}, \bibinfo
  {author} {\bibfnamefont {Erum}\ \bibnamefont {Mansoor}}, \bibinfo {author}
  {\bibfnamefont {Sam~F.}\ \bibnamefont {Manzer}}, \bibinfo {author}
  {\bibfnamefont {Shan-Ping}\ \bibnamefont {Mao}}, \bibinfo {author}
  {\bibfnamefont {Aleksandr~V.}\ \bibnamefont {Marenich}}, \bibinfo {author}
  {\bibfnamefont {Thomas}\ \bibnamefont {Markovich}}, \bibinfo {author}
  {\bibfnamefont {Stephen}\ \bibnamefont {Mason}}, \bibinfo {author}
  {\bibfnamefont {Simon~A.}\ \bibnamefont {Maurer}}, \bibinfo {author}
  {\bibfnamefont {Peter~F.}\ \bibnamefont {McLaughlin}}, \bibinfo {author}
  {\bibfnamefont {Maximilian F. S.~J.}\ \bibnamefont {Menger}}, \bibinfo
  {author} {\bibfnamefont {Jan-Michael}\ \bibnamefont {Mewes}}, \bibinfo
  {author} {\bibfnamefont {Stefanie~A.}\ \bibnamefont {Mewes}}, \bibinfo
  {author} {\bibfnamefont {Pierpaolo}\ \bibnamefont {Morgante}}, \bibinfo
  {author} {\bibfnamefont {J.~Wayne}\ \bibnamefont {Mullinax}}, \bibinfo
  {author} {\bibfnamefont {Katherine~J.}\ \bibnamefont {Oosterbaan}}, \bibinfo
  {author} {\bibfnamefont {Garrette}\ \bibnamefont {Paran}}, \bibinfo {author}
  {\bibfnamefont {Alexander~C.}\ \bibnamefont {Paul}}, \bibinfo {author}
  {\bibfnamefont {Suranjan~K.}\ \bibnamefont {Paul}}, \bibinfo {author}
  {\bibfnamefont {Fabijan}\ \bibnamefont
  {Pavo{\ifmmode\check{s}\else\v{s}\fi}evi{\ifmmode\acute{c}\else\'{c}\fi}}},
  \bibinfo {author} {\bibfnamefont {Zheng}\ \bibnamefont {Pei}}, \bibinfo
  {author} {\bibfnamefont {Stefan}\ \bibnamefont {Prager}}, \bibinfo {author}
  {\bibfnamefont {Emil~I.}\ \bibnamefont {Proynov}}, \bibinfo {author}
  {\bibfnamefont
  {{\ifmmode\acute{A}\else\'{A}\fi}d{\ifmmode\acute{a}\else\'{a}\fi}m}\
  \bibnamefont {R{\ifmmode\acute{a}\else\'{a}\fi}k}}, \bibinfo {author}
  {\bibfnamefont {Eloy}\ \bibnamefont {Ramos-Cordoba}}, \bibinfo {author}
  {\bibfnamefont {Bhaskar}\ \bibnamefont {Rana}}, \bibinfo {author}
  {\bibfnamefont {Alan~E.}\ \bibnamefont {Rask}}, \bibinfo {author}
  {\bibfnamefont {Adam}\ \bibnamefont {Rettig}}, \bibinfo {author}
  {\bibfnamefont {Ryan~M.}\ \bibnamefont {Richard}}, \bibinfo {author}
  {\bibfnamefont {Fazle}\ \bibnamefont {Rob}}, \bibinfo {author} {\bibfnamefont
  {Elliot}\ \bibnamefont {Rossomme}}, \bibinfo {author} {\bibfnamefont {Tarek}\
  \bibnamefont {Scheele}}, \bibinfo {author} {\bibfnamefont {Maximilian}\
  \bibnamefont {Scheurer}}, \bibinfo {author} {\bibfnamefont {Matthias}\
  \bibnamefont {Schneider}}, \bibinfo {author} {\bibfnamefont {Nickolai}\
  \bibnamefont {Sergueev}}, \bibinfo {author} {\bibfnamefont {Shaama~M.}\
  \bibnamefont {Sharada}}, \bibinfo {author} {\bibfnamefont {Wojciech}\
  \bibnamefont {Skomorowski}}, \bibinfo {author} {\bibfnamefont {David~W.}\
  \bibnamefont {Small}}, \bibinfo {author} {\bibfnamefont {Christopher~J.}\
  \bibnamefont {Stein}}, \bibinfo {author} {\bibfnamefont {Yu-Chuan}\
  \bibnamefont {Su}}, \bibinfo {author} {\bibfnamefont {Eric~J.}\ \bibnamefont
  {Sundstrom}}, \bibinfo {author} {\bibfnamefont {Zhen}\ \bibnamefont {Tao}},
  \bibinfo {author} {\bibfnamefont {Jonathan}\ \bibnamefont {Thirman}},
  \bibinfo {author} {\bibfnamefont {G{\ifmmode\acute{a}\else\'{a}\fi}bor~J.}\
  \bibnamefont {Tornai}}, \bibinfo {author} {\bibfnamefont {Takashi}\
  \bibnamefont {Tsuchimochi}}, \bibinfo {author} {\bibfnamefont {Norm~M.}\
  \bibnamefont {Tubman}}, \bibinfo {author} {\bibfnamefont {Srimukh~Prasad}\
  \bibnamefont {Veccham}}, \bibinfo {author} {\bibfnamefont {Oleg}\
  \bibnamefont {Vydrov}}, \bibinfo {author} {\bibfnamefont {Jan}\ \bibnamefont
  {Wenzel}}, \bibinfo {author} {\bibfnamefont {Jon}\ \bibnamefont {Witte}},
  \bibinfo {author} {\bibfnamefont {Atsushi}\ \bibnamefont {Yamada}}, \bibinfo
  {author} {\bibfnamefont {Kun}\ \bibnamefont {Yao}}, \bibinfo {author}
  {\bibfnamefont {Sina}\ \bibnamefont {Yeganeh}}, \bibinfo {author}
  {\bibfnamefont {Shane~R.}\ \bibnamefont {Yost}}, \bibinfo {author}
  {\bibfnamefont {Alexander}\ \bibnamefont {Zech}}, \bibinfo {author}
  {\bibfnamefont {Igor~Ying}\ \bibnamefont {Zhang}}, \bibinfo {author}
  {\bibfnamefont {Xing}\ \bibnamefont {Zhang}}, \bibinfo {author}
  {\bibfnamefont {Yu}~\bibnamefont {Zhang}}, \bibinfo {author} {\bibfnamefont
  {Dmitry}\ \bibnamefont {Zuev}}, \bibinfo {author} {\bibfnamefont
  {Al{\ifmmode\acute{a}\else\'{a}\fi}n}\ \bibnamefont {Aspuru-Guzik}}, \bibinfo
  {author} {\bibfnamefont {Alexis~T.}\ \bibnamefont {Bell}}, \bibinfo {author}
  {\bibfnamefont {Nicholas~A.}\ \bibnamefont {Besley}}, \bibinfo {author}
  {\bibfnamefont {Ksenia~B.}\ \bibnamefont {Bravaya}}, \bibinfo {author}
  {\bibfnamefont {Bernard~R.}\ \bibnamefont {Brooks}}, \bibinfo {author}
  {\bibfnamefont {David}\ \bibnamefont {Casanova}}, \bibinfo {author}
  {\bibfnamefont {Jeng-Da}\ \bibnamefont {Chai}}, \bibinfo {author}
  {\bibfnamefont {Sonia}\ \bibnamefont {Coriani}}, \bibinfo {author}
  {\bibfnamefont {Christopher~J.}\ \bibnamefont {Cramer}}, \bibinfo {author}
  {\bibfnamefont {Gy{\ifmmode\ddot{o}\else\"{o}\fi}rgy}\ \bibnamefont
  {Cserey}}, \bibinfo {author} {\bibfnamefont {A.~Eugene}\ \bibnamefont
  {DePrince}}, \bibinfo {author} {\bibfnamefont {Robert~A.}\ \bibnamefont
  {DiStasio}}, \bibinfo {author} {\bibfnamefont {Andreas}\ \bibnamefont
  {Dreuw}}, \bibinfo {author} {\bibfnamefont {Barry~D.}\ \bibnamefont
  {Dunietz}}, \bibinfo {author} {\bibfnamefont {Thomas~R.}\ \bibnamefont
  {Furlani}}, \bibinfo {author} {\bibfnamefont {William~A.}\ \bibnamefont
  {Goddard}}, \bibinfo {author} {\bibfnamefont {Sharon}\ \bibnamefont
  {Hammes-Schiffer}}, \bibinfo {author} {\bibfnamefont {Teresa}\ \bibnamefont
  {Head-Gordon}}, \bibinfo {author} {\bibfnamefont {Warren~J.}\ \bibnamefont
  {Hehre}}, \bibinfo {author} {\bibfnamefont {Chao-Ping}\ \bibnamefont {Hsu}},
  \bibinfo {author} {\bibfnamefont {Thomas-C.}\ \bibnamefont {Jagau}}, \bibinfo
  {author} {\bibfnamefont {Yousung}\ \bibnamefont {Jung}}, \bibinfo {author}
  {\bibfnamefont {Andreas}\ \bibnamefont {Klamt}}, \bibinfo {author}
  {\bibfnamefont {Jing}\ \bibnamefont {Kong}}, \bibinfo {author} {\bibfnamefont
  {Daniel~S.}\ \bibnamefont {Lambrecht}}, \bibinfo {author} {\bibfnamefont
  {WanZhen}\ \bibnamefont {Liang}}, \bibinfo {author} {\bibfnamefont
  {Nicholas~J.}\ \bibnamefont {Mayhall}}, \bibinfo {author} {\bibfnamefont
  {C.~William}\ \bibnamefont {McCurdy}}, \bibinfo {author} {\bibfnamefont
  {Jeffrey~B.}\ \bibnamefont {Neaton}}, \bibinfo {author} {\bibfnamefont
  {Christian}\ \bibnamefont {Ochsenfeld}}, \bibinfo {author} {\bibfnamefont
  {John~A.}\ \bibnamefont {Parkhill}}, \bibinfo {author} {\bibfnamefont
  {Roberto}\ \bibnamefont {Peverati}}, \bibinfo {author} {\bibfnamefont
  {Vitaly~A.}\ \bibnamefont {Rassolov}}, \bibinfo {author} {\bibfnamefont
  {Yihan}\ \bibnamefont {Shao}}, \bibinfo {author} {\bibfnamefont
  {Lyudmila~V.}\ \bibnamefont {Slipchenko}}, \bibinfo {author} {\bibfnamefont
  {Tim}\ \bibnamefont {Stauch}}, \bibinfo {author} {\bibfnamefont {Ryan~P.}\
  \bibnamefont {Steele}}, \bibinfo {author} {\bibfnamefont {Joseph~E.}\
  \bibnamefont {Subotnik}}, \bibinfo {author} {\bibfnamefont {Alex J.~W.}\
  \bibnamefont {Thom}}, \bibinfo {author} {\bibfnamefont {Alexandre}\
  \bibnamefont {Tkatchenko}}, \bibinfo {author} {\bibfnamefont {Donald~G.}\
  \bibnamefont {Truhlar}}, \bibinfo {author} {\bibfnamefont {Troy}\
  \bibnamefont {Van~Voorhis}}, \bibinfo {author} {\bibfnamefont {Tomasz~A.}\
  \bibnamefont {Wesolowski}}, \bibinfo {author} {\bibfnamefont {K.~Birgitta}\
  \bibnamefont {Whaley}}, \bibinfo {author} {\bibfnamefont {H.~Lee}\
  \bibnamefont {Woodcock}}, \bibinfo {author} {\bibfnamefont {Paul~M.}\
  \bibnamefont {Zimmerman}}, \bibinfo {author} {\bibfnamefont {Shirin}\
  \bibnamefont {Faraji}}, \bibinfo {author} {\bibfnamefont {Peter M.~W.}\
  \bibnamefont {Gill}}, \bibinfo {author} {\bibfnamefont {Martin}\ \bibnamefont
  {Head-Gordon}}, \bibinfo {author} {\bibfnamefont {John~M.}\ \bibnamefont
  {Herbert}}, \ and\ \bibinfo {author} {\bibfnamefont {Anna~I.}\ \bibnamefont
  {Krylov}},\ }\bibfield  {title} {\enquote {\bibinfo {title} {{Software for
  the frontiers of quantum chemistry: An overview of developments in the Q-Chem
  5 package}},}\ }\href {\doibase 10.1063/5.0055522} {\bibfield  {journal}
  {\bibinfo  {journal} {J. Chem. Phys.}\ }\textbf {\bibinfo {volume} {155}},\
  \bibinfo {pages} {084801} (\bibinfo {year} {2021})}\BibitemShut {NoStop}%
\bibitem [{\citenamefont {Lippert}\ \emph {et~al.}(1997)\citenamefont
  {Lippert}, \citenamefont {Hutter},\ and\ \citenamefont
  {Parrinello}}]{lippert1997hybrid}%
  \BibitemOpen
  \bibfield  {author} {\bibinfo {author} {\bibfnamefont {Gerald}\ \bibnamefont
  {Lippert}}, \bibinfo {author} {\bibfnamefont {J{\"u}rg}\ \bibnamefont
  {Hutter}}, \ and\ \bibinfo {author} {\bibfnamefont {Michele}\ \bibnamefont
  {Parrinello}},\ }\bibfield  {title} {\enquote {\bibinfo {title} {A hybrid
  gaussian and plane wave density functional scheme},}\ }\href@noop {}
  {\bibfield  {journal} {\bibinfo  {journal} {Mol. Phys.}\ }\textbf {\bibinfo
  {volume} {92}},\ \bibinfo {pages} {477--488} (\bibinfo {year}
  {1997})}\BibitemShut {NoStop}%
\bibitem [{\citenamefont {VandeVondele}\ \emph {et~al.}(2005)\citenamefont
  {VandeVondele}, \citenamefont {Krack}, \citenamefont {Mohamed}, \citenamefont
  {Parrinello}, \citenamefont {Chassaing},\ and\ \citenamefont
  {Hutter}}]{vandevondele2005quickstep}%
  \BibitemOpen
  \bibfield  {author} {\bibinfo {author} {\bibfnamefont {Joost}\ \bibnamefont
  {VandeVondele}}, \bibinfo {author} {\bibfnamefont {Matthias}\ \bibnamefont
  {Krack}}, \bibinfo {author} {\bibfnamefont {Fawzi}\ \bibnamefont {Mohamed}},
  \bibinfo {author} {\bibfnamefont {Michele}\ \bibnamefont {Parrinello}},
  \bibinfo {author} {\bibfnamefont {Thomas}\ \bibnamefont {Chassaing}}, \ and\
  \bibinfo {author} {\bibfnamefont {J{\"u}rg}\ \bibnamefont {Hutter}},\
  }\bibfield  {title} {\enquote {\bibinfo {title} {Quickstep: Fast and accurate
  density functional calculations using a mixed gaussian and plane waves
  approach},}\ }\href@noop {} {\bibfield  {journal} {\bibinfo  {journal}
  {Comput. Phys. Commun.}\ }\textbf {\bibinfo {volume} {167}},\ \bibinfo
  {pages} {103--128} (\bibinfo {year} {2005})}\BibitemShut {NoStop}%
\bibitem [{\citenamefont {F{\ifmmode\ddot{u}\else\"{u}\fi}sti-Molnar}\ and\
  \citenamefont {Pulay}(2002)}]{Fusti-Molnar2002May}%
  \BibitemOpen
  \bibfield  {author} {\bibinfo {author} {\bibfnamefont
  {L{\ifmmode\acute{a}\else\'{a}\fi}szl{\ifmmode\acute{o}\else\'{o}\fi}}\
  \bibnamefont {F{\ifmmode\ddot{u}\else\"{u}\fi}sti-Molnar}}\ and\ \bibinfo
  {author} {\bibfnamefont {Peter}\ \bibnamefont {Pulay}},\ }\bibfield  {title}
  {\enquote {\bibinfo {title} {{Accurate molecular integrals and energies using
  combined plane wave and Gaussian basis sets in molecular electronic structure
  theory}},}\ }\href {\doibase 10.1063/1.1467901} {\bibfield  {journal}
  {\bibinfo  {journal} {J. Chem. Phys.}\ }\textbf {\bibinfo {volume} {116}},\
  \bibinfo {pages} {7795--7805} (\bibinfo {year} {2002})}\BibitemShut {NoStop}%
\bibitem [{\citenamefont
  {F{\ifmmode\ddot{u}\else\"{u}\fi}sti-Moln{\ifmmode\acute{a}\else\'{a}\fi}r}\
  and\ \citenamefont {Pulay}(2002)}]{Fusti-Molnar2002Nov}%
  \BibitemOpen
  \bibfield  {author} {\bibinfo {author} {\bibfnamefont
  {L{\ifmmode\acute{a}\else\'{a}\fi}szl{\ifmmode\acute{o}\else\'{o}\fi}}\
  \bibnamefont
  {F{\ifmmode\ddot{u}\else\"{u}\fi}sti-Moln{\ifmmode\acute{a}\else\'{a}\fi}r}}\
  and\ \bibinfo {author} {\bibfnamefont {Peter}\ \bibnamefont {Pulay}},\
  }\bibfield  {title} {\enquote {\bibinfo {title} {{The Fourier transform
  Coulomb method: Efficient and accurate calculation of the Coulomb operator in
  a Gaussian basis}},}\ }\href {\doibase 10.1063/1.1510121} {\bibfield
  {journal} {\bibinfo  {journal} {J. Chem. Phys.}\ }\textbf {\bibinfo {volume}
  {117}},\ \bibinfo {pages} {7827--7835} (\bibinfo {year} {2002})}\BibitemShut
  {NoStop}%
\bibitem [{\citenamefont {Rozzi}\ \emph {et~al.}(2006)\citenamefont {Rozzi},
  \citenamefont {Varsano}, \citenamefont {Marini}, \citenamefont {Gross},\ and\
  \citenamefont {Rubio}}]{Rozzi2006May}%
  \BibitemOpen
  \bibfield  {author} {\bibinfo {author} {\bibfnamefont {Carlo~A.}\
  \bibnamefont {Rozzi}}, \bibinfo {author} {\bibfnamefont {Daniele}\
  \bibnamefont {Varsano}}, \bibinfo {author} {\bibfnamefont {Andrea}\
  \bibnamefont {Marini}}, \bibinfo {author} {\bibfnamefont {Eberhard K.~U.}\
  \bibnamefont {Gross}}, \ and\ \bibinfo {author} {\bibfnamefont {Angel}\
  \bibnamefont {Rubio}},\ }\bibfield  {title} {\enquote {\bibinfo {title}
  {{Exact Coulomb cutoff technique for supercell calculations}},}\ }\href
  {\doibase 10.1103/PhysRevB.73.205119} {\bibfield  {journal} {\bibinfo
  {journal} {Phys. Rev. B}\ }\textbf {\bibinfo {volume} {73}},\ \bibinfo
  {pages} {205119} (\bibinfo {year} {2006})}\BibitemShut {NoStop}%
\bibitem [{\citenamefont {Rom{\'a}n-P{\'e}rez}\ and\ \citenamefont
  {Soler}(2009)}]{roman2009efficient}%
  \BibitemOpen
  \bibfield  {author} {\bibinfo {author} {\bibfnamefont {Guillermo}\
  \bibnamefont {Rom{\'a}n-P{\'e}rez}}\ and\ \bibinfo {author} {\bibfnamefont
  {Jos{\'e}~M}\ \bibnamefont {Soler}},\ }\bibfield  {title} {\enquote {\bibinfo
  {title} {Efficient implementation of a van der waals density functional:
  application to double-wall carbon nanotubes},}\ }\href@noop {} {\bibfield
  {journal} {\bibinfo  {journal} {Phys. Rev. Lett.}\ }\textbf {\bibinfo
  {volume} {103}},\ \bibinfo {pages} {096102} (\bibinfo {year}
  {2009})}\BibitemShut {NoStop}%
\bibitem [{zen(2021)}]{zenodo}%
  \BibitemOpen
  \href {\doibase 10.5281/zenodo.5130020} {\enquote {\bibinfo {title} {Data
  repository for `approaching the basis set limit in gaussian-orbital-based
  periodic calculations with transferability: Performance of pure density
  functionals for simple semiconductors'},}\ } (\bibinfo {year}
  {2021})\BibitemShut {NoStop}%
\bibitem [{\citenamefont {Monkhorst}\ and\ \citenamefont
  {Pack}(1976)}]{monkhorst1976special}%
  \BibitemOpen
  \bibfield  {author} {\bibinfo {author} {\bibfnamefont {Hendrik~J}\
  \bibnamefont {Monkhorst}}\ and\ \bibinfo {author} {\bibfnamefont {James~D}\
  \bibnamefont {Pack}},\ }\bibfield  {title} {\enquote {\bibinfo {title}
  {Special points for brillouin-zone integrations},}\ }\href@noop {} {\bibfield
   {journal} {\bibinfo  {journal} {Phys. Rev. B}\ }\textbf {\bibinfo {volume}
  {13}},\ \bibinfo {pages} {5188} (\bibinfo {year} {1976})}\BibitemShut
  {NoStop}%
\bibitem [{\citenamefont {Sun}\ \emph {et~al.}(2020)\citenamefont {Sun},
  \citenamefont {Zhang}, \citenamefont {Banerjee}, \citenamefont {Bao},
  \citenamefont {Barbry}, \citenamefont {Blunt}, \citenamefont {Bogdanov},
  \citenamefont {Booth}, \citenamefont {Chen}, \citenamefont {Cui} \emph
  {et~al.}}]{sun2020recent}%
  \BibitemOpen
  \bibfield  {author} {\bibinfo {author} {\bibfnamefont {Qiming}\ \bibnamefont
  {Sun}}, \bibinfo {author} {\bibfnamefont {Xing}\ \bibnamefont {Zhang}},
  \bibinfo {author} {\bibfnamefont {Samragni}\ \bibnamefont {Banerjee}},
  \bibinfo {author} {\bibfnamefont {Peng}\ \bibnamefont {Bao}}, \bibinfo
  {author} {\bibfnamefont {Marc}\ \bibnamefont {Barbry}}, \bibinfo {author}
  {\bibfnamefont {Nick~S}\ \bibnamefont {Blunt}}, \bibinfo {author}
  {\bibfnamefont {Nikolay~A}\ \bibnamefont {Bogdanov}}, \bibinfo {author}
  {\bibfnamefont {George~H}\ \bibnamefont {Booth}}, \bibinfo {author}
  {\bibfnamefont {Jia}\ \bibnamefont {Chen}}, \bibinfo {author} {\bibfnamefont
  {Zhi-Hao}\ \bibnamefont {Cui}},  \emph {et~al.},\ }\bibfield  {title}
  {\enquote {\bibinfo {title} {Recent developments in the pyscf program
  package},}\ }\href@noop {} {\bibfield  {journal} {\bibinfo  {journal} {J.
  Chem. Phys.}\ }\textbf {\bibinfo {volume} {153}},\ \bibinfo {pages} {024109}
  (\bibinfo {year} {2020})}\BibitemShut {NoStop}%
\bibitem [{\citenamefont {Mardirossian}\ and\ \citenamefont
  {Head-Gordon}(2013)}]{mardirossian2013characterizing}%
  \BibitemOpen
  \bibfield  {author} {\bibinfo {author} {\bibfnamefont {Narbe}\ \bibnamefont
  {Mardirossian}}\ and\ \bibinfo {author} {\bibfnamefont {Martin}\ \bibnamefont
  {Head-Gordon}},\ }\bibfield  {title} {\enquote {\bibinfo {title}
  {Characterizing and understanding the remarkably slow basis set convergence
  of several minnesota density functionals for intermolecular interaction
  energies},}\ }\href@noop {} {\bibfield  {journal} {\bibinfo  {journal} {J.
  Chem. Theory Comput.}\ }\textbf {\bibinfo {volume} {9}},\ \bibinfo {pages}
  {4453--4461} (\bibinfo {year} {2013})}\BibitemShut {NoStop}%
\bibitem [{\citenamefont {Civalleri}\ \emph {et~al.}(2012)\citenamefont
  {Civalleri}, \citenamefont {Presti}, \citenamefont {Dovesi},\ and\
  \citenamefont {Savin}}]{Civalleri2012Oct}%
  \BibitemOpen
  \bibfield  {author} {\bibinfo {author} {\bibfnamefont {Bartolomeo}\
  \bibnamefont {Civalleri}}, \bibinfo {author} {\bibfnamefont {Davide}\
  \bibnamefont {Presti}}, \bibinfo {author} {\bibfnamefont {Roberto}\
  \bibnamefont {Dovesi}}, \ and\ \bibinfo {author} {\bibfnamefont {Andreas}\
  \bibnamefont {Savin}},\ }\bibfield  {title} {\enquote {\bibinfo {title} {{On
  choosing the best density functional approximation}},}\ }in\ \href {\doibase
  10.1039/9781849734790-00168} {\emph {\bibinfo {booktitle} {{Chemical
  Modelling: Applications and Theory Volume 9}}}},\ Vol.~\bibinfo {volume} {9}\
  (\bibinfo  {publisher} {The Royal Society of Chemistry},\ \bibinfo {year}
  {2012})\ pp.\ \bibinfo {pages} {168--185}\BibitemShut {NoStop}%
\bibitem [{\citenamefont {Ramberger}\ \emph {et~al.}(2019)\citenamefont
  {Ramberger}, \citenamefont {Sukurma}, \citenamefont {Sch{\"a}fer},\ and\
  \citenamefont {Kresse}}]{ramberger2019rpa}%
  \BibitemOpen
  \bibfield  {author} {\bibinfo {author} {\bibfnamefont {Benjamin}\
  \bibnamefont {Ramberger}}, \bibinfo {author} {\bibfnamefont {Zoran}\
  \bibnamefont {Sukurma}}, \bibinfo {author} {\bibfnamefont {Tobias}\
  \bibnamefont {Sch{\"a}fer}}, \ and\ \bibinfo {author} {\bibfnamefont {Georg}\
  \bibnamefont {Kresse}},\ }\bibfield  {title} {\enquote {\bibinfo {title} {Rpa
  natural orbitals and their application to post-hartree-fock electronic
  structure methods},}\ }\href@noop {} {\bibfield  {journal} {\bibinfo
  {journal} {J. Chem. Phys.}\ }\textbf {\bibinfo {volume} {151}},\ \bibinfo
  {pages} {214106} (\bibinfo {year} {2019})}\BibitemShut {NoStop}%
\bibitem [{\citenamefont {Irmler}\ \emph {et~al.}(2018)\citenamefont {Irmler},
  \citenamefont {Burow},\ and\ \citenamefont {Pauly}}]{irmler2018robust}%
  \BibitemOpen
  \bibfield  {author} {\bibinfo {author} {\bibfnamefont {Andreas}\ \bibnamefont
  {Irmler}}, \bibinfo {author} {\bibfnamefont
  {Asbj{\ifmmode\ddot{o}\else\"{o}\fi}rn}\ \bibnamefont {Burow}}, \ and\
  \bibinfo {author} {\bibfnamefont {Fabian}\ \bibnamefont {Pauly}},\ }\bibfield
   {title} {\enquote {\bibinfo {title} {{Robust Periodic Fock Exchange with
  Atom-Centered Gaussian Basis Sets}},}\ }\href {\doibase
  10.1021/acs.jctc.8b00122} {\bibfield  {journal} {\bibinfo  {journal} {J.
  Chem. Theory Comput.}\ }\textbf {\bibinfo {volume} {14}},\ \bibinfo {pages}
  {4567--4580} (\bibinfo {year} {2018})}\BibitemShut {NoStop}%
\bibitem [{\citenamefont {Gillan}\ \emph {et~al.}(2008)\citenamefont {Gillan},
  \citenamefont {Alf{\ifmmode\grave{e}\else\`{e}\fi}}, \citenamefont
  {de~Gironcoli},\ and\ \citenamefont {Manby}}]{Gillan2008Oct}%
  \BibitemOpen
  \bibfield  {author} {\bibinfo {author} {\bibfnamefont {M.~J.}\ \bibnamefont
  {Gillan}}, \bibinfo {author} {\bibfnamefont {D.}~\bibnamefont
  {Alf{\ifmmode\grave{e}\else\`{e}\fi}}}, \bibinfo {author} {\bibfnamefont
  {S.}~\bibnamefont {de~Gironcoli}}, \ and\ \bibinfo {author} {\bibfnamefont
  {F.~R.}\ \bibnamefont {Manby}},\ }\bibfield  {title} {\enquote {\bibinfo
  {title} {{High-precision calculation of Hartree-Fock energy of crystals}},}\
  }\href {\doibase 10.1002/jcc.21033} {\bibfield  {journal} {\bibinfo
  {journal} {J. Comput. Chem.}\ }\textbf {\bibinfo {volume} {29}},\ \bibinfo
  {pages} {2098--2106} (\bibinfo {year} {2008})}\BibitemShut {NoStop}%
\bibitem [{\citenamefont {Marsman}\ \emph {et~al.}(2009)\citenamefont
  {Marsman}, \citenamefont {Gr{\ifmmode\ddot{u}\else\"{u}\fi}neis},
  \citenamefont {Paier},\ and\ \citenamefont {Kresse}}]{Marsman2009May}%
  \BibitemOpen
  \bibfield  {author} {\bibinfo {author} {\bibfnamefont {M.}~\bibnamefont
  {Marsman}}, \bibinfo {author} {\bibfnamefont {A.}~\bibnamefont
  {Gr{\ifmmode\ddot{u}\else\"{u}\fi}neis}}, \bibinfo {author} {\bibfnamefont
  {J.}~\bibnamefont {Paier}}, \ and\ \bibinfo {author} {\bibfnamefont
  {G.}~\bibnamefont {Kresse}},\ }\bibfield  {title} {\enquote {\bibinfo {title}
  {{Second-order M{\o}ller{\textendash}Plesset perturbation theory applied to
  extended systems. I. Within the projector-augmented-wave formalism using a
  plane wave basis set}},}\ }\href {\doibase 10.1063/1.3126249} {\bibfield
  {journal} {\bibinfo  {journal} {J. Chem. Phys.}\ }\textbf {\bibinfo {volume}
  {130}},\ \bibinfo {pages} {184103} (\bibinfo {year} {2009})}\BibitemShut
  {NoStop}%
\bibitem [{\citenamefont {Paier}\ \emph {et~al.}(2009)\citenamefont {Paier},
  \citenamefont {Diaconu}, \citenamefont {Scuseria}, \citenamefont {Guidon},
  \citenamefont {VandeVondele},\ and\ \citenamefont {Hutter}}]{Paier2009Nov}%
  \BibitemOpen
  \bibfield  {author} {\bibinfo {author} {\bibfnamefont {Joachim}\ \bibnamefont
  {Paier}}, \bibinfo {author} {\bibfnamefont {Cristian~V.}\ \bibnamefont
  {Diaconu}}, \bibinfo {author} {\bibfnamefont {Gustavo~E.}\ \bibnamefont
  {Scuseria}}, \bibinfo {author} {\bibfnamefont {Manuel}\ \bibnamefont
  {Guidon}}, \bibinfo {author} {\bibfnamefont {Joost}\ \bibnamefont
  {VandeVondele}}, \ and\ \bibinfo {author} {\bibfnamefont
  {J{\ifmmode\ddot{u}\else\"{u}\fi}rg}\ \bibnamefont {Hutter}},\ }\bibfield
  {title} {\enquote {\bibinfo {title} {{Accurate Hartree-Fock energy of
  extended systems using large Gaussian basis sets}},}\ }\href {\doibase
  10.1103/PhysRevB.80.174114} {\bibfield  {journal} {\bibinfo  {journal} {Phys.
  Rev. B}\ }\textbf {\bibinfo {volume} {80}},\ \bibinfo {pages} {174114}
  (\bibinfo {year} {2009})}\BibitemShut {NoStop}%
\bibitem [{\citenamefont {Civalleri}\ \emph {et~al.}(2010)\citenamefont
  {Civalleri}, \citenamefont {Orlando}, \citenamefont {Zicovich-Wilson},
  \citenamefont {Roetti}, \citenamefont {Saunders}, \citenamefont {Pisani},\
  and\ \citenamefont {Dovesi}}]{Civalleri2010Mar}%
  \BibitemOpen
  \bibfield  {author} {\bibinfo {author} {\bibfnamefont {Bartolomeo}\
  \bibnamefont {Civalleri}}, \bibinfo {author} {\bibfnamefont {Roberto}\
  \bibnamefont {Orlando}}, \bibinfo {author} {\bibfnamefont {Claudio~M.}\
  \bibnamefont {Zicovich-Wilson}}, \bibinfo {author} {\bibfnamefont {Carla}\
  \bibnamefont {Roetti}}, \bibinfo {author} {\bibfnamefont {Victor~R.}\
  \bibnamefont {Saunders}}, \bibinfo {author} {\bibfnamefont {Cesare}\
  \bibnamefont {Pisani}}, \ and\ \bibinfo {author} {\bibfnamefont {Roberto}\
  \bibnamefont {Dovesi}},\ }\bibfield  {title} {\enquote {\bibinfo {title}
  {{Comment on ``Accurate Hartree-Fock energy of extended systems using large
  Gaussian basis sets''}},}\ }\href {\doibase 10.1103/PhysRevB.81.106101}
  {\bibfield  {journal} {\bibinfo  {journal} {Phys. Rev. B}\ }\textbf {\bibinfo
  {volume} {81}},\ \bibinfo {pages} {106101} (\bibinfo {year}
  {2010})}\BibitemShut {NoStop}%
\bibitem [{\citenamefont {Usvyat}\ \emph {et~al.}(2011)\citenamefont {Usvyat},
  \citenamefont {Civalleri}, \citenamefont {Maschio}, \citenamefont {Dovesi},
  \citenamefont {Pisani},\ and\ \citenamefont
  {Sch{\ifmmode\ddot{u}\else\"{u}\fi}tz}}]{Usvyat2011Jun}%
  \BibitemOpen
  \bibfield  {author} {\bibinfo {author} {\bibfnamefont {Denis}\ \bibnamefont
  {Usvyat}}, \bibinfo {author} {\bibfnamefont {Bartolomeo}\ \bibnamefont
  {Civalleri}}, \bibinfo {author} {\bibfnamefont {Lorenzo}\ \bibnamefont
  {Maschio}}, \bibinfo {author} {\bibfnamefont {Roberto}\ \bibnamefont
  {Dovesi}}, \bibinfo {author} {\bibfnamefont {Cesare}\ \bibnamefont {Pisani}},
  \ and\ \bibinfo {author} {\bibfnamefont {Martin}\ \bibnamefont
  {Sch{\ifmmode\ddot{u}\else\"{u}\fi}tz}},\ }\bibfield  {title} {\enquote
  {\bibinfo {title} {{Approaching the theoretical limit in periodic local MP2
  calculations with atomic-orbital basis sets: The case of LiH}},}\ }\href
  {\doibase 10.1063/1.3595514} {\bibfield  {journal} {\bibinfo  {journal} {J.
  Chem. Phys.}\ }\textbf {\bibinfo {volume} {134}},\ \bibinfo {pages} {214105}
  (\bibinfo {year} {2011})}\BibitemShut {NoStop}%
\end{thebibliography}%
\end{document}